\documentclass[12pt]{article}
\usepackage{amscd,amssymb,amsmath,latexsym,enumerate}
\usepackage[mathscr]{euscript}
\usepackage{mathrsfs}
\usepackage{epsfig}
\usepackage{fancybox}
\usepackage{verbatim}
\usepackage{tikz}
\usepackage{tikz-cd}
\usepackage{todonotes}
\usepackage{multicol}
\usepackage{graphicx}
\usepackage{graphbox}
\usepackage{mathtools}

\usepackage{color}

\usepackage{xcolor}
\definecolor{MyBlue}{cmyk}{1,0.13,0,0.63}
\definecolor{MyGreen}{cmyk}{0.91,0,0.88,0.52}
\newcommand{\mylinkcolor}{MyBlue}
\newcommand{\mycitecolor}{MyGreen}
\newcommand{\myurlcolor}{black}

\usepackage{hyperref}
\hypersetup{%
  bookmarksnumbered=true,bookmarksopen=false,%
  plainpages=false,
  linktocpage=true,%
  colorlinks=true,breaklinks=true,%
  linkcolor=\mylinkcolor,citecolor=\mycitecolor,urlcolor=\myurlcolor,%
  pdfpagelayout=OneColumn,%
  pageanchor=true,%
}

\title{Transfer matrix analysis of non-hermitian Hamiltonians: \\
asymptotic spectra and topological eigenvalues
} 

\author{Lars Koekenbier, Hermann Schulz-Baldes
\\
\\
{\small Friedrich-Alexander-Universit\"at Erlangen-N\"urnberg}
\\
{\small Department Mathematik, Cauerstr.~11, D-91058 Erlangen, Germany}
}

\date{ }

\newtheorem{theorem}{Theorem}
\newtheorem{proposition}[theorem]{Proposition}
\newtheorem{lemma}[theorem]{Lemma}
\newtheorem{corollary}[theorem]{Corollary}
\newtheorem{definition}[theorem]{Definition}
\newtheorem{remark}[theorem]{Remark}
\newtheorem{example}[theorem]{Example}

\newcommand{\BM}{{\mathbb B}}
\newcommand{\CM}{{\mathbb C}}
\newcommand{\NM}{{\mathbb N}}
\newcommand{\RM}{{\mathbb R}}

\newcommand{\TM}{{\mathbb T}}
\newcommand{\ZM}{{\mathbb Z}}

\newcommand{\EM}{{\mathbb E}}

\newcommand{\Ee}{{\cal E}}
\newcommand{\Pp}{{\cal P}}

\newcommand{\Dd}{{\cal D}}
\newcommand{\Ff}{{\cal F}}

\newcommand{\Ss}{{\cal S}}
\newcommand{\Oo}{{\cal O}}
\newcommand{\Tt}{{\cal T}}
\newcommand{\Rr}{{\cal R}}

\newcommand{\Mm}{{\cal M}}

\newcommand{\Ii}{{\cal I}}

\newcommand{\one}{{\bf 1}}

\newcommand{\spec}{\mbox{\rm spec}}

\newcommand{\Ch}{{\rm Ch}} 
\newcommand{\Log}{{\rm Log}} 
\newcommand{\Ind}{{\rm Ind}} 
\newcommand{\Ker}{{\rm Ker}} 
\newcommand{\Ran}{{\rm Ran}}

\newcommand{\Sig}{{\rm Sig}} 
\newcommand{\diag}{{\rm diag}} 
 
\newcommand{\Wind}{{\rm Wind}}

\newcommand{\ess}{{\mbox{\rm\tiny ess}}}

\newcommand{\HR}{H^R}
\newcommand{\HRs}{H^{s,R}}
\newcommand{\HLs}{H^{s,L}}
\newcommand{\HL}{H^L}
\newcommand{\HTR}{\widetilde{H}^R}
\newcommand{\HTL}{\widetilde{H}^L}
\newcommand{\sca}{s}
\newcommand{\scar}{r}
\newcommand{\IndSet}{I}
\newcommand{\QFct}{Q}
\newcommand{\LimitSet}{\Lambda}
\newcommand{\Outliers}{\Gamma}

\begin{document}

\maketitle

\begin{abstract}
Transfer matrix techniques are used to {provide} a new proof of  Widom's results on the asymptotic spectral theory of finite block Toeplitz matrices. Furthermore, a  rigorous treatment of {the} skin effect, spectral outliers, the generalized Brillouin zone and the bulk-boundary correspondence in such systems is given. This covers chiral Hamiltonians with topological eigenvalues close to zero, but no line-gap.
\hfill {MSC2020:} 15B05, 19K56, 81Q99






\end{abstract}



\vspace{1cm}

\section{Overview}
\label{sec-Intro}

This paper is about the spectral theory of translation invariant block-tridiagonal operators of the form
\begin{equation}
\label{eq-HamGen}
H
\;=\;
\begin{pmatrix}
\ddots & \ddots & & & 
\\
\ddots & V & T & &
\\
& R & \ddots & \ddots &
\\
& & \ddots & \ddots  &
\end{pmatrix}
\;,
\end{equation}
where $R,V,T\in \CM^{L\times L}$ are all $L\times L$ complex matrices. This non-hermitian Hamiltonian $H$ is viewed as a bounded operator on the Hilbert space $\ell^2(\ZM,\CM^L)$. It is a widely studied model in non-hermitian quantum mechanics \cite{HN,YM,AGU} in which, more specifically, the skin effect is discussed \cite{YM,ZYF,ZZLC} as well as {the} existence of robust topological eigenvalues via a so-called {bulk-boundary correspondence} \cite{YW,VJP,PS,YZFH,ZYF,KSUS,OKSS,BKS,BBK,PeS}. These latter two physical effects are studied on finite, but large volume compressions $H_N$ of $H$, namely the restrictions of $H$ to $\ell^2(\{1,\ldots,N\},\CM^L)\cong\CM^{NL}$. It is well-known that the spectra of $H$ and $H_N$ have little in common, and also differ from the spectrum of the (right) half-space restriction $\HR$ of $H$ to $\ell^2(\NM,\CM^L)$. The operator $H$ is often also called a Laurent operator, $\HR$ a Toeplitz operator and $H_N$ a finite Toeplitz matrix, also a compression or section of $H$.

\vspace{.2cm}

There is an extensive mathematical literature, including the books  \cite{GF,BS,HRS,Nic1,BG,Sim,Nic2}, about the spectra of Toeplitz operators and matrices. These sources mainly deal with general (fully occupied) Toeplitz operators with focus on the scalar case $L=1$, but also cover some results on the block case $L\geq 2$. Reference \cite{BG} provides a detailed study of banded scalar Toeplitz matrices, a case that can be cast in the tridiagonal block form \eqref{eq-HamGen} and is hence recovered by the present analysis (see Example~\ref{ex-ScalarBand} below). Topological non-hermitian systems, however, can only be described by the general matrix-valued case $L\geq 2$. The asymptotic spectral theory of $H_N$ with $L\geq 2$ was studied 50 years ago by Widom in the breakthrough work \cite{Wid1}, but there seem to be only few further developments since (let us mention Delvaux's work \cite{Del} though; also remarkable is Widom's work on Szego asymptotics of the Toeplitz determinants \cite{Wid2} with numerous follow-ups nicely described in \cite{BBE}, but this does not seem to be directly relevant for the asymptotic spectral theory). 

\vspace{.2cm}

This work provides an elementary approach based on transfer matrix methods which are well-known for selfadjoint block Jacobi matrices. For the scalar case $L=1$, there are countless contributions by both the physics and mathematics community, starting with work of Dyson, Schmidt and others in the 1950's. The block Jacobi case $L\geq 2$ is heavily used in the solid state physics community where it is at the basis for numerical studies of Anderson localization, see the review \cite{KMK}. This is tightly connected to the field of products of (symplectic) random matrices, {\it e.g.} \cite{BL}. Important elements of the mathematical theory (restricted to the hermitian case) are matrix-valued Sturm-Liouville-type oscillation theory \cite{SB1} and Weyl theory \cite{SB0} (these works contain references to various earlier contributions, in particular, from the Russian school). These techniques also transpose to block CMV matrices \cite{DPS}  and the slightly more general scattering zippers \cite{MS}. As to transfer matrix techniques for the non-hermitian case \eqref{eq-HamGen}, there is the work \cite{GK} for the scalar case $L=1$, but for $L\geq 2$, we are only aware of the non-rigorous works of the physics community \cite{KEBB,Lie,KSUS,YM,KD,OKSS,BB}.

\vspace{.2cm}

In this work transfer matrix methods are implemented for non-hermitian block Toeplitz matrices. They lead to new self-contained proofs of Widom's main results on asymptotic spectra  \cite{Wid1}, albeit under somewhat different hypotheses. The arguments also invoke Widom's determinant formula (which is re-derived below), but it is then used to explicitly construct eigenfunctions of the finite volume operators $H_N$ rather than appealing to potential theory.  Along the way it is also shown that non-hermitian Bloch theory is essentially a rediscovery of Widom's results by several theoretical physicists \cite{YW,KD,YM}. Furthermore, a new criterion for topological zero-energy eigenvalues (bound states) is derived for Hamiltonians having a chiral symmetry. This is rooted in an interplay of Widom's results with index-theoretic statements (essentially only the well-known Noether-Gohberg-Krein index theorem). In particular, this allows to show that such chiral Hamiltonians can have topological approximate zero modes (namely eigenvalues close to $0$) without having a line-gap, a situation that to the best of our knowledge is not covered by the physics literature. Summing up, we hope that this work opens up an alternative perspective for mathematicians interested in the spectral theory of Toeplitz matrices, and on the other hand brings some of the existing deep mathematical results to the attention of the theoretical physics community.

\vspace{.2cm}

Let us now start with a short overview of the main results. Crucial is the following hypothesis, that is assumed throughout the whole text.

\vspace{.2cm}

\noindent {\bf Hypothesis A:} {\it Both $T$ and $R$ are invertible matrices.} 

\vspace{.2cm}

\noindent It allows for the construction of the $2L\times 2L$ transfer matrix at complex energy $E\in\CM$ by
\begin{equation}
\label{eq-TraDef}
\Tt^E
\;=\;
\begin{pmatrix}
(E\,\one-V)T^{-1} & - R \\ T^{-1} & 0 
\end{pmatrix}
\;,
\end{equation}
and implies that $\Tt^E$ is invertible. In Section~\ref{sec-Operators} it is recalled how the transfer matrix allows for the construction of solutions of the eigenvalue equations for $H$, $\HR$ and $H_N$. Of crucial importance are the eigenvalues $z_1(E),\ldots,z_{2L}(E)$ of $\Tt^E$, ordered such that $|z_l(E)|\leq |z_{l+1}(E)|$ for $l=1,\ldots,2L-1$. Degeneracies are collected in the set
\begin{equation}
\label{eq-FDef}
\Ff
\;=\;
\big\{E\in\CM\,:\,{\Tt}^E\mbox{ has a degenerate eigenvalue}\big\}
\;.
\end{equation}

\vspace{.1cm}

\noindent {\bf Hypothesis B:} {\it The set $\mathcal{F}$ is assumed to be finite.}

\vspace{.2cm}

\noindent The spectra of $\Tt^E$ furthermore allow for the introduction of the following two subsets of $\CM$:
\begin{align}
\Sigma
& 
\;=\;
\big\{E\in\CM\,:\,{\Tt}^E\mbox{ has at least one eigenvalue of modulus }1\big\}
\;,
\label{eq-SigmaDef}
\\
\LimitSet 
&
\;=\;
\big\{E\in\CM\,:\,\mbox{middle two eigenvalues }z_L(E),\,z_{L+1}(E)\;\mbox{ of }{\Tt}^E\mbox{ have the same modulus}\big\} .
\label{eq-LambdaDef}
\end{align}
It is relatively elementary to see that $\Sigma=\spec(H)$ is precisely the spectrum of the Laurent operator $H$. Theorem~\ref{thm-LimitSpecIntro} below will show that $\LimitSet$ is the attractor of the bulk states of $H_N$, under suitable further hypotheses. 

\vspace{.2cm}

\noindent {\bf Hypothesis C:} {\it $|z_{L-1}(E)|<|z_L(E)|=|z_{L+1}(E)|<|z_{L+2}(E)|$ for all but finitely many $E \in \partial\LimitSet$.}

\vspace{.2cm}

\noindent Here $\partial\LimitSet$ denotes the boundary of $\LimitSet$. It is shown in Sections~\ref{sec-Subsets} and~\ref{sec-SpectraLimitFinite} that Hypothesis C implies that $\LimitSet=\partial\LimitSet$ and that it is given by finite union of analytic arcs. To formulate the last hypothesis, consider the sets $I_0=\{1,\ldots,L\}$ and $I_1=\{1,\ldots,L-1,L+1\}$ and let ${\Rr}_{\IndSet_j}^E$ denote the Riesz projection of $\Tt^E$ onto the $L$ eigenvalues $z_l(E)$ with $l\in \IndSet_j$ where $j=0,1$. (It is irrelevant in the sequel that the definition is ambiguous for $E\in\LimitSet$; the notation ${\Rr}_{\IndSet_j}^E$ instead of ${\Rr}_{j}^E$ may seem awkward and clumsy at this point, but fits nicely in the general framework developed below.) Then introduce
$$
q_{\IndSet_j}(E)
\;=\;
{\det}_L\Big(
\binom{0}{\one}^*
{\Rr}_{\IndSet_j}^E
\binom{\one}{0}\Big)
\;,
\qquad
E\in\CM\setminus\Ff
\;.
$$

\noindent {\bf Hypothesis D:} {\it The functions $q_{\IndSet_0}$ and $q_{\IndSet_1}$ are not constantly zero locally near points in $\partial\LimitSet \setminus \mathcal{F}$.} 

\vspace{.2cm}

\noindent  Finally let us introduce the discrete set:
\begin{align}
\Outliers
\;=\;
\big\{E\in\CM \setminus \LimitSet  \,:\,q_{\IndSet_0}(E)=0  \big\}
\;.
\label{eq-OutliersDef}
\end{align}
It is referred to as the set of spectral outliers. The next theorem is the main result about spectral asymptotics of the $H_N$, essentially contained in \cite{Wid1} because it is shown in Section~\ref{sec-WidomRevisited} that Hypothesis C implies Condition A and B in \cite{Wid1}.

\begin{theorem}
\label{thm-LimitSpecIntro}
Suppose that {\rm Hypotheses A, B, C} and {\rm D} hold. Then for all $E\in\partial\LimitSet\cup\Outliers$ but a finite number of $E$, there exist $E_N\in\spec(H_N)$ such that $\lim_{N\to\infty}E_N=E$. 
\end{theorem}

\begin{figure}
\centering
\includegraphics[width=8.3cm,height=8.3cm]{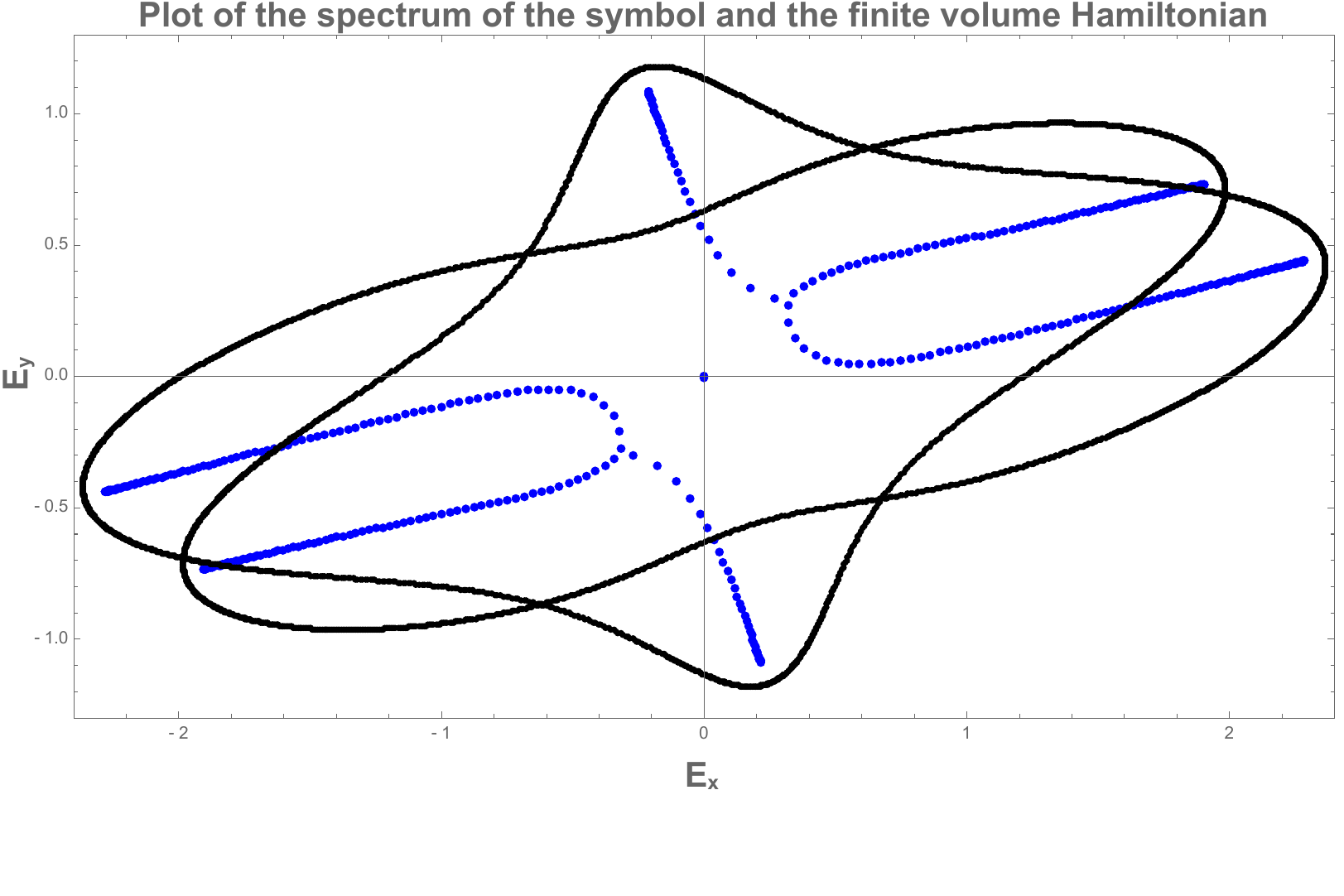}
\hspace{.1cm}
\includegraphics[width=8.3cm,height=8.3cm]{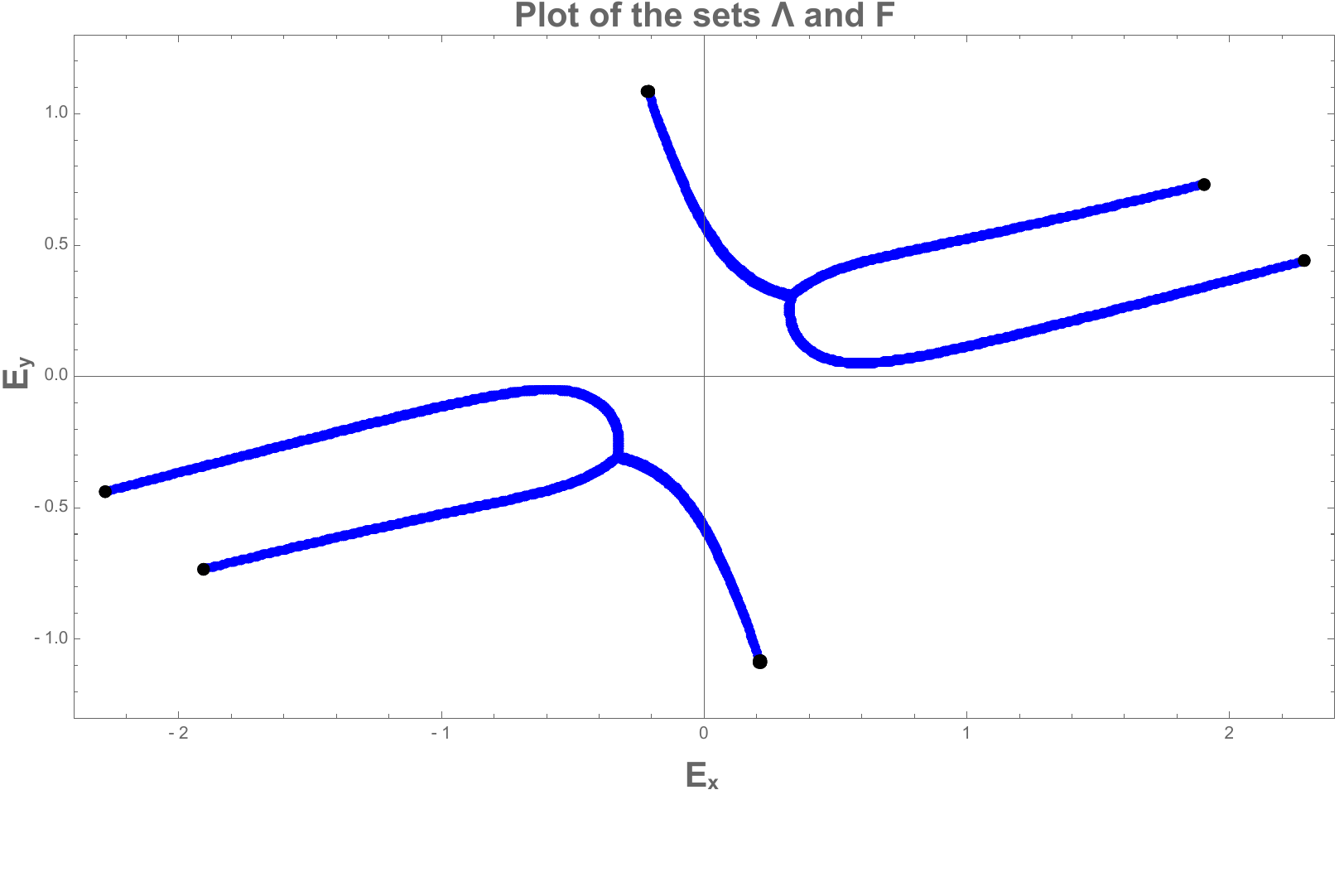}
\vspace{-1.8cm}
\caption{\sl 
Plots for the same model and parameters described in {\rm Figs.~\ref{fig-Ser2} and \ref{fig-Ser2bis}}, except $s=1.4$. On the left, the black curve shows $\Sigma$ and in blue are the eigenvalues $E = E_x + \imath E_y$, $E_x , E_y \in \mathbb{R}$, of $H_N$ for $N=200$. The right plot shows the set $\LimitSet$. There is a twice degenerate approximate zero mode visible in the left plot, as predicted by {\rm Theorem~\ref{thm-ChiralIntro}}.
}
\label{fig-Ser2Intro}
\end{figure}

Section~\ref{sec-SpectraLimitFinite} contains a more precise formulation of this result. It also provides an example of a Hamiltonian for which Hypotheses C and D are violated and the set $\LimitSet$ has non-empty interior. Such non-generic models typically have some symmetries leading to degeneracies in the spectrum of $\Tt^E$. Figure~\ref{fig-Ser2Intro} illustrates Theorem~\ref{thm-LimitSpecIntro}. It contains numerical plots for a Hamiltonian $H$ with $L=2$ that is described in detail in Section~\ref{sec-TopoZero}. Here the focus is on the qualitative aspects. The continuous curve in the left plot is the spectrum $\Sigma=\spec(H)$ of the periodic operator, and the dots show the eigenvalues of $H_N$ for $N=200$ computed with the Mathematica eigenvalue solver (there are thus $400$ points, counting multiplicities). The right plot shows that set $\LimitSet$ computed from the $4\times 4$ transfer matrix by implementing \eqref{eq-LambdaDef}, together with bigger dots showing the points in $\Ff$ (at the ends of the arcs). One readily observes the good agreement of $\LimitSet$ with $\spec(H_N)$ as predicted by Theorem~\ref{thm-LimitSpecIntro}. Further examples are dealt with in Section~\ref{sec-Applications} which also contains a discussion of the skin effect and the generalized Brillouin zone.

\vspace{.2cm}

A more careful inspection of Figure~\ref{fig-Ser2Intro} also shows that there is an eigenvalue of $H_N$ close to $0$, actually there are two such eigenvalues. These are the approximate zero modes advertised above. In fact, the Hamiltonian in question has a chiral symmetry. For even $L$, such a symmetry is implemented by the matrix $K=\binom{\one \;\;\; 0}{0 \; -\one}$ with blocks of size $\frac{L}{2}$. Then $H$ is said to have a chiral symmetry if $KRK=-R$, $KVK=-V$ and $KTK=-T$. The symmetry immediately implies that $\Sigma=-\Sigma$, $\LimitSet=-\LimitSet$ and $\Outliers=-\Outliers$, which is  also visible in Figure~\ref{fig-Ser2Intro}. Furthermore, one can readily check that $R$, $V$ and $T$ are block off-diagonal in the grading of $K$ and we will denote the upper right entries by $R_+$, $V_+$ and $T_+$ and the lower left ones by $R_-$, $V_-$ and $T_-$. The following result gives a criterion for $0\in\Outliers$ involving two winding numbers associated to merely these latter six matrices (in physics jargon, bulk invariants). It is proved in Section~\ref{sec-ZeroOutlier}. By Theorem~\ref{thm-LimitSpecIntro} this then implies that $H_N$ has an eigenvalue near $0$, hence establishing a bulk-boundary correspondence.

\begin{theorem}
\label{thm-ChiralIntro}
Let  $H$ be chiral and satisfy {\rm Hypothesis A}. Further let $W_\pm(s)\in\ZM$ denote the winding numbers of $z\in\TM \mapsto \det\big((sz)^{-1}R_\pm+V_\pm+(sz)T_\pm\big)$.  If $W_+(s)=-W_-(s)\not=0$ for some $s>0$, then $0\in\Outliers$. 
\end{theorem}

To apply the criterion, one first has to search for an $s>0$ such that $W_+(s)=-W_-(s)$, and then check that this number is non-vanishing. This is carried out in Section~\ref{sec-TopoZero}. The criterion in Theorem~\ref{thm-ChiralIntro} differs from that in other works mentioned above, in particular, also the one in the supplementary material of \cite{OKSS}. Let us also stress that $\Sigma=\spec(H)$ may encircle $0$ so that, again in physics jargon, the system has a point gap, but no line gap. Therefore standard index-theoretic arguments assuring the existence of zero modes \cite{PS,KSUS} do not apply in this situation. Once available, these zero modes can potentially be used for technological innovations, {\it e.g.} trapped light modes \cite{OPA} or topological sensors \cite{BB}.

\vspace{.2cm}

The paper is organized as follows. In Section~\ref{sec-Operators} transfer matrices and their spectra are introduced. Section~\ref{sec-IndexTh} is essentially index-theoretic, but also already establishes links to the Riesz projections. Then in Section~\ref{sec-WidomRevisited} the sets $\LimitSet$ and $\Outliers$ are analyzed in detail and Theorems~\ref{thm-LimitSpecIntro} and~\ref{thm-ChiralIntro} are proven. Finally Section~\ref{sec-Applications} offers a physics perspective on these results and provides varies numerical examples.

\vspace{.2cm}

\noindent {\bf Acknowledgements:} The authors thank Carolyn Echter, Flore Kunst and Tom Stoiber for useful and inspiring discussions at an early stage of this work. This work was supported by the DFG grant SCHU 1358/8-1.


\section{Set-up and basic properties}
\label{sec-Operators}

In this section, many of the objects are introduced and some basic properties are described.

\subsection{Hamiltonians and their symbols}
\label{sec-Hamiltonians}

Throughout this text, the non-hermitian Hamiltonian $H$ on the Hilbert space $\ell^2(\ZM,\CM^L)$ is the Laurent operator given by \eqref{eq-HamGen}. It is hence specified by the three $L\times L$ matrices $R$, $V$ and $T$. The operator $H$ is self-adjoint if and only if $V^*=V$ and $R=T^*$. The finite volume approximation of $H$ on $\ell^2(\{1,\ldots,N\},\CM^L)$ will be denoted by $H_N\in\CM^{NL\times NL}$ (in the mathematics literature, $H_N$ is called a section or a compression of $H$). Moreover, $\HR$ and $\HL$ are the right and left half-space restrictions of $H$ to the half-line Hilbert spaces $\ell^2(\NM,\CM^L)$ and $\ell^2(\NM_\leq,\CM^L)$ respectively. The operators $\HR$ and $\HL$ are called Toeplitz operators in the mathematics literature. Note that $\ell^2(\ZM,\CM^L)=\ell^2(\NM_\leq,\CM^L)\oplus \ell^2(\NM,\CM^L)$ and that one has
\begin{equation}
\label{eq-HamDiff}
H
\,-\,\HL\oplus\HR
\;=\;
|1\rangle \,R\,\langle 0|\;+\;|0\rangle \,T\,\langle 1|
\;,
\end{equation}
where the Dirac bra-ket notation was used for partial isometries $|n\rangle:\CM^L\to\ell^2(\ZM,\CM^L)$ onto the matrix degrees of freedom over the lattice sites $n\in\ZM$. In particular, \eqref{eq-HamDiff} implies that $\HL\oplus\HR$ is a compact perturbation of $H$. The next example shows under which conditions banded scalar Laurent and Toeplitz operators fit into the present framework.

\begin{example}
\label{ex-ScalarBand}
{\rm
A Laurent operator with scalar entries and finite band width can be cast into the form \eqref{eq-HamGen} and satisfy Hypothesis A, provided it is balanced with as many upper as lower diagonals. More precisely, let $t_{-L},t_{1-L},\ldots, t_{L-1},t_L\in\CM$ with $t_{\pm L}\not=0$ and consider the Hamiltonian $H$ on $\ell^2(\ZM)$ with entries $H_{l,k}=\langle l|H|k\rangle=\chi(|l-k|\leq L)\,t_{l-k}$ where $\chi$ denotes the indicator function. Then set
$$
R
=
\begin{pmatrix}
t_L & t_{L-1}  & \cdots & t_1
\\
0 & t_L & &  \vdots
\\
\vdots &  & \ddots & t_{L-1}
\\
0 & \cdots & 0 &  t_L
\end{pmatrix}
\,,
\;\;
V
=
\begin{pmatrix}
t_0 & t_{-1}  & \cdots & t_{-L}
\\
t_1 & t_0 & &  \vdots
\\
\vdots &  & \ddots & t_{-1}
\\
t_{L+1} & \cdots & t_1 &  t_0
\end{pmatrix}
\,,
\;\;
T
=
\begin{pmatrix}
t_{-L} & 0  & \cdots & 0
\\
t_{1-L} & t_{-L} & &  \vdots
\\
\vdots &  & \ddots & 0
\\
t_{-1} & \cdots & t_{1-L} &  t_{-L}
\end{pmatrix}
\,.
$$
Hence clearly $R$ and $T$ are invertible due to $t_{\pm L}\not=0$.
}
\hfill $\diamond$
\end{example}

\begin{example}
\label{ex-ScalarPeriodic}
{\rm
A special class of periodic scalar banded operators $H$ can also be dealt with using Hamiltonians of the form \eqref{eq-HamGen}. Suppose that the matrix entries $H_{l,k}=\langle l|H|k\rangle$ satisfy $H_{l+p,k+p}=H_{l,k}$ for all $k,l\in\ZM$ and for an integer $p$ called the period. If, moreover, $H_{l,k}=0$ for $|k-l|>p$, that is, the band width is $2p+1$, then one can set
$$
R
=
\begin{pmatrix}
H_{1,-p+1} & H_{1,-p+2}  & \cdots & H_{1,0}
\\
0 & H_{2,-p+2} & &  \vdots
\\
\vdots &  & \ddots & H_{p-1,0}
\\
0 & \cdots & 0 &  H_{p,0}
\end{pmatrix}
\,,
\;\;
\;\;
T
=
\begin{pmatrix}
H_{1,p+1} & 0  & \cdots & 0
\\
H_{2,p+1} & H_{2,p+2} & &  \vdots
\\
\vdots &  & \ddots & 0
\\
H_{p,p+1} & \cdots & H_{p,2p } &  H_{p,2p+1}
\end{pmatrix}
\,.
$$
If furthermore $V$ is simply the compression of $H$ onto the sites $\{1,\ldots,p\}$, then $H$ coincides with \eqref{eq-HamGen}. If all the diagonal entries of $R$ and $T$  are non-vanishing, then Hypothesis A is satisfied and the transfer matrix methods can be used. Note that the entries $H_{l,k}$ may also be square matrices instead of scalars, without altering the reasoning above. This example may seem somewhat particular because the band width and periodicity have to coincide. It is, however, possible to deal with more general periodic banded matrices by applying the reduced transfer matrices similarly as in \cite{DC,Sad,KD,SB2}. This will be developed elsewhere.
} 
\hfill $\diamond$
\end{example}

The symbol of $H$ is a matrix-valued meromorphic function given by
\begin{equation}
\label{eq-SymbDef}
H(z)
\;=\;
R\,z^{-1}\,+\,V\,+\,T\,z
\;,
\end{equation}
and furthermore the associated symbol is defined by
\begin{equation}
\label{eq-AssSymbDef}
\widetilde{H}(z)
\;=\;
H(z^{-1})
\;.
\end{equation}
Let us stress that these definitions differ from a large part of the literature where $H$ and $\widetilde{H}$ are exchanged \cite{Wid1,Wid2,BS,BG,HRS,Del}. The reason for the modification is that it leads to nice consistent notations in connection with the transfer matrix, most notably in \eqref{eq-TransferResol} and \eqref{eq-TransferDet} below. The associated symbol also specifies an operator $\widetilde{H}$ which is defined as in \eqref{eq-HamGen}, but with $R$ and $T$ exchanged. One then also has corresponding half-space operators $\HTR$ and $\HTL$, as well as finite volume restrictions $\widetilde{H}_N$. Using a spatial flip implemented by a selfadjoint unitary, one sees that $\widetilde{H}$ and $\widetilde{H}_N$ are unitarily equivalent to $H$ and $H_N$ respectively. This does not apply to the half-sided operator though. Indeed, consider the unitary $W:\ell^2(\NM_\leq,\CM^L)\to \ell^2(\NM,\CM^L)$ given by $W|-n\rangle=|n+1\rangle$ for $n\geq 0$, for which one then has 
\begin{equation}
\label{eq-HRHL}
W\HL W^*
\;=\;
\HTR
\;,
\qquad
W\HR W^*
\;=\;
\HTL
\;.
\end{equation}
Also the adjoint $H^*$ will be of relevance and its associated symbol will be denoted by
$$
\widehat{H}(z)
\;=\;
T^*\,z^{-1}\,+\,V^*\,+\,R^*\,z
\;.
$$
For selfadjoint $H$, one has $\widehat{H}(z)=H(z)$. For the half-sided operators, one then checks
\begin{equation}
\label{eq-HRHL2}
\widehat{H}^R\;=\;(\HR)^*\;=\;(H^*)^R\;,
\qquad
\widehat{H}^L\;=\;(\HL)^*\;=\;(H^*)^L
\;.
\end{equation}

\vspace{.2cm}

Of crucial importance will be the resolvent of the symbol $E\in\CM\mapsto (H(z)-E\,\one)^{-1}$ and the associated symbol $E\in\CM\mapsto (\widetilde{H}(z)-E\,\one)^{-1}$ which are meromorphic functions in $z$ and $E$ with values in $\CM^{L\times L}$.  The determinants
\begin{equation}
\label{eq-DetSymb}
D(z,E)\;=\;{\det}_L(H(z)-E\,\one)
\;,
\qquad
\widetilde{D}(z,E)\;=\;{\det}_L(\widetilde{H}(z)-E\,\one)
\end{equation}
can be factorized as
$$
D(z,E)
\;=\;
z^{-L}\,d(z,E)
\;,
\qquad
\widetilde{D}(z,E)
\;=\;
z^{-L}\,\widetilde{d}(z,E)
\;.
$$
where 
$$
d(z,E)\;=\;{\det}_L(z\,H(z)-z\,E\,\one)
\;,
\qquad
\widetilde{d}(z,E)\;=\;{\det}_L(z\,\widetilde{H}(z)-z\,E\,\one)
\;,
$$ 
are polynomials in $z$ of degree $2L$.

\subsection{Spectra of periodic (Laurent) operators}
\label{sec-SpectraPeriodic}

The spectrum of the periodic operator $H$ can readily be computed by the discrete Fourier transform $\Ff:\ell^2(\ZM)\to L^2(\TM)$ where $\TM\subset\CM$ is the unit circle equipped with the Lebesgue measure, trivially extended to the vector-valued case. It shows that the spectrum $\Sigma=\spec(H)$ of $H$ satisfies
\begin{equation}
\label{eq-SigmaSpec}
\Sigma
\;=\;
\bigcup_{z\in\TM}
\,\spec(H(z))
\;.
\end{equation}
As $z\in\TM\mapsto H(z)\in\CM^{L\times L}$ is a real analytic function (with an analytic extension to a neighborhood of $\TM$ in $\CM$), analytic perturbation theory implies that all eigenvalues of $H(z)$ are real analytic in $z$ except at level crossings, where the branches can be chosen analytic in the $L$th root (Puiseux expansion) \cite{Kat}. In conclusion, $\spec(H)$ is given by a set of closed curves in $\CM$.

\subsection{Transfer matrices}
\label{sec-Transfer}

Another standard element of the theory of hermitian block Jacobi operators is the transfer matrix. It allows for the construction of the formal solutions of the Sch\"odinger equation $H\psi=E\psi$ for a complex energy $E$ and a state $\psi=(\psi_n)_{n\in\ZM}$ with $\psi_n\in\CM^L$ which is not necessarily in the Hilbert space. For the periodic operator $H$ given in \eqref{eq-HamGen} the Schr\"odinger equation is
$$
T\psi_{n+1}\,+\,V\psi_n\,+\,R\psi_{n-1}\;=\;E\psi_n
\;,
\qquad
n\in\ZM
\;.
$$
As $T$ is invertible (see Hypothesis A), this can be rewritten as
\begin{equation}
\label{eq-SchrTra}
\begin{pmatrix}
T\psi_{n+1} \\ \psi_{n}
\end{pmatrix}
\;=\;
\begin{pmatrix}
(E\,\one-V)T^{-1} & - R \\ T^{-1} & 0 
\end{pmatrix}
\begin{pmatrix}
T\psi_n \\ \psi_{n-1}
\end{pmatrix}
\;,
\end{equation}
where the second line of this equation is tautological. Therefore one introduces the transfer matrix at complex energy $E$ by \eqref{eq-TraDef} which can also be factorized as
\begin{equation}
\label{eq-TraDef2}
\Tt^E
\;=\;
\begin{pmatrix}
E\,\one-V & - \one \\ \one & 0 
\end{pmatrix}
\begin{pmatrix}
T^{-1} & 0 \\ 0 & R 
\end{pmatrix}
\;.
\end{equation}
Then one can iterate \eqref{eq-SchrTra} to obtain
\begin{equation}
\label{eq-SchrTra2}
\begin{pmatrix}
T\psi_{n+1} \\ \psi_{n}
\end{pmatrix}
\;=\;
(\Tt^E)^n\begin{pmatrix}
T\psi_1 \\ \psi_{0}
\end{pmatrix}
\;.
\end{equation}
If $T$ is not invertible, one can construct a reduced transfer matrix of size $2r$ where $r$ is the rank of $T$ \cite{KD}. For sake of simplicity, this is not carried out here, but let us stress that this allows to analyze many more Laurent and Toeplitz operators of finite band width by transfer matrix techniques. Note that $\det_{2L}(\Tt^E)=\det_L(T)^{-1}\det_L(R)$ so that the transfer matrix is invertible for all $E\in\CM$ as also $R$ is invertible (see Hypothesis A). The inverse of $\Tt^E$ is explicitly given by
$$
(\Tt^E)^{-1}
\;=\;
\begin{pmatrix}
T & 0 \\ 0 & R^{-1} 
\end{pmatrix}
\begin{pmatrix}
0 & \one \\ - \one &E\,\one-V 
\end{pmatrix}
\;=\;
\begin{pmatrix}
0 & T \\
-R^{-1}& R^{-1}(E\,\one-V) 
\end{pmatrix}
\;.
$$

\vspace{.2cm}

Associated to $\widetilde{H}(z)$ and $\widehat{H}(z)$, there are two further naturally associated transfer matrices:
$$
\widetilde{\Tt}^E\;=\;
\begin{pmatrix}
(E\,\one-V)R^{-1} & - T \\ R^{-1} & 0 
\end{pmatrix}
\;,
\qquad
\widehat{\Tt}^E\;=\;
\begin{pmatrix}
(E\,\one-V^*)(R^*)^{-1} & - T^* \\ (R^*)^{-1} & 0 
\end{pmatrix}
\;.
$$
Straightforward algebra shows:
\begin{equation}
\label{eq-Iunitarity}
(\widehat{\Tt}^{\overline{E}})^*
\,\Ii\,
\Tt^E
\;=\;
\Ii
\;,
\qquad
\Ii
\;=\;
\begin{pmatrix}
0 & -\one\\ \one & 0
\end{pmatrix}
\;.
\end{equation}
Note that this implies $(\Tt^E)^{-1}=\Ii^*(\widehat{\Tt}^{\overline{E}})^*\Ii$. 
If the Hamiltonian is hermitian and the energy $E\in\RM$ is real, the transfer matrix thus satisfies the $\Ii$-unitary relation $\Tt^*\Ii\Tt=\Ii$.  The above identity is hence the non-hermitian analogue of the $\Ii$-unitarity relation. There is yet another identity that will be used below, namely one readily checks
\begin{equation}
\label{eq-TransferConjugate}
\begin{pmatrix}
0 & R \\ T^{-1} & 0
\end{pmatrix}^{-1}
\widetilde{\Tt}^E
\begin{pmatrix}
0 & R \\ T^{-1} & 0
\end{pmatrix}
\;=\;
(\Tt^E)^{-1}
\;.
\end{equation}

\vspace{.2cm}

Of great importance will be the resolvent of the transfer matrix. It can be computed by the Schur complement formula:
\begin{align*}
(\Tt^E\,-\,z\,\one)^{-1}
&
\;=\;
\begin{pmatrix}
(E\,\one-V)T^{-1} \,-\,z\,\one & - R \\ T^{-1} & \,-\,z\,\one 
\end{pmatrix}^{-1}
\\
&
\;=\;
\begin{pmatrix}
S^{-1} & -z^{-1}\,S^{-1}\,R \\
z^{-1}\,T^{-1}\,S^{-1} & -z^{-1}\,\one-z^{-2} \,T^{-1}\,S^{-1}\,R
\end{pmatrix}
\end{align*}
where 
$$
S
\;=\;
(E\,\one-V)T^{-1}\,-\,z\,\one\,-\,z^{-1}\,R\,T^{-1}
\;=\;
\big(E\,\one\,-\,{H}(z)\big)\,T^{-1}
\;.
$$
Replacing the latter condition shows
\begin{align}
&
(\Tt^E\,-\,z\,\one)^{-1}
\nonumber
\\
&
\;=\;
\begin{pmatrix}
T & 0 \\ 0 & \one
\end{pmatrix}
\begin{pmatrix}
\big(E\,\one\,-\,{H}(z)\big)^{-1} & -z^{-1}\,\big(E\,\one\,-\,{H}(z)\big)^{-1}
\\
z^{-1}\,\big(E\,\one\,-\,{H}(z)\big)^{-1} & -\,z^{-1}\,R^{-1}\,-\, z^{-2}\,\big(E\,\one\,-\,{H}(z)\big)^{-1}
\end{pmatrix}
\begin{pmatrix}
\one & 0 \\ 0 & R
\end{pmatrix}
\;.
\label{eq-TransferResol}
\end{align}

Next let us compute the characteristic polynomical of the transfer matrix. Recalling the identity
$$
\det\begin{pmatrix} A & B \\ C & D \end{pmatrix}
\;=\;
\det(D)\det(A-BD^{-1}C)
\;,
$$
one finds
\begin{align}
{\det}_{2L}(\Tt^E-z\,\one)
&
\;=\;
(-z)^L\,{\det}_L\big( (E-V)T^{-1}-z\,\one\,-\,z^{-1}RT^{-1}\big)
\nonumber
\\
&
\;=\;
z^L\,{\det}_L(T)^{-1}\,{\det}_L(V-E\,\one+zT+z^{-1}R)
\nonumber
\\
&
\;=\;
z^L\,{\det}_L(T)^{-1}\,{\det}_L(H(z)-E\,\one)
\label{eq-TransferDet}
\\
&\;=\;
z^L\,{\det}_L(T)^{-1}\,{D}(z,E)
\nonumber
\\
&\;=\;
{\det}_L(T)^{-1}\,{d}(z,E)
\;.
\nonumber
\end{align}
This identity also follows by taking the determinant of \eqref{eq-TransferResol}. This shows that $\Sigma=\spec(H)$ is indeed given by both \eqref{eq-SigmaDef} and \eqref{eq-SigmaSpec}. Together with \eqref{eq-TransferConjugate}, one also deduces
\begin{equation}
\label{eq-TransferDet2}
{\det}_{2L}(\Tt^E-z\,\one)
\;=\;
{\det}_{2L}\big((\widetilde{\Tt}^E )^{-1}-z\,\one\big)
\;.
\end{equation}
%

\subsection{Spectral propertes of transfer matrices}
\label{sec-SpectraTransfer}

Let $z_1(E),\ldots,z_{2L}(E)$ be the $2L$ eigenvalues of $\Tt^E$, listed with their multiplicity and ordered such that $|z_l(E)|\leq |z_{l+1}(E)|$. As they are the zeros of characteristic polynomial and $\Tt^E$ is invertible, it follows from \eqref{eq-TransferDet} that they are also the zeros of $z\mapsto D(z,E)$ (also of  $z\mapsto d(z,E)$). Similarly, the eigenvalues $\widetilde{z}_1(E),\ldots,\widetilde{z}_{2L}(E)$ of $\widetilde{\Tt}^E$, ordered such that $|\widetilde{z}_l(E)|\leq |\widetilde{z}_{l+1}(E)|$, are the zeros of $z\mapsto \widetilde{D}(z,E)$. From \eqref{eq-TransferDet2}, one then deduces the equalities
\begin{equation}
\label{eq-zSym}
\widetilde{z}_k(E)
\;=\;
\frac{1}{z_{2L+1-k}(E)}
\;.
\end{equation}
In particular,
\begin{equation}
\label{eq-TransferConjugateSpec}
\spec(\widetilde{\Tt}^E)
\;=\;
\spec(\Tt^E)^{-1}
\;,
\end{equation}
which also follows directly from \eqref{eq-TransferConjugate}. Moreover, it follows from identity \eqref{eq-Iunitarity} that
$$
\spec(\Tt^E)^{-1}
\;=\;
\overline{\spec(\widehat{\Tt}^{\overline{E}})}
\;.
$$
If the Hamiltonian is hermitian and the energy $E\in\RM$ is real, then the spectrum satisfies $\spec(\Tt^E)^{-1}=\overline{\spec(\Tt^E)}$. At several instances it will be useful to diagonalize $\Tt^E$. Of course, this is not possible for all $E$ as, in general, there may be Jordan blocks at level crossings. If all Jordan blocks are trivial, let the basis change $\Mm^E$ be chosen such that
\begin{equation}
\label{eq-MBasisChange}
(\Mm^E)^{-1}\Tt^E\Mm^E
\;=\;
\diag\big({z}_1(E),\ldots,{z}_{2L}(E)\big)
\;=\;
\diag\big(\widetilde{z}_{2L}(E)^{-1},\ldots,\widetilde{z}_{1}(E)^{-1}\big)
\;.
\end{equation}
Recall that $\Ff\subset\CM$ defined in \eqref{eq-FDef} consists of the set of $E\in\CM$ for which $\Tt^E$ has a degenerate eigenvalue. Away from the set $\Ff$, diagonalization is always possible with a matrix $\Mm^E$ which is unique up to normalization (right multiplication by an invertible diagonal matrix).

\begin{lemma}
\label{lem-FinateF}
$\mathcal{F}$ is either finite or $\mathcal{F}=\CM$.
\end{lemma}

\noindent {\bf Proof.}
Recall that the eigenvalues of $\Tt^E$ are precisely the zeros of the polynomial $p_E(z)=d (z,E) = {\det}_L (z(H(z)-E\one))$. Further recall that the polynomial $p_E$ has zeros of order greater than one if and only if the resultant $r_E$ of $p_E$ and its derivative $p'_E$ is zero. The resultant is the determinant of the Sylvester matrix and hence a polynomial expression in the coefficients of the polynomials $p_E$ and its derivative $p'_E$ which in turn are polynomials in $E$. Therefore the resultant $r_E$ is also polynomial in $E$. It is thus either zero for all or just for a finite set of $E$.
\hfill $\Box$

\vspace{.2cm}

Hypothesis B states that we only consider the first of the two cases in Lemma~\ref{lem-FinateF}.

\begin{lemma}
\label{lem-ConstantModulus}
Let $z_i (E)$ and $z_j (E)$ be two continuous choices of eigenvalues of ${\Tt}^E$. If $|z_i (E)|=|z_j (E)|$ for all $E$ in some open subset of $\CM$, then there exists a scalar $c\in\TM $ such that $z_i (E)=c\,z_j (E)$ for all $E\in\CM$.
\end{lemma}

\noindent {\bf Proof.}
Note that the $z_i (E)$ are the zeros of the polynomial $p_E (z) = {\det}_L (z(H(z)-E\one))$. Since zeros of polynomials are continuous in the coefficients of the polynomial and the coefficients of $p_E (z)$ are polynomial in $E$, the $z_i (E)$ can indeed be chosen continuous in $E$. Moreover, for such a choice the $z_i (E)$ are well-known to be holomorphic in $E$ on $\CM\setminus V$, where $V$ is some branch cut \cite{Kat}. 

\vspace{.1cm}

Let now $U\subset\CM$ be the open subset on which $|z_i (E)|=|z_j (E)|$. Then $z_i (E) / z_j (E)$ is holomorphic on $\CM \setminus V$ and has constant modulus equal to $1$ on the open non-empty set $U \setminus V$. By the maximum modulus principle it follows that $z_i (E) / z_j (E)$ is constant on $\CM \setminus V$, notably there is a scalar $c\in\TM $ such that $z_i (E)=c\,z_j (E)$ for all $E\in\CM\setminus V$. By continuity this also extends to $V$.
\hfill $\Box$

\section{Spectra of half-sided (Toeplitz) operators}
\label{sec-IndexTh}

This section collects results on the spectrum of the half-sided (Toeplitz) operators $\HR$ and $\HL$. This involves two technical elements: the Noether-Gohberg-Krein index theorem and the study of eigenvalues using the intersection theory of the boundary conditions with the contracting directions of the transfer matrices.

\subsection{Winding number and index theorem}
\label{sec-FredHalfSided}

According to \eqref{eq-HamDiff}, the direct sum $\HR\oplus\HL$ differs from $H$ only by a compact operator. Hence $\Sigma=\spec(H)\subset\spec(\HR)\cup\spec(\HL)$ by a Weyl sequence argument (which is feasible because $\Sigma=\spec(H)$ has empty interior). Furthermore, a direct consequence of analytic Fredholm theory ({\it e.g.} Appendix D in \cite{SB}) shows that the components of $\CM\setminus\Sigma$ either only contain discrete spectrum of $\HR\oplus\HL$ or each point in $\CM\setminus\Sigma$ is an eigenvalue of $\HR\oplus\HL$ (each of these eigenvalues are eigenvalues of either $\HR$ or $\HL$, but not necessarily both). Which one of the two cases applies can be read off the winding number of the determinant of the symbol shifted by a complex energy $E\in\CM\setminus\Sigma$, more precisely defined by
$$
\Wind^E(H)
\;=\;
\int_{\TM}\frac{dz}{2\pi\imath}\;\partial_z\;\log\big({\det}_L(H(z)-E\,\one)\big)
\;.
$$
Due to \eqref{eq-TransferDet} and \eqref{eq-TransferResol}, this is also given by 
\begin{align}
\Wind^E(H)
&
\;=\;
\int_{\TM}\frac{dz}{2\pi\imath}\;\partial_z\;\log\big(z^{-L}\,{\det}_L(R)\,{\det}_{2L}({\Tt}^E-z\,\one)\big)
\;=\;
-L\;+\;\sum_{|z|<1}\,m_z({\Tt}^E)
\label{eq-WindComp}
\;,
\end{align}
where $m_z(\Tt)$ denotes the algebraic multiplicity of $z$ as eigenvalue of $\Tt$ so that the sums are actually finite.  Indeed, a classical theorem (due to Noether, Gohberg and Krein and holding even for continuous symbols, see {\it e.g.} Theorem~6.5 in \cite{BS} where there is a different sign due to the altered definition \eqref{eq-SymbDef} of the symbol) is the following: 

\begin{theorem}
\label{theo-NoetherGoberg}
For all $E\in\CM$, 
\begin{equation}
\label{eq-FredChar}
\HR-E\,\one\;\;\mbox{\rm Fredholm }
\quad\Longleftrightarrow\quad
{\det}_L(H(z)-E\,\one)\,\not=\,0\quad\mbox{\rm for all }z\in\TM 
\;,
\end{equation}
and, if this holds, its index is given by
\begin{equation}
\label{eq-FredIndWind}
\Ind(\HR-E\,\one)\;=\;\Wind^E(H)
\;.
\end{equation}
\end{theorem}

This shows that $\Sigma=\spec(H)$ is actually equal to the essential spectrum $\spec_\ess(\HR)$ of $\HR$ (which by definition consists of all $E$ such that $\HR-E\,\one$ is a Fredholm operator, so we do {\it not} use Weyl's notion of essential spectrum here). Moreover,  $\Sigma=\spec(H)$ consists of curves in $\CM$ that split $\CM$ in several connected components; picking a point $E$ in one of these components, one has a winding number $\Wind^E(H)$; if it is non-zero, then the whole component of $\CM\setminus\Sigma$ belongs to the spectrum of $\HR$; if the winding vanishes, one has to analyze the spectrum in that component by other means; the unbounded component of $\CM\setminus\Sigma$ will never be in the essential spectrum of $\HR$ though. Thus:

\begin{corollary}
\label{coro-NoetherGoberg}
One has
$$
\spec(\HR)
\;\supset\;
\Sigma \cup \big\{E\in\CM\setminus\Sigma\,:\,\Wind^E(H)\not =0\big\}
\;.
$$
\end{corollary}

The same arguments apply to $\HTR=W\HL W^*$. Moreover, \eqref{eq-FredChar} and \eqref{eq-zSym} also imply
\begin{align}
\label{eq-FredPropEquiv}
\HR-E\,\one\;\;\mbox{\rm Fredholm }
\quad\Longleftrightarrow\quad
\HTR-E\,\one\;\;\mbox{\rm Fredholm }
\quad\Longleftrightarrow\quad
\HL-E\,\one\;\;\mbox{\rm Fredholm }
.
\end{align}
As 
\begin{equation}
\label{eq-IndSum}
0
\;=\;
\Ind(H-E\,\one)
\;=\;
\Ind(\HR-E\,\one)\,+\,\Ind(\HL-E\,\one)
\end{equation}
by \eqref{eq-HamDiff},  one has
$$
\spec(\HR)\cap \spec(\HL)
\;\supset\;
\Sigma \cup \big\{E\in\CM\setminus\Sigma\,:\,\Wind^E(H)\not =0\big\}
\;.
$$
The discrete spectra of $\HR$ and $\HL$ lie outside of this set and are, in general, not the same. It is one of the objects of this work to study these so-called {\it outliers} of the spectrum. Under suitable conditions, zero energy outliers will be shown to be stable and of topological nature.

\subsection{Uniform point spectrum}
\label{sec-EigenvalHalf}

The spectrum of half-sided (Toeplitz) operators $\HR$ and $\HL$ can have open subsets consisting entirely of eigenvalues, and we will refer to such regions as uniform point spectrum. This is well-known to occur for the example of the half-sided shift operator on $\ell^2(\NM)$, and the arguments transpose to other cases. Here several mathematical perspectives on the phenomenon are offered. First of all, the index theorem can also be used to determine eigenvalues for the half-sided operators $\HR$ and $\HL$. Indeed, for $E\not\in\Sigma=\spec(H)$, Theorem~\ref{theo-NoetherGoberg} implies 
\begin{align}
\Wind^E(H)
&
\;=\;
\dim\big(\Ker(\HR-E\,\one)\big)\,-\,\dim\big(\Ker((\HR)^*-\overline{E}\,\one)\big)
\label{eq-WindKer1}
\\
&
\;=\;
\dim\big(\Ker((\HL)^*-\overline{E}\,\one)\big)
\,-\,
\dim\big(\Ker(\HL-E\,\one)\big)
\;,
\label{eq-WindKer2}
\end{align}
where the first equality follows directly from \eqref{eq-FredIndWind} and the definition of the index and the second one, moreover, uses \eqref{eq-HRHL} as well as the important fact
\begin{equation}
\label{eq-WindTilde}
\Wind^E(H)\;=\;-\,\Wind^E(\widetilde{H})
\;.
\end{equation}
For non-vanishing winding number, \eqref{eq-WindKer1} and \eqref{eq-WindKer2} imply a dichotomy (with defect): 

\begin{proposition}
\label{prop-FilledHoles}
If  $\Wind^E(H)>0$, then $\Ker(\HR-E\,\one)\not=\{0\}$ and $E$ is an eigenvalue of $\HR$.
If $\Wind^E(H)<0$, then $\Ker(\HL-E\,\one)\not=\{0\}$ and $E$ is an eigenvalue of $\HL$. 
\end{proposition}

Note that the winding number is locally constant on $\CM\setminus\Sigma$ and only changes on $\Sigma$. As, moreover, $\spec((\HR)^*)=\overline{\spec(\HR)}$, a connected component of $\CM\setminus\Sigma$ belongs entirely to the spectrum of $\HR$ if the winding number is non-vanishing in this component. The same statements hold for $\HL$. Note also that these facts align exactly with analytic Fredholm theory as described in \cite{SB}.

\vspace{.2cm}

Another perspective on \eqref{eq-WindKer1} and \eqref{eq-WindKer2} (actually only a disguise of the above) invokes the bulk-boundary correspondence \cite{PS} for an associated hermitian Hamiltonian given by 
\begin{equation}
\label{eq-hDouble}
h^E
\;=\;
\begin{pmatrix}
0 & H\,-\,E\,\one \\
H^*\,-\,\overline{E}\,\one & 0
\end{pmatrix}
\;.
\end{equation}
By construction, $h^E$ has chiral symmetry, namely $Jh^EJ=-h^E$ where $J=\binom{\one \;\;\; 0}{0\;-\one}$ is the third Pauli matrix in the grading of \eqref{eq-hDouble}. It is possible to write $h^E$ again in the form \eqref{eq-HamGen} with blocks of doubled size $2L$. The off-diagonal entry $H-E\one$ has a winding number $\Wind^E(H)$ which in this context is also called the first Chern number $\Ch_1(h^E)$ of the chiral Hamiltonian $h^E$ \cite{PS}. The bulk-boundary correspondence for one-dimensional chiral systems ({\it e.g.} \cite{PS}, here applied to a periodic and not merely covariant system) states that the so-called bulk invariant $\Wind^E(H)$ is connected to boundary states via 
\begin{equation}
\label{eq-BBC}
\Wind^E(H)
\;=\;
-\,\Sig\big(J|_{\Ker(h^{E,R})}\big)
\;.
\end{equation}
Here $\Sig$ denotes the signature of a quadratic form, given by the restriction of $J$ to the subspace $\Ker(h^{E,R})$ where $h^{E,R}$ is the half-space restriction of $h^E$ on $\ell^2(\NM,\CM^{2L})$ (with Dirichlet boundary conditions). Because $(H^*)^R=(H^R)^*$ by \eqref{eq-HRHL2}, one thus deduces
$$
h^{E,R}
\;=\;
\begin{pmatrix}
0 & H^R\,-\,E\,\one \\
(H^R)^*\,-\,\overline{E}\,\one & 0
\end{pmatrix}
\;,
$$
and using this in \eqref{eq-BBC} then directly leads to \eqref{eq-WindKer1}. The second identity \eqref{eq-WindKer2} then follows from $\Wind^E(\widetilde{H})= \Sig\big(J|_{\Ker(\widetilde{h}^{E,R})}\big)$ combined with $\HTR=W^* \HL W$, see \eqref{eq-HRHL}.

\vspace{.2cm}

Yet another proof of Proposition~\ref{prop-FilledHoles} is rooted in the spectral analysis of the transfer matrix, and this is hence more in line with the remainder of the paper. Let us first focus on $\HR$. It has Dirichlet boundary conditions at $0$. Then the geometric multiplicity of $E$ as eigenvalue of $\HR$ is given by
\begin{equation}
\label{eq-EVHR}
\dim\big(\Ker(\HR\,-\,E\,\one)\big)
\;=\;
\dim\big(\Ee^E_{1}\cap\Dd_0\big)
\;,
\end{equation}
where $\Ee^E_{\scar}\subset\CM^{2L}$ is the span of the eigenvectors of $\Tt^E$ with modulus strictly less than $\scar$ and $\Dd_0\subset\CM^{2L}$ is the range of $\binom{\one}{0}$, namely fixing the Dirichlet boundary conditions at $0$. Indeed, the Schr\"odinger equation $\HR\psi=E\psi$ for $\psi=(\psi_n)_{n\geq 1}$ becomes on the first site  $V\psi_1+T\psi_2=E\psi_1$ which can be rewritten as $\Tt^E\binom{\one}{0}T=\Tt^E\binom{T}{0}=\binom{E\,\one-V }{\one}$ just as in \eqref{eq-SchrTra}. Hence if $\binom{v}{0}\in \Ee^E_{1}\cap\Dd_0$, then $(\Tt^E)^n\binom{v}{0}$ is exponentially decreasing and $\psi_n=\binom{0}{\one}^*(\Tt^E)^n\binom{v}{0}$ is thus a square integrable eigenstate of $\HR$ with energy $E$. Similar statements hold for the eigenvalues of $\HL$. Now as $\dim(\Dd_0)=L$, the intersection $\Ee^E_{1}\cap\Dd_0$ is always non-trivial if $\dim(\Ee^E_1)>L$. However, according to \eqref{eq-WindComp}, one has 
\begin{equation}
\label{eq-E1Wind}
\dim(\Ee^E_1)
\;=\;
L\,+\,\Wind^E(H)
\;.
\end{equation}
Hence if $\Wind^E(H)>0$, then $\dim(\Ee^E_1\cap\Dd_0)\geq 1$ and $E$ is an eigenvalue of $\HR$. In a similar manner, one can show that $\Wind^E(H)<0$ implies that $E$ is an eigenvalue of $\HL$.

\subsection{Outliers of half-sided operators}
\label{sec-OutliersHalf}

This section is about so-called spectral outliers of $\HR$ and $\HL$, namely eigenvalues $E$ in regions where the winding number vanishes, but the intersection in \eqref{eq-EVHR} is nevertheless non-trivial. It is hence of interest to introduce the associated (skew) Riesz projections 
\begin{equation}
\label{eq-DefP}
\Rr^E_\scar
\;=\;
\oint_{\TM_{\scar}}\frac{dz}{2\pi\imath}\;(z\,\one-\Tt^E)^{-1}
\;,
\end{equation}
provided that $\TM_{\scar}\cap\spec(\Tt^E)=\emptyset$.  Then $\Ee^E_\scar=\Ran(\Rr^E_{\scar})$ still for $\TM_{\scar}\cap\spec(\Tt^E)=\emptyset$, and 
$$
\mbox{rk}(\Rr^E_\scar)
\;=\;
\dim\big(\Ran(\Rr^E_{\scar})\big)
\;=\;
\sum_{|z|<\scar}m_z(\Tt^E)
\;.
$$
Moreover, using the basis change $\Mm^E$ as in \eqref{eq-MBasisChange},
$$
\Rr^E_\scar
\;=\;
\Mm^E
\begin{pmatrix}
\one & 0 \\ 0 & 0
\end{pmatrix}
(\Mm^E)^{-1}
\;,
$$
where $\one$ is of the size $\mbox{rk}(\Rr^E_\scar)$.
Changing variables, one also has 
\begin{equation}
\label{eq-RieszScale}
\Rr^E_\scar
\;=\;
\oint_{\TM}\frac{dz}{2\pi\imath}\;(z\,\one-\scar^{-1}\Tt^{E})^{-1} 
\;,
\end{equation}
To further compute $\Rr^E_\scar$, let us replace \eqref{eq-TransferResol} for the resolvent of $\Tt^E$. One finds
\begin{equation}
\label{eq-PRep}
\Rr^E_\scar
\;=\;
\begin{pmatrix}
T & 0 \\ 0 & \one
\end{pmatrix}
\begin{pmatrix}
{\QFct}^{(0)}_\scar(E)  & -{\QFct}^{(1)}_\scar(E) \\
{\QFct}^{(1)}_\scar(E) & R^{-1}- {\QFct}^{(2)}_\scar(E)
\end{pmatrix}
\begin{pmatrix}
\one & 0 \\ 0 & R
\end{pmatrix}
\;,
\end{equation}
where
\begin{equation}
\label{eq-QDef}
{\QFct}^{(j)}_\scar(E)
\;=\;
\oint_{\TM_{\scar}}\frac{dz}{2\pi\imath\,z^j}\;\big({H}(z)\,-\,E\,\one\big)^{-1}
\;,
\qquad
j\in\ZM
\;.
\end{equation}
Then
$$
{\QFct}^{(1)}_\scar(E)
\;=\;
\binom{0}{\one}^*\Rr^E_\scar\binom{\one}{0}
\;=\;
-\,T^{-1}\binom{\one}{0}^*\Rr^E_\scar\binom{0}{\one}R^{-1}
\;,
$$
and similarly for the other entries. Changing variables, ${\QFct}^{(j)}_\scar(E)$ can also be rewritten as 
$$
{\QFct}^{(j)}_\scar(E)
\;=\;
\scar^{1-j} \oint_{\TM}\frac{dz}{2\pi\imath\,z^j}\;\big({H}(\scar\,z)\,-\,E\,\one\big)^{-1} 
\;,
\qquad
j\in\ZM
\;.
$$

There are similar objects $\widetilde{\Rr}^E_\scar$, $\widetilde{\QFct}^{(j)}_\scar(E)$, etc., constructed from $\widetilde{H}$ instead of ${H}$, {\it e.g.}
$$
\widetilde{\QFct}^{(j)}_\scar(E)
\;=\;
\oint_{\TM_{\scar}}\frac{dz}{2\pi\imath\,z^j}\;\big(\widetilde{H}(z)\,-\,E\,\one\big)^{-1}
\;,
\qquad
j\in\ZM\,,\;\;\TM_{\scar}\cap\spec(\widetilde{\Tt}^E)=\emptyset
\;.
$$
As above,
$$
\widetilde{\Rr}^E_\scar
\;=\;
\begin{pmatrix}
R & 0 \\ 0 & \one
\end{pmatrix}
\begin{pmatrix}
\widetilde{\QFct}^{(0)}_\scar(E)  & -\widetilde{\QFct}^{(1)}_\scar(E) \\
\widetilde{\QFct}^{(1)}_\scar(E) & T^{-1}- \widetilde{\QFct}^{(2)}_\scar(E)
\end{pmatrix}
\begin{pmatrix}
\one & 0 \\ 0 & T
\end{pmatrix}
\;.
$$
A change of variable $z\mapsto z^{-1}$ also shows that 
\begin{equation}
\label{eq-JInvert}
\widetilde{\QFct}^{(j)}_\scar(E)
\;=\;
{\QFct}^{(2-j)}_{\scar^{-1}}(E)
\;.
\end{equation}
In particular, for $j=1$ and $\scar=1$, one has
$$
\widetilde{\QFct}^{(1)}_1(E)
\;=\;
{\QFct}^{(1)}_{1}(E)
\;.
$$

Let us now consider $\HL$. As $W\HL W^*=\HTR$ with the unitary reflection $W$, one has  
\begin{equation}
\label{eq-EVHL}
\dim\big(\Ker(\HL\,-\,E\,\one)\big)
\;=\;
\dim\big(\Ker(\HTR\,-\,E\,\one)\big)
\;=\;
\dim\big(\widetilde{\Ee}^E_{1}\cap\Dd_0\big)
\;,
\end{equation}
where now $\widetilde{\Ee}^E_\scar=\Ran(\widetilde{\Rr}_\scar^E)$. Using \eqref{eq-TransferConjugate}, one also finds
\begin{align*}
\one\,-\,\widetilde{\Rr}^E_\scar
&
\;=\;
\oint_{\TM_{\scar}}\frac{dz}{2\pi\imath}\;
\Big[z^{-1}\one\,-\,
(z\,\one-\widetilde{\Tt}^E)^{-1}\Big]
\\
&
\;=\;
\oint_{\TM_{\scar}}\frac{dz}{2\pi\imath}\;
z^{-2}
\big(z^{-1}\one-(\widetilde{\Tt}^E)^{-1}\big)^{-1}
\\
&
\;=\;
\oint_{\TM_{\scar^{-1}}}\frac{dz}{2\pi\imath}\;
\big(z\,\one-(\widetilde{\Tt}^E)^{-1}\big)^{-1}
\\
&
\;=\;
\oint_{\TM_{\scar^{-1}}}\frac{dz}{2\pi\imath}\;
\begin{pmatrix}
0 & R \\ T^{-1} & 0
\end{pmatrix}
\big(z\,\one-{\Tt}^E\big)^{-1}
\begin{pmatrix}
0 & R \\ T^{-1} & 0
\end{pmatrix}^{-1}
\\
&
\;=\;
\begin{pmatrix}
0 & R \\ T^{-1} & 0
\end{pmatrix}
\Rr^E_{\scar^{-1}}
\begin{pmatrix}
0 & R \\ T^{-1} & 0
\end{pmatrix}^{-1}
\;.
\end{align*}
Actually this identity also follows from \eqref{eq-JInvert} replaced into the representation \eqref{eq-PRep}, or even directly from \eqref{eq-TransferConjugate}. Equivalently
\begin{equation}
\label{eq-PPRange}
\one\,-\,{\Rr}^E_\scar
\;=\;
\begin{pmatrix}
0 & T \\ R^{-1} & 0
\end{pmatrix}
\widetilde{\Rr}^E_{\scar^{-1}}
\begin{pmatrix}
0 & T \\ R^{-1} & 0
\end{pmatrix}^{-1}
\;.
\end{equation}
From this, one deduces
\begin{equation}
\label{eq-PPRange2}
\Ran(\one\,-\,{\Rr}^E_\scar)
\;=\;
\begin{pmatrix}
0 & T \\ R^{-1} & 0
\end{pmatrix}
\Ran\big(\widetilde{\Rr}^E_{\scar^{-1}}\big)
\;.
\end{equation}
By essentially the same arguments, one finds from \eqref{eq-Iunitarity} that
$$
\Ii^*\,({\Rr}^E_\scar)^*\,\Ii
\;=\;
\one\,-\,\widehat{\Rr}^{\overline{E}}_{\scar^{-1}}
\;,
$$
with $\widehat{\Rr}^E_\scar$ defined as in \eqref{eq-DefP} with $\Tt^E$ replaced by $\widetilde{\Tt}^E$.

\begin{lemma}
\label{lem-SpaceIntersect}
Recall that $\Dd_0$ is the range of $\binom{\one}{0}$. Then for all $\scar>0$ with $\TM_{\scar}\cap\spec(\Tt^E)=\emptyset$,
$$
\Ran(\widetilde{\Rr}^E_{\scar^{-1}})\cap\Dd_0 
\;=\; 
\Ran(\one-\Rr^E_{\scar})\cap\Dd_0
\;.
$$
\end{lemma}

\noindent {\bf Proof.} 
Let $v\in\CM^N$ be such that
$$
\binom{R v}{0} 
\;\in\; 
\Ran(\widetilde{\Rr}^E_{\scar^{-1}})\cap\Dd_0
\;.
$$
Due to \eqref{eq-PPRange2} one then has 
$$
\binom{0}{v}
\;=\;
\begin{pmatrix}
 0 & T \\ R^{-1} & 0
\end{pmatrix}
\binom{R v}{0} 
\;\in\;
\Ran(\one-\Rr^E_{\scar})
\;.
$$ 
As the latter subspace is a direct sum of generalized eigenspaces of $ \mathcal{T}^E$, it is invariant under $\mathcal{T}^E$. Therefore
$$
\mathcal{T}^E \binom{0}{v} 
\;=\; 
\binom{-R v}{0}
\;\in\;
\Ran(\one-\Rr^E_{\scar})
\;.
$$
As the latter vector is clearly also in $\Dd_0$, one concludes that $\Ran(\widetilde{\Rr}^E_{\scar^{-1}})\cap\Dd_0\subset \Ran(\one-\Rr^E_{\scar})\cap\Dd_0$. The inverse inclusion is shown in the same manner.
\hfill $\Box$

\begin{proposition}
\label{prop-KernelCalc}
Suppose that $\spec(\Tt^E)\cap\TM=\emptyset$, or equivalently $E\not\in\Sigma$. Then $E$ is an eigenvalue of $\HR$ or $\HL$ if and only if $\Ker({\QFct}^{(1)}_{1}(E))\not=\{0\}$. More precisely,
$$
\dim\big(\Ker({\QFct}^{(1)}_{1}(E))\big) 
\;=\; 
\dim\big(\Ker(\HR-E\,\one)\big)
\,+\,
\dim\big(\Ker(\HL - E\,\one)\big)
\;.
$$ 
\end{proposition}

\noindent {\bf Proof.} 
Suppose $E$ is an eigenvalue of $\HR$ with (non-zero and square integrable) eigenvector $\psi=(\psi_n)_{n\geq 1}$. Then $\psi$ has to decay, {\it i.e.} the vector $\binom{T \psi_{n+1}}{\psi_n}$ lies in the decaying subspace $\Ee^E_{1}$ of $\Tt^E$. In particular, 
$$
\binom{T \psi_1}{0} \;\in\; \Ran(\Rr^E_{1})\cap\Dd_0
\;.
$$
Moreover, since $\Rr^E_{1}$ is an idempotent, one has
$$
\binom{T \psi_1}{0} 
\;=\; 
\Rr^E_1 \binom{T \psi_1}{0} 
\;=\; 
\binom{T {\QFct}^{(0)}_1(E) T \psi_1}{{\QFct}^{(1)}_1(E) T \psi_1} 
\;.
$$
Thus indeed $T \psi_1 \in \Ker({\QFct}^{(1)}_{1}(E))$ and $T \psi_1$ is non-zero, since otherwise successive application of the transfer matrix would imply $\psi = 0$. Similarly, supposing $E$ is an eigenvalue of $\widetilde{H}^R$ with eigenvector $\widetilde{\psi}$, one has
$$
\binom{R \widetilde{\psi}_1}{0} 
\;\in\; 
\Ran(\widetilde{\Rr}^E_1)\cap\Dd_0
$$
and
$$
\binom{R \widetilde{\psi}_1}{0} 
\;=\; 
\widetilde{\Rr}^E_1 \binom{R \widetilde{\psi}_1}{0} 
\;=\; 
\binom{R \widetilde{\QFct}^{(0)}_{1}(E) R \psi_1}{\widetilde{\QFct}^{(1)}_{1}(E) R \psi_1}
\;=\; 
\binom{R \QFct^{(2)}_1 (E) R \widetilde{\psi}_1}{\QFct^{(1)}_1(E) R \widetilde{\psi}_1}
\;.
$$
Hence $0\neq R \widetilde{\psi}_1 \in \Ker(\QFct^{(1)}_1(E))$. Moreover, $\Ran(\widetilde{\Rr}^E_1)\cap\Dd_0 =\Ran(\one-\Rr^E_1)\cap\Dd_0$ by Lemma~\ref{lem-SpaceIntersect}. As $\Ran(\Rr^E_1) $ and $\Ran(\one-\Rr^E_1)$ are linearly independent, it follows that also $\binom{T \psi_1}{0}$ and $\binom{R \widetilde{\psi}_1}{0}$ are linearly independent. As both first components lie in  $\Ker({\QFct}^{(1)}_{1}(E))$, one concludes together with \eqref{eq-EVHR} and \eqref{eq-EVHL} that
$$
\dim\big(\Ker({\QFct}^{(1)}_{1}(E))\big) 
\;\geq\; 
\dim\big(\Ker(\HR-E\,\one)\big)
\,+\,
\dim\big(\Ker(\HL - E\,\one)\big)
\;.
$$
Now let $0 \neq v \in \Ker(\QFct^{(1)}_\scar(E))$. Then
$$
\Rr^E_1 \binom{v}{0} 
\;=\; 
\binom{T{\QFct}^{(0)}_1(E) v}{{\QFct}^{(1)}_1(E) v} 
\;=\; 
\binom{ T{\QFct}^{(0)}_1(E) v}{0} 
\;\in\; 
\Ran({\Rr}^E_1)\cap\Dd_0
$$
and similarly
$$
(\one-\Rr^E_1) \binom{v}{0} 
\;=\; 
\binom{(\one - T{\QFct}^{(0)}_1(E)) v}{0} 
\;\in\; 
\Ran(\one-\Rr^E_1)\cap\Dd_0\,=\,\Ran(\widetilde{\Rr}^E_1)\cap\Dd_0
\;,
$$
where the last claim follows from  Lemma~\ref{lem-SpaceIntersect}. As at least one of the last two vectors is non-vanishing, one concludes that $E$ is an eigenvalue of $H^R$ or of $\widetilde{H}^R$ due to \eqref{eq-EVHR} and \eqref{eq-EVHL}. Moreover, if $\Ss^E=\binom{\Ker(\QFct^{(1)}_1(E))}{0}\subset\Dd_0$, then 
\begin{align*}
\dim\big(\Ker({\QFct}^{(1)}_{1}(E))\big) 
&\; = \;
\dim\big(\Ran(\Rr^E_1)\cap\Ss^E\big)
\,+\,
\dim\big(\Ran(\one-\Rr^E_1)\cap\Ss^E\big)
\\
&\; \leq \;
\dim\big(\Ran(\Rr^E_1)\cap\Dd_0\big)
\,+\,
\dim\big(\Ran(\one-\Rr^E_1)\cap\Dd_0\big)
\\
&
\;=\;
\dim\big(\Ker(\HR-E\,\one)\big)
\,+\,
\dim\big(\Ker(\widetilde{H}^R - E\,\one)\big)
\;,
\end{align*}
concluding the proof due to \eqref{eq-EVHL}.
\hfill $\Box$

\begin{corollary}
\label{coro-KernelCalc}
The sum of the geometric multiplicities of $E$ as eigenvalue of $\HR$ or $\HL$ is at most $L$:
$$
\dim\big(\Ker(\HR-E\,\one)\big)
\,+\,
\dim\big(\Ker(\HL - E\,\one)\big)
\;\leq\;
L\;.
$$ 
\end{corollary}

\begin{example}
{\rm
Let us consider $L=1$, namely a scalar periodic operators with $T,V,R\in\CM$ with $T,R\not=0$ (this is the Hatano-Nelson model, see Section~\ref{sec-HNmodel}). Then 
$$
\spec(H)
\;=\;\{Re^{-\imath k}+V+Te^{\imath k}\,:\;k\in[-\pi,\pi)\}
$$ 
is the boundary of an open ellipse $\EM\subset\CM$. Using the residue theorem, one has
$$
\QFct^{(1)}_1(E)
\;=\;
\frac{1}{2\pi\imath \,T}\oint_{\TM}dz\;\frac{1}{z^2+(V-E)T^{-1}z+RT^{-1}}
\;=\;
\left\{
\begin{array}{cc}
0\;, & E\in\EM\;,
\\
c_E\;, & E\in\CM\setminus\overline{\EM}\;,
\end{array}
\right. 
$$
with $c_E=\big((E-V)^2-4RT\big)^{-\frac{1}{2}}\not=0$ with the root taken using the first branch. Hence Proposition~\ref{prop-KernelCalc} tells us that for $E\in\EM$ either $\HR$ or $\HL$ has an eigenvalue. Of course, as $|\Wind^E(H)|=1$ for $E\in\EM$, this merely reproduces the implication of Theorem~\ref{theo-NoetherGoberg} (which is actually stronger because it also determines whether $\HR$ or $\HL$ must have an eigenvalue).
\hfill $\diamond$
}
\end{example}

\begin{figure}
\centering
\includegraphics[width=7.5cm,height=2.5cm]{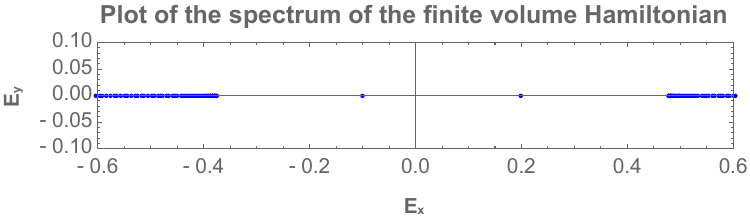}
\hspace{.3cm}
\includegraphics[width=7.5cm,height=2.5cm]{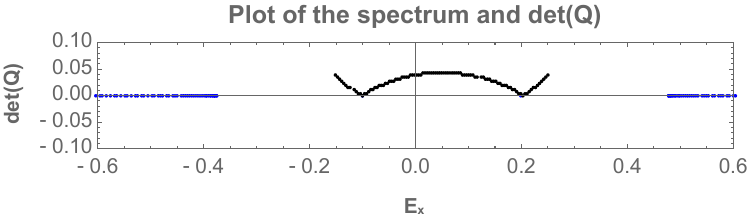}
\caption{\sl Plot of the central part of the spectrum of the selfadjoint Hamiltonian described in {\rm Example~\ref{ex-SelfAd}} with parameters are $t_{0,1}=1.0$, $t_{1,0}=0.1$, $t_{0,0}=t_{1,1}=0$ and $v_{0,0}=0.2$, $v_{0,1}=0.5\,\imath =\overline{v_{1,0}}$, $v_{1,1}=-0.1$. The right plot also shows the value of $|\det(Q^{(1)}_1(E))|$ as a function of the $E\in\RM$. One clearly sees the two zeros, in accordance with {\rm Proposition~\ref{prop-KernelCalc}}.
}
\label{fig-Ser4}
\end{figure}

\begin{example}
\label{ex-SelfAd}
{\rm
Let us consider a selfadjoint Hamiltonian with $L=2$. It is specified by two matrices $T=R^*=\binom{t_{0,0}\;\;t_{0,1}}{t_{1,0}\;\;t_{1,1}}$ and $V=\binom{v_{0,0}\;\;v_{0,1}}{v_{1,0}\;\;v_{1,1}}=V^*$. For suitable parameters inspired by the so-called SSH model \cite{PS}, the Hamiltonian has two bands and and two eigenvalues (bound states) in the gap. One can then compute the $2\times 2$ matrix $\QFct^{(1)}_1(E)$ and its determinant numerically and its zeros lie exactly at these eigenvalues. This is illustrated in Figure~\ref{fig-Ser4}, and can also be understood using intersection theory of Lagrangian subspaces, see \cite{SB1}.
\hfill $\diamond$
}
\end{example}

\subsection{Topological zero modes of half-sided chiral operators}
\label{sec-ChiralOutliersHalf}

This section first defines what a chiral symmetry is and provides some structural information that it induces on the objects introduced so far. It then shows how index-theoretic arguments imply the existence of zero eigenvalues for suitable half-sided (Toeplitz) operators with chiral symmetry. Throughout the dimension $L$ will be supposed to be even. Then on each fiber there is selfadjoint unitary of vanishing signature given by $K=\binom{\one \;\;\; 0}{0 \; -\one}$. Its fiberwise extension $\one\otimes K$ will also be denoted by $K$. The chiral symmetry of $H$ then reads 
\begin{equation}
\label{eq-ChiralSym}
K\,H\,K\;=\;-\,H
\;.
\end{equation} 
Due to the chiral symmetry \eqref{eq-ChiralSym}, it follows that $\Sigma=\spec(H) = -\Sigma$. 
Moreover, one deduces $KVK=-V$, $KTK=-T$ and $KRK=-R$. Hence each of these matrices is off-diagonal in the grading of $K$. Let us introduce the notations
$$
V
\;=\;
\begin{pmatrix}
0 & V_+ \\ V_- & 0
\end{pmatrix}
\;,
\qquad
T
\;=\;
\begin{pmatrix}
0 & T_+ \\ T_- & 0
\end{pmatrix}
\;,
\qquad
R
\;=\;
\begin{pmatrix}
0 & R_+ \\ R_- & 0
\end{pmatrix}
\;.
$$
It is also natural and useful to introduce a notation for the off-diagonal entries of the symbol:
\begin{equation}
\label{eq-HChrial}
H(z)
\;=\;
\begin{pmatrix}
0  & H_+(z) \\  H_-(z) & 0
\end{pmatrix}
\;,
\qquad
H_\pm(z)
\;=\;
R_\pm\,z^{-1}\,+\,V_\pm\,+\,T_\pm\,z
\;,
\end{equation}
and then consider $H_\pm$ also as block operators of the form \eqref{eq-HamGen}, but on the Hilbert space $\ell^2(\ZM,\CM^{\frac{L}{2}})$.  As above one also has $\widetilde{H}_\pm(z)$ and $\widehat{H}_\pm(z)$. Next let us consider the winding numbers. As $\det\big(K ( H(z) - E\,\one ) K \big)=(-1)^L \det(H(z) + E\,\one)$ and $L$ is even,  one has  for $E\in\CM\setminus\Sigma$
\begin{equation}
\Wind^E(H) 
\;=\; \oint_{\TM}\frac{dz}{2\pi\imath}\; \partial_z \log\det\big(K ( H(z) - E\,\one ) K \big) 
\label{eq-WindChiral}
\;=\; \Wind^{-E}(H)
\;.
\end{equation}
Of particular importance is the winding number at zero energy. Due to the off-diagonal form of $H(z)$, one has
\begin{equation}
\label{eq-WindAd}
\Wind^0(H)
\;=\;
\Wind^0(H_+)
\,+\,
\Wind^0(H_-)
\;.
\end{equation}
These latter two winding numbers imply the following extension of the bulk-boundary correspondence as described in Section~\ref{sec-EigenvalHalf}.

\begin{proposition}
\label{prop-ChiralKernel}
Let $H$ have the chiral symmetry \eqref{eq-ChiralSym}. Then $\Wind^0(H_\pm)>0$ implies that $\dim(\Ker(\HR_\pm))\geq 1$ and hence also $\dim(\Ker(\HR))\geq 1$. Similarly, $\Wind^0(H_\pm)<0$ implies $\dim(\Ker(\HL_\mp))\geq 1$ and $\dim(\Ker(\HL))\geq 1$.
\end{proposition}

\noindent {\bf Proof.} As in \eqref{eq-hDouble}, one has
\begin{equation}
\label{eq-h0Doubled}
h^0
\;=\;
\begin{pmatrix}
0 & 0 & 0 & H_+ \\
0 & 0 & H_- & 0 \\
0 & (H_-)^* & 0 & 0 \\
(H_+)^* & 0 & 0 & 0 
\end{pmatrix}
\;.
\end{equation}
Now \eqref{eq-BBC} holds and implies that  $\Wind^0(H)=-\Sig\big(J|_{\Ker(h^{0,R})}\big)$. Next let us recall  the identities $((H_\pm)^*)^R=\widehat{H}_\pm^R$ and $((H_\pm)^*)^L=\widehat{H}_\pm^L$, see \eqref{eq-HRHL2}. Together with \eqref{eq-WindAd}, one therefore has
\begin{align*}
\Wind^0  (H_+)&
\,+\,\Wind^0(H_-)
\\
&
\;=\;
\dim\big(\Ker(\HR_-)\big)\,+\,\dim\big(\Ker(\HR_+)\big)
\,-\,\dim\big(\Ker(\widehat{H}_+^R)\big)\,-\,\dim\big(\Ker(\widehat{H}_-^R)\big)
\;.
\end{align*}
Moreover, using $J'=\diag(\one,-\one,\one,-\one)$ in the grading of \eqref{eq-h0Doubled}, the selfadjoint Hamiltonian $h^0$ also has the further chiral symmetry $J' h^0J'=-h^0$. For this symmetry, \eqref{eq-BBC} combined with $\Wind^0((H_-)^*)=-\Wind^0(H_-)$ leads to
\begin{align*}
\Wind^0(H_+) & -\Wind^0(H_-)
\\
&
\;=\;
\dim\big(\Ker(\widehat{H}_-^R)\big)
\,+\,\dim\big(\Ker(\HR_+)\big)
\,-\,
\dim\big(\Ker(\widehat{H}_+^R)\big)\,-\,\dim\big(\Ker(\HR_-)\big)
\;.
\end{align*}
Adding and subtracting these identities then leads to the two identities
$$
\Wind (H_\pm,0)
\;=\;
\dim\big(\Ker(\HR_\pm)\big)
\,-\,
\dim\big(\Ker(\widehat{H}_\pm^R)\big)
\;,
$$
which imply the first claim. For the second one, the same reasoning is applied to $\widetilde{H}$ for which the winding number is given by \eqref{eq-WindTilde}. Because of the relations $W\HL_\pm W^*= \HTR_\pm$ and $W\HR_\pm W^*=\HTL_\pm$ as in \eqref{eq-HRHL} hold, one obtains
\begin{align*}
\Wind^0(H_+)&
\,+\,\Wind^0(H_-)
\\
&
\;=\;
\dim\big(\Ker(\widehat{H}_+^L)\big)\,+\,\dim\big(\Ker(\widehat{H}_-^L)\big)
\,-\,\dim\big(\Ker(\HL_+)\big)\,-\,\dim\big(\Ker(\HL_-)\big)
\end{align*}
and
\begin{align*}
\Wind^0 (H_+)&
\,-\,\Wind^0(H_-)
\\
&
\;=\;
\dim\big(\Ker(\widehat{H}_+^L)\big)\,+\,\dim\big(\Ker(\HL_+)\big)
\,-\,\dim\big(\Ker(\widehat{H}_-^L)\big)\,-\,\dim\big(\Ker(\HL_-)\big)
\;.
\end{align*}
Hence
$$
\Wind^0(H_\pm)
\;=\;
\dim\big(\Ker(\widehat{H}_\pm^L)\big)
\,-\,
\dim\big(\Ker(\HL_\mp)\big)
\;,
$$
implying the second claim.
\hfill $\Box$

\vspace{.2cm}

On first sight, the statements of Proposition~\ref{prop-ChiralKernel} and Proposition~\ref{prop-FilledHoles} resemble each other. There is, however, a crucial difference: Proposition~\ref{prop-FilledHoles} is stable under perturbations in $E$ (because the winding number $\Wind^E(H)$ is so), while Proposition~\ref{prop-ChiralKernel} merely concerns the eigenvalue $0$, the so-called zero modes or zero energy bound states. Let us discuss two distinct scenarios for the application of Proposition~\ref{prop-ChiralKernel} (which in physical terminology explained in Section~\ref{sec-Applications} are compatible with line-gapped and point-gapped systems respectively). If $\Wind^0(H)=0$, then the component of $\CM\setminus\Sigma$ containing $0$ only contains discrete spectrum (by analytic Fredholm theory) and the zero eigenvalues resulting from Proposition~\ref{prop-ChiralKernel} (in the case $\Wind^0(H_+)\not=0$) are robust under perturbations conserving the chiral symmetry (and are merely moved away from the origin if the perturbations break the chiral symmetry). On the other hand, if $\Wind^0(H)\not=0$, then Proposition~\ref{prop-FilledHoles} implies that the component of $\CM\setminus\Sigma$ containing $0$ consists entirely of spectrum of $\HR$ (or $\HL$) and then the supplementary eigenvalues resulting from Proposition~\ref{prop-ChiralKernel} are covered by this spectrum and seem to be of little interest. It will, however, be shown in Section~\ref{sec-TopoZero} that finite-volume approximations do have approximate zero modes resulting from $\Wind^0(H_+)\not=0$. This is based on Widom's theory of asymptotic spectra described in Section~\ref{sec-WidomRevisited}.

\vspace{.2cm}

The remainder of this section will provide an alternative proof of Proposition~\ref{prop-ChiralKernel} based on transfer matrix methods (similar to the arguments towards the end of Section~\ref{sec-EigenvalHalf}, leading to Proposition~\ref{prop-FilledHoles}). This is crucially based on the following property of the transfer matrix of a chiral Hamiltonian:
$$
(K \otimes \sigma_3 )\, \Tt^E\, (K \otimes \sigma_3 )
\; =\; 
\begin{pmatrix}
K (E-V) T^{-1} K & KRK \\ -K T^{-1} K & 0
\end{pmatrix} \\
\;=  \;
\begin{pmatrix}
-(E+V) T^{-1} & -R \\ T^{-1} & 0
\end{pmatrix} \\
= \; 
\Tt^{-E}
\;.
$$
This implies ${\det}_{2L}( \Tt^E - z\one) ={\det}_{2L}(\Tt^{-E} - z\one)$ so that $\spec( \Tt^E ) = \spec(\Tt^{-E})$. Furthermore, $(K \otimes \sigma_3 ) \,\Tt^0\, (K \otimes \sigma_3 )=\Tt^0$, notably the zero energy transfer matrix has invariant subspaces. Indeed, one readily checks
$$
\Tt^0 
\;=\; 
\begin{pmatrix}
-V T^{-1} & -R \\ T^{-1} & 0
\end{pmatrix} 
\;=\; 
\begin{pmatrix}
- V_{+} T_{+}^{-1} & 0 & 0 & - R_{+} \\
0 & - V_{-} T_{-}^{-1} & - R_{-} & 0 \\
0 & T_{-}^{-1} & 0 & 0 \\
T_{+}^{-1} & 0 & 0 & 0 
\end{pmatrix}
\;=\;
\Tt_+^0\,\hat{\oplus}\,\Tt^0_-
\;,
$$
where the direct sum is understood in the obvious way (regrouping the first and fourth columns and lines, as well as the second and third ones) and these $2\times 2$ blocks are given by
\begin{align*}
\Tt_{+}^0  
\;=\; 
\begin{pmatrix}
-V_{+} T_{+}^{-1} & -R_{+} \\ T_{+}^{-1} & 0
\end{pmatrix} \;,
\qquad
\Tt_{-}^0 
\;=\; 
\begin{pmatrix}
-V_{-} T_{-}^{-1} & -R_{-} \\ T_{-}^{-1} & 0
\end{pmatrix} \;.
\end{align*}
Note that $\Tt_{+}^0$ and $\Tt_{-}^0$ are the transfer matrices at $E=0$ for $H_+$ and $H_-$ respectively. In complete analogy to \eqref{eq-EVHR} for $E=0$, one now has for these two subsystems
\begin{equation}
\label{eq-EVHRChiral}
\dim\big(\Ker(\HR_\pm)\big)
\;=\;
\dim\big(\Ee^0_{1,\pm}\cap\Dd_0\big)
\;,
\end{equation}
where $\Ee^0_{1,\pm}\subset\CM^L$ is the generalized eigenspace of all eigenvalues of $\Tt^0_\pm$ with modulus less than $1$. Moreover, similarly as in \eqref{eq-E1Wind}, one has
\begin{equation}
\label{eq-E1WindChiral}
\dim(\Ee^0_{1,\pm})
\;=\;
\frac{L}{2}\,+\,\Wind^0(H_\pm)
\;.
\end{equation}
Hence if $\Wind^0(H_\pm)>0$ is positive, then $\dim(\Ee^0_{1,\pm})>\frac{L}{2}$ which enforces $\dim(\Ee^0_{1,\pm}\cap\Dd_0)\geq 1$ so that, due to \eqref{eq-EVHRChiral}, $0$ is an eigenvalue of $\HR_\pm$. This provides the promised alternative proof of Proposition~\ref{prop-ChiralKernel}. 

\begin{remark}
{\rm
Let us briefly show how Proposition~\ref{prop-ChiralKernel} relates to Proposition~\ref{prop-KernelCalc}. For $0\not\in\Sigma=\spec(H)$ the latter result states that
$$
\dim\big(\Ker({\QFct}^{(1)}_{1}(0))\big) 
\;=\; 
\dim\big(\Ker(\HR)\big)
\,+\,
\dim\big(\Ker(\HL)\big)
\;.
$$
The matrix-valued function $E\mapsto {\QFct}^{(1)}_{r}(E)$ satisfies
$$
K \,\QFct_r^{(1)}(E)\, K \,=\,
\oint_{\TM_{\scar}}\frac{dz}{2\pi\imath\,z}\; K\, \big({H}(z)\,-\,E\,\one\big)^{-1} K \,=\, 
\oint_{\TM_{\scar}}\frac{dz}{2\pi\imath\,z}\; \big(-\,{H}(z)\,-\,E\,\one\big)^{-1} \,=\, 
- \,\QFct_r^{(1)}(-E) \,,
$$
implying, in particular, that $\QFct_r^{(1)}(0)$ is chiral. It is hence off-diagonal in the grading of $K$ with two off-diagonal entries $\QFct_{r,\pm}^{(1)}(0)$ given by \eqref{eq-QDef} from $H_\pm$:
\begin{align*}
{\QFct}^{(1)}_\scar(0)
\;=\;
\oint_{\TM_{\scar}}\frac{dz}{2\pi\imath\,z}\;
\begin{pmatrix}
0 & H_+(z) \\
H_-(z) & 0
\end{pmatrix}^{-1}
\;=\;
\begin{pmatrix}
0 &  \QFct_{r,-}^{(1)}(0) \\
\QFct_{r,+}^{(1)}(0) & 0
\end{pmatrix}
\;.
\end{align*}
One hence has
$$
\dim\big(\Ker({\QFct}^{(1)}_{r}(0))\big) 
\;=\; 
\dim\big(\Ker({\QFct}^{(1)}_{r,+}(0))\big) 
\,+\,
\dim\big(\Ker({\QFct}^{(1)}_{r,-}(0))\big) 
\;.
$$
Furthermore, $0\not\in\Sigma$ is equivalent to $\det(H(e^{\imath k}))=\det(H_+(e^{\imath k}))\det(H_-(e^{\imath k}))\not=0$ for all $e^{\imath k}\in\TM $, which is hence also equivalent to $0\not\in\spec(H_+)$ and $0\not\in\spec(H_-)$. Therefore one can apply Proposition~\ref{prop-KernelCalc} also separately to $H_\pm$, giving
$$
\dim\big(\Ker({\QFct}^{(1)}_{1,\pm}(0))\big) 
\;=\; 
\dim\big(\Ker(\HR_\pm)\big)
\,+\,
\dim\big(\Ker(\HL_\pm)\big)
\;.
$$ 
Together these equations merely reflect that  $\Ker(\HR)=\Ker(\HR_+)\oplus\Ker(\HR_-)$. 
\hfill $\diamond$
}
\end{remark}

\section{Widom's theory of asymptotic spectra revisited}
\label{sec-WidomRevisited}

This section is the mathematical core of the paper, providing the proofs of Theorems~\ref{thm-LimitSpecIntro} and \ref{thm-ChiralIntro}. 

\subsection{The set $\LimitSet $ of energies with middle eigenvalues of equal modulus}
\label{sec-Subsets}

Let us first note that the set $\Sigma=\spec(H)$ can be written as
\begin{align}
\Sigma
\;=\;
\big\{E\in\CM\,:\,|z_l(E)|= 1\mbox{ for some }l=1,\ldots,2L \big\}
\;,
\label{eq-SigmaTransfer}
\end{align}
see also \eqref{eq-SigmaDef}. In the hermitian case in which the transfer matrix is $\Ii$-unitary, the existence of one eigenvalue on the unit circle implies that there is also a second eigenvalue on the unit circle and in the corresponding two-dimensional subspace one can construct eigenfunctions of finite volume approximations $H_N$ at nearby energies (by well-known techniques described in the proof of Proposition~\ref{prop-TransferEigenfct},  or by the procedure in Section~\ref{sec-EVConstruct}). Hence the points in $\Sigma$ are indeed near the spectrum of $H_N$.  However, for a non-hermitian $H$ one has generically only one eigenvalue on the unit circle and the fundamental solutions constructed by the transfer matrices do not allow to construct eigenfunctions for $H_N$ satisfying both boundary conditions. From this perspective, it is natural to introduce the set $\LimitSet$ of complex energies $E$ where the middle two eigenvalues $|z_L(E)|\leq |z_{L+1}(E)|$ of the transfer matrix $\Tt^E$ (or equivalently zeros  of $D(z,E)={\det}_L(H(z)-E\,\one)$) coincide:
\begin{align}
\LimitSet 
&
\;=\;\big\{E\in\CM\,:\,|z_L(E)|= |z_{L+1}(E)|\big\}
\;,
\label{eq-Gamma0}
\end{align}
see \eqref{eq-LambdaDef}. (On p.~312 of Widom's work \cite{Wid1}, this set is denoted by $C_2$.) Due to \eqref{eq-TransferConjugateSpec}, one can also write this set as
$$
\LimitSet 
\;=\;
\big\{E\in\CM\,:\,\mbox{middle two eigenvalues of }\widetilde{\Tt}^E\mbox{ are of same modulus}\big\}
\;.
$$
This section is about proving topological and analytical properties of this set $\LimitSet$. A first property is the following: if $E$ is in $\Lambda$, then either $E\in\Sigma$ or $E\notin\Sigma$. In the latter case, $\det(H(e^{\imath k})-E\,\one)\not=0$ for all $e^{\imath k}\in\TM $ so that $\Wind^E(H)$ is well-defined and, due to \eqref{eq-WindComp}, non-vanishing. Hence
\begin{equation}
\label{eq-GammaWind}
\LimitSet 
\;\subset\; \Sigma\,\cup\, \big\{ E\in\CM \,:\, \Wind^E(H)\neq 0\big\} 
\;.
\end{equation}

\begin{proposition}
\label{prop-Gamma0Include}
The set $\LimitSet $ satisfies $\LimitSet \subset\spec(\HR)$ and $\LimitSet \subset\spec(\HL)$.
\end{proposition}

\noindent {\bf Proof.} 
Rewriting \eqref{eq-GammaWind} gives together with Theorem~\ref{theo-NoetherGoberg} that
\begin{align*}
\LimitSet 
&
\;\subset\;  \big\{ E\in\CM \colon {\det}_L(H(z)-E\;\one)=0 \text{ for some } z\in\mathbb{S}^1 \text{ or } \Wind^E(H)\neq 0\big\} 
\\
&
\;=\;  \big\{ E\in\CM \colon H^R -E\;\one \text{ is not Fredholm or is Fredholm with } \Ind (H^R -E\;\one)\neq 0\big\} 
\\
&\;\subset\; \spec (H^R)
\;.
\end{align*}
The same holds for $H^L$.
\hfill $\Box$

\begin{remark}
\label{rem-Gamma0L1}
{\rm 
Let us consider the special case of hermitian $H$ (namely $T=R^*$ and $V=V^*$). Then the spectrum of the $\Ii$-unitary matrix $\Tt^E$ has a reflection symmetry on $\TM$ and therefore the middle two eigenvalues are of the same modulus if and only if they lie on the unit circle, notably
$$
\LimitSet 
\;=\;
\big\{E\in\CM\,:\,\spec(\Tt^E)\cap\TM\not = \emptyset\big\}
\;.
$$
This is a well-known characterization of the spectrum of periodic hermitian operators, namely $\Sigma=\LimitSet $. There are close connections to the standard Brillouin zone, see the discussion in Section~\ref{sec-Brillouin} below. Furthermore, the inclusion $\LimitSet  \subset \lim_{N\to\infty} \spec(H_N)$ follows from a standard Weyl sequence argument. For systems without any bound states (also called outliers), one has $\LimitSet  = \lim_{N\to\infty} \spec(H_N)$, namely the set $\Outliers$ introduced in Section~\ref{sec-Lambda} is empty.  On the other hand, for  hermitian chiral SSH-type topological insulators (of the type described in Section~\ref{sec-TopoZero}), it is known that there are spectral outliers at zero energy, similar to the eigenvalues already encountered in Example~\ref{ex-SelfAd}. This paper provides non-hermitian chiral Hamiltonians with zero energy outliers. 
\hfill $\diamond$
}
\end{remark}

In the remainder of this section, various other properties of the set $\LimitSet$ are proved.

\begin{lemma}
\label{lem-GammaComp}
The set $\LimitSet$ is compact.
\end{lemma}

\noindent {\bf Proof.}
As the $z_j(E)$ are continuous in $E$, the set $\LimitSet$ is closed. Moreover, $\LimitSet \subset\spec(H^R)$ by Proposition~\ref{prop-Gamma0Include}. As $H^R$ is bounded, this also shows that $\LimitSet$ is a bounded set.
\hfill $\Box$

\begin{lemma}
\label{lem-LimitSetNoIso}
The set $\LimitSet$ does not contain isolated points.
\end{lemma}

\noindent {\bf Proof.}
Suppose $E$ is an isolated point of $\LimitSet \setminus \mathcal{F}$ so that $|z_L(E)|= |z_{L+1} (E)|$ and, in particular, $z_L(E)\not=z_{L+1} (E)$. These two eigenvalues are isolated and hence holomorphic at $E$. Let us denote these two holomorphic eigenvalues by $\zeta_L(E)$ and $\zeta_{L+1}(E)$. This eliminates discontinuities of the functions $z_L(E)$ and $z_{L+1}(E)$ resulting from labelling according to the modulus. Note that, if there are only two eigenvalues with modulus $|z_L(E)|$ at $E$, then $\{z_L(E+\epsilon),z_{L+1} (E+\epsilon)\}=\{\zeta_L(E+\epsilon),\zeta_{L+1}(E+\epsilon)\}$ locally, namely for $\epsilon$ small enough. Note now, that the labeling of the eigenvalues is determined only up to permutation of the labels among eigenvalues of the same modulus. Since $E \in \Lambda$ is isolated, (because $\mathcal{F}$ is finite,) it follows that one always could have chosen the labeling such that $|\zeta_L(E+\epsilon)|\not=|\zeta_{L+1}(E+\epsilon)|$ for all $\epsilon\not=0$ sufficiently small. Now $\epsilon = 0$ is a local maximum or minimum of the function $\epsilon\mapsto |\zeta_L (E+\epsilon) / \zeta_{L+1} (E+\epsilon)|$, because $E$ is isolated in $\LimitSet$ so that the quotient does not take the value $1$ in a some pointed neighborhood $B_{\delta}(0)\setminus \{0\}$. But then the maximum modulus principle would imply that $|\zeta_L (E+\epsilon)| = |\zeta_{L+1} (E+\epsilon)|$ for all $\epsilon$ in an open neighborhood of $0$. Then also $|z_L (E+\epsilon)| = |z_{L+1} (E+\epsilon)|$ for $\epsilon\in B_{\delta}(0)$, a contradiction to the isolation of $E$. 

\vspace{.1cm}

Suppose now $E\in \LimitSet \cap \mathcal{F}$ is an isolated point of $\LimitSet$. Then not necessarily all of the $z_l$ are holomorphic in a neighborhood of $E$. However, since they are the zeros of a polynomial whose coefficients are polynomials in $E$, they can be given by a Puiseux series in $E$ with non-zero convergence radius. Hence there exist  $p_{l,E}\in \NM$ (compare \cite{Kat}) such that the functions
\begin{equation}
\label{eq-Puiseux}
\epsilon\,\mapsto\,\tau_l^E (\epsilon) = \zeta_l (E+\epsilon^{p_{l,E}})
\end{equation}
are holomorphic in a neighborhood of $\epsilon = 0$, where the $\zeta_l$ are again a choice of eigenvalue making the $\tau_{l}^{E}$ holomorphic. Repeating the previous argument for the pair of functions $\epsilon \mapsto \tau_L^E(\epsilon^{p_{L+1,E}})$ and $\epsilon \mapsto \tau_{L+1}^E(\epsilon^{p_{L,E}})$ shows that also a point $E\in \LimitSet \cap \mathcal{F}$ cannot be isolated in $\LimitSet$.
\hfill $\Box$

\begin{lemma}
\label{lem-Boundary}
The set $\partial\LimitSet$ locally consists of a finite union of analytic arcs.
\end{lemma}

\noindent {\bf Proof.} Let $E\in\partial\LimitSet$. Then
$$
|z_{L-a-1}(E)| \,<\,|z_{L-a}(E)|\, = \,\ldots \,= \,|z_{L+b}(E)| \,<\, |z_{L+b+1}(E)|
$$
for some integers $a\geq 0$ and $b\geq 1$. Using the function $\tau_l^E$ defined in \eqref{eq-Puiseux}, let us introduce  the holomorphic functions
$$
\chi_{l,k}^E(\epsilon) 
\;=\; 
\frac{\tau_l^E (\epsilon^{p_{k,E}})}{\tau_k^E (\epsilon^{p_{l,E}})}
\;=\; 
\frac{\zeta_l(E+\epsilon^{p_{l,E}p_{k,E}})}{\zeta_k (E+\epsilon^{p_{k,E}p_{l,E}})}
\;.
$$
(Instead of the power $p_{l,E}p_{k,E}$ it is also possible to work with $q_{l,k}=\mbox{lcm}(p_{l,E},p_{k,E})$, namely the lowest common multiple of $p_{l,E}$ and $p_{k,E}$.)
There is an open neighborhood $V\subset\CM$ of $0$ such that for all $\epsilon\in V$ one has 
$$
|\tau_1^E(\epsilon)|\, , \,\ldots \,,\, |\tau_{L-a-1}^E(\epsilon)|\, <\, |\tau_{L-a}^E(\epsilon)| \,, \,\ldots\, ,\, |\tau_{L+b}^E(\epsilon)|\, < \, |\tau_{L+b+1}^E(\epsilon)|\, ,\, \ldots\, ,\, |\tau_{2L}^E(\epsilon)|
\;.
$$
Let now $\Pp\subset \{L-a,\dots , L+b\}^2$ be the set of pairs $(l,k)$ of indices $l<k$ such that $\chi_{l,k}$ is not locally constant around $0$ and $\zeta_l,\zeta_k$ is the pair of middle eigenvalues somewhere on $E+V\setminus\{0\}$. Note that $\Pp$ is non-empty, since $E\in \partial\LimitSet$ (because then in some direction of $\epsilon$ the middle two eigenvalues have to have different modulus so that the corresponding $\chi$ cannot be constant). For all $(l,k)\in \Pp$ let us set
$$
\gamma_{l,k} 
\;=\; 
\{ \epsilon \in V \colon |\chi^E_{l,k}(\epsilon)| = 1 \}
\;.
$$
Since all $\chi^E_{l,k}$ are holomorphic and non-constant around $0$, there is an $n\in\NM$ such that
\begin{equation}
\label{eq-Behavior}
\chi^E_{l,k}(\epsilon) 
\;=\;
\chi^E_{l,k}(0)\big(1\,+\,c_n(E)\,\epsilon^n\,+\,\mathcal{O}(\epsilon^{n+1}) \big)
\;,
\end{equation}
with $c_n(E)\not=0$.  Note that $E\mapsto c_1(E)$ is analytic on $\mathbb{C}\setminus\Ff$ and cannot vanish identically around points $E\in \partial \LimitSet \setminus \Ff$, for the following reason. For all $\epsilon$ and $\delta$ small enough,
$$
\chi_{l,k}^{E}(\epsilon + \delta) 
\;=\; 
\frac{z_{l}(E+\epsilon+\delta)}{z_{k}(E+\epsilon+\delta)} 
\;=\; 
\chi_{l,k}^{E+\delta}(\epsilon) \;.
$$
Since both $\chi_{l,k}^{E}$ and $\chi_{l,k}^{E+\delta}$ are analytic around $0$, they have a convergent power series expression there. Comparing the terms of order $\epsilon$, one finds
$$
c_1(E+\delta) 
\;=\; 
\sum_{m > 0} m c_m(E) \delta^{m-1} 
\;.
$$
This is the power series expansion of $c_1$ around $E$. Now if $c_1$ would vanish identically around $E$, it follows from the expansion that $c_m(E)=0$ for all $m$. However, this means that $\chi_{l,k}^{E}$ has to be constant around $0$, which it is not. Therefore $c_1$ does not vanish identically around any $E\in \partial \LimitSet \setminus \Ff$. Hence the map $E\mapsto c_1(E)$ has a discrete set of zeros. (Recall that $\Ff$ is discrete by Hypothesis B.) One simply has $n=1$ except on a discrete set of points $E$, but these points with $n\geq 2$ are singular points of interest. Namely, \eqref{eq-Behavior} implies that $\gamma_{l,k}$ is the union of $n$ analytic curves in $V$ going through $0\in V$, possibly after choosing $V$ sufficiently smaller. Consider now $W=\bigcap_{(l,k)\in \Pp} V^{p_{l,E} p_{k,E}}$ where $V^q=\{\epsilon^q:\epsilon\in V\}$. Then $E+W$ is a neighborhood of $E$. 
Recalling that $\tau_l^E (\epsilon) = \zeta_l (E+\epsilon^{p_{l,E}})$ with $p_{l, E}>0$, one finds that on the open neighborhood $E+W$, the points in $\partial\LimitSet$ are locally near $E$ precisely given by $E + W\cap\bigcup_{(l,k)\in \Pp}(\gamma_{l,k})^{p_{l,E} p_{k,E}}$. This latter set consists of a finite union of analytic arcs.
\hfill $\Box$

\begin{remark}
\label{rem-EndJordan}
{\rm
Let us consider a point $E\in\LimitSet\cap\Ff=\{E\in\LimitSet\,:\,z_L(E)=z_{L+1}(E)\}$ for which the Puiseux expansion of some of the middle eigenvalues is non-trivial, namely that $p>1$ in \eqref{eq-Puiseux}. Then $\Tt^E$ necessarily has a Jordan block for one of the middle eigenvalues (which is hence degenerate, reflecting again that $E\in\Ff$). If there are only two middle eigenvalues and the coefficient  in \eqref{eq-Behavior} is $n=1$, then $\Pp$ consists of just this one pair and there is just one set $\gamma$ given by an analytic line. Taking the power the gives a set $\gamma^{p^2}$, that is a cusp or half-line. This is observed in numerical examples in Figures~\ref{fig-HN} and \ref{fig-Ser2} at the extremities of $\LimitSet$. 
}
\hfill $\diamond$
\end{remark}

\begin{corollary}
\label{coro-Arcs}
If the interior of $\LimitSet$ is empty, the set $\LimitSet$ locally consists of a finite union of analytic arcs.
\end{corollary}

\noindent {\bf Proof.} As $\LimitSet$ is closed and the closure is always given by the disjoint union $\overline{\LimitSet}=\partial\LimitSet\,\mathring{\cup}\,\mathring{\LimitSet} $ of the boundary and the interior, it follows from the hypothesis that $\LimitSet=\partial\LimitSet$ so that the claim holds by Lemma~\ref{lem-Boundary}.
\hfill $\Box$

\vspace{.2cm}

Let us provide a simple criterion assuring that $\LimitSet$ has empty interior so that the above corollary applies.

\begin{lemma}
\label{lem-EmpInt}
If $T$ and $R$ have no eigenvalues of same modulus, the set $\LimitSet\subset\CM$ has empty interior. If $T$ and $R$ have no double eigenvalues, then {\rm Hypothesis B} holds. 
\end{lemma}

\noindent {\bf Proof.} 
This will follow from an asymptotic analysis of the spectrum of the transfer matrices $\Tt^E$ and $\widetilde{\Tt}^E$ at $|E|\to\infty$.  For this purpose, let us set $\xi=\frac{1}{E}$ and $\Ss^\xi=\frac{1}{E} \Tt^E$. The function $\xi\in\CM\setminus\{0\}\mapsto \Ss^\xi$ is analytic and has a removable singularity at $0$, namely
$$
\Ss^\xi\;=\;
\begin{pmatrix}
T^{-1} & 0 \\ 0 & 0
\end{pmatrix}
\;+\;\Oo(\xi)
\;.
$$
By analytic perturbation theory \cite{Kat},  $\Ss^\xi$ has $L$ non-vanishing eigenvalues $\mu_1(\xi),\ldots,\mu_{L}(\xi)$ of modulus of order $1$, and the others are of order $\xi$. More precisely, if $\lambda_1,\ldots,\lambda_L$ are the eigenvalues of $T$, then for an appropriate choice of the labelling one has $|\mu_l(\xi)-\frac{1}{\lambda_l}|=\Oo(\xi)$ for $l=1,\ldots, L$. As $\spec(\Tt^E)=E\,\spec(\Ss^E)$, the spectrum of $\Tt^E$ has $L$ eigenvalues $E\mu_1(\frac{1}{E}),\ldots,E\mu_L(\frac{1}{E})$ of order $E$ and $L$ eigenvalues of order $1$. As $E\mu_l(\frac{1}{E})=\frac{E}{\lambda_l}+\Oo(1)$ for $l=1,\ldots,L$, the hypothesis that all $\lambda_l$ have different moduli implies that these $L$ eigenvalues all have different modulus for $E$ sufficiently large. Moreover, one can run the same argument for $\widetilde{\Tt}^E$ based on the hypothesis on the spectrum of $R$. Due to \eqref{eq-TransferConjugateSpec} this shows that none of the eigenvalues of $\Tt^E$ have equal modulus for $E$ sufficiently large. Due to Lemma~\ref{lem-ConstantModulus} this implies the first claim. The second follows in a similar manner.
\hfill $\Box$

\subsection{Scaling operation}
\label{sec-Scaling}

Let us introduce an invertible scaling matrix by
$$
S_N
\;=\;
\diag(\sca,\sca^2,\ldots,\sca^{N})
\;,
\qquad
\sca\not=0
\;.
$$
Then
\begin{equation}
\label{eq-HamGenScale}
S_N^{-1}\,H_N\,S_N
\;=\;
\begin{pmatrix}
V & \sca T & & 
\\
\sca^{-1} R & \ddots & \ddots & 
\\
& \ddots & \ddots &  \sca T
\\
& & \sca^{-1} R & V
\end{pmatrix}
\;.
\end{equation}
In particular, for all $\sca>0$,
$$
\spec(H_N)
\;=\;
\spec(S_N^{-1}\,H_N\,S_N)
\;.
$$
This suggests to introduce the half-sided block-Toeplitz matrix $\HRs$ with entries as in \eqref{eq-HamGenScale}. Let us stress that for $\sca\not=1$, the operators $\HRs$ and $\HR$ are not conjugate to each other, and are therefore typically {\it not} isospectral.  The symbol of $\HRs$ will be denoted by $H^\sca(z)$. Comparing with \eqref{eq-SymbDef}, it is hence obtained by scaling 
$$
H^\sca(z)
\;=\;
\sca^{-1}\,R\,z^{-1}\,+\,V\,+\,\sca\,T\,z
\;=\;
H(\sca\, z)
\;.
$$
Its associated symbol is according to \eqref{eq-AssSymbDef} given by
$$
\widetilde{H^\sca}(z)
\;=\;
H^\sca(\tfrac{1}{z})
\;=\;
H(\tfrac{\sca}{z})
\;=\;
\widetilde{H}(\tfrac{z}{\sca})
\;=\;
\widetilde{H}^{\frac{1}{\sca}}(z)
\;.
$$
Then the corresponding determinants defined by $D^s(z,E)={\det}_L(H^s(z)-E\one)$ as in \eqref{eq-DetSymb} satisfy 
$$
D^\sca(z,E)
\;=\;
D(\sca z,E)
\;,
\qquad
\widetilde{D}^\sca(z,E)
\;=\;
\widetilde{D}(\sca z,E)
\;.
$$
Moreover, let  $\Tt^{\sca,E}$ and $\widetilde{\Tt}^{\sca,E}$ denote the transfer matrices associated to the rescaled Hamiltonian symbols $H^\sca$ and $\widetilde{H^\sca}$. One finds
\begin{equation}
\label{eq-TransferScale}
\Tt^{\sca,E}\;=\;\sca^{-1}\,\Tt^{E}
\;,
\qquad
\widetilde{\Tt}^{\sca,E}\;=\;\sca\,\widetilde{\Tt}^{E}
\;.
\end{equation}
Then there are also associated Riesz projections 
$$
\Rr^{\sca,E}_\scar
\;=\;
\oint_{\TM_\scar}\frac{dz}{2\pi\imath}\;(z\,\one-\Tt^{\sca,E})^{-1} 
\;.
$$
They do not contain supplementary information though because the range and kernel of $\Tt^{\sca,E}$ is independent of $\sca$ due to \eqref{eq-TransferScale}. Actually, also the change of variable formula \eqref{eq-RieszScale} shows
$$
\Rr^{\sca,E}_\scar
\;=\;
\Rr^{E}_{\sca \scar}
\;.
$$
Therefore Proposition~\ref{prop-KernelCalc} immediately implies the following:

\begin{corollary}
\label{coro-JKernelCalc}
Suppose that $\spec(\Tt^E)\cap\TM_{\sca }=\emptyset$. Then 
$$
\dim\big(\Ker({\QFct}^{(1)}_{\sca}(E))\big) 
\;=\; 
\dim\big(\Ker(\HRs-E\,\one)\big)
\,+\,
\dim\big(\Ker(\HLs - E\,\one)\big)
\;.
$$ 
\end{corollary}

Let us now start with the application of the scaling to the spectral analysis. The argument of Proposition~\ref{prop-Gamma0Include} applies to each $\HRs$ and therefore
$$
\LimitSet \;\subset\;\bigcap_{\sca>0}\spec(\HRs)
\;.
$$
The inverse implication holds for $L=1$, see Corollary~\ref{coro-SchmidtSpitzer} below. In the following it will be shown that a modification is necessary in the matrix-valued case $L\geq 2$. Due to \eqref{eq-WindComp} applied to $H^\sca$ the scaling \eqref{eq-TransferScale} implies that, as long as $\spec(\sca\,\Tt^E)\cap \TM^1=\emptyset$ or equivalently $\HRs-E\,\one$ is a Fredholm operator (see Theorem~\ref{theo-NoetherGoberg}), 
\begin{align*}
\Ind(\HRs-E\,\one)
&
\;=\;
\Wind^E(H^\sca )
\;=\;
-\,L\;+\;\sum_{|z|<1}\,m_z({\Tt}^{\sca,E})
\;=\;
-\,L\;+\;\sum_{|z|<\sca}\,m_z({\Tt}^E)
\;.
\end{align*}
This implies that, one has $\spec(s\,\mathcal{T}^E)\cap \mathbb{T}^1 \neq\emptyset$ or $\Wind^E(H^s ) \neq 0$ for all $s > 0$ if and only if the two middle eigenvalues of ${\Tt}^E$ are of same modulus. One therefore has the characterization
\begin{align}
\LimitSet 
&
\;=\;
\big\{E\in\CM\,:\,
\forall\,\,\sca>0 \mbox{ either }\HRs-E\,\one \;\mbox{ {\sl not} Fredholm or }\Wind^E(H^s)\not=0\big\}
\label{eq-Gamma0CharBis}
\\
&
\;=\;
\big\{E\in\CM\,:\,
\forall\,\,\sca>0 \mbox{ either }\HRs-E\,\one \;\mbox{ {\sl not} Fredholm or }\Ind(\HRs-E\,\one)\not=0\big\}
\;.
\label{eq-Gamma0Char}
\end{align}
One can carry out the same argument with $\HL$ (or simply use \eqref{eq-FredPropEquiv} and \eqref{eq-IndSum}) to deduce
\begin{equation}
\label{eq-Gamma0Char2}
\LimitSet 
\;=\;
\big\{E\in\CM\,:\,
\forall\,\,\sca>0 \mbox{ either }\HLs-E\,\one \;\mbox{ {\sl not} Fredholm or }\Ind(\HLs-E\,\one)\not=0\big\}
\;.
\end{equation}
Hence the set $\LimitSet$ is equivalently characterized by either $\HR$ or $\HL$ and their scaling.

\vspace{.2cm}

Now let  $E\in\bigcap_{\sca>0}\spec(\HRs)$. Then for all $\sca>0$, one of the following three mutually exclusive cases holds true:
\begin{enumerate}
\item[{\rm (i)}] $\HRs-E\,\one$ is not Fredholm;
\vspace{-.2cm}
\item[{\rm (ii)}] $\HRs-E\,\one$ is Fredholm with non-vanishing index;
\vspace{-.2cm}
\item[{\rm (iii)}] $\HRs-E\,\one$ is Fredholm with vanishing index, but $\Ker(\HRs-E\,\one)\not=\{0\}$ is non-trivial.
\end{enumerate}
In the last point (iii), one can also replace $\Ker(\HRs-E\,\one)\not=\{0\}$ by the statement that $\HRs-E\,\one$ is not invertible (or, as matter of fact, that  $\Ker((\HRs-E\,\one)^*)\not=\{0\}$). Comparing with \eqref{eq-Gamma0Char}, one sees that $\LimitSet $ contains all $E\in\bigcap_{\sca>0}\spec(\HRs)$ satisfying either (i) or (ii). It is hence natural to introduce the set of $E$ for which (iii) holds, which is done in the next section.

\subsection{The set $\Outliers$ of spectral outliers}
\label{sec-Lambda}

As motivated in the end of the last section, let us introduce the set
$$
\Outliers^R 
\;=\; 
\big\{
E\in\CM \setminus \LimitSet  \;:\;\forall\; \sca>0 \mbox{ with } \Ind(\HRs-E\,\one)=0 \text{ one has } \Ker(\HRs-E\,\one)\not=\{0\}
\big\}
\;,
$$
where the condition $\Ind(\HRs-E\,\one)=0$ includes the Fredholm property of $\HRs-E\,\one$. Alternatively
$$
\Outliers^R 
\;=\; 
\big\{
E\in\CM \setminus \LimitSet  \;:\;\forall\; \sca>0 \mbox{ with } \Wind^E(H^\sca )=0 \text{ one has } \Ker(\HRs-E\,\one)\not=\{0\}
\big\}
\;,
$$
where again the condition $\Wind^E(H^\sca )=0$ contains that the winding number is well-defined. By construction, the discussion at the end of Section~\ref{sec-Scaling} thus implies
\begin{equation}
\label{eq-PreSS}
\bigcap_{\sca>0}\spec(\HRs) \;=\; \LimitSet  \cup \Outliers^R
\;.
\end{equation}
Moreover, if the defining property for $E$ being in $\Outliers^R$ holds for some $\sca$, it actually holds for all $\sca$ in some interval, as shows the proof of the following lemma:

\begin{lemma}
\label{lem-scaIndependence}
$$
\Outliers^R 
\,=\,
\big\{E\in\CM \setminus \LimitSet  \,:\,\exists\; \sca>0\; \mbox{\rm with } \Ind(\HRs-E\,\one)=0\; \text{\rm and } \HRs-E\,\one \;\text{\rm not invertible}\big\}
\;.
$$
\end{lemma}

\noindent {\bf Proof.}
By definition,
$$
\Outliers^R \,\subset\, \big\{E\in\CM \setminus \LimitSet  \,:\,\exists\; s>0 \mbox{ with } \Ind(\HRs-E\,\one)=0 \text{ and } \HRs-E\,\one \text{ not invertible}\big\}
\,.
$$
Let now $E$ be an element of the r.h.s. and let $\sca>0$ be such that $\HRs-E\,\one$ is Fredholm with index zero, but also not invertible. This means that $\HRs-E\,\one$ has a non-trivial kernel. This can be reformulated in terms of the $L$-dimensional subspace $\Ee^{\sca,E}_1\subset\CM^{2L}$ containing all generalized eigenvectors of $\Tt^{\sca,E}$ with eigenvalues of modulus strictly less than $1$. In fact, just as in \eqref{eq-EVHR},
$$
\dim\big(\Ker(\HRs-E\,\one)\big)
\;=\;
\dim\big(\Ee^{\sca,E}_1\cap\Dd_0\big)
\;.
$$
Now $\Tt^{\sca,E}=\sca^{-1}\,\Tt^E$ and thus $\Ee^{\sca,E}_1$ is independent of $s$ as long as no eigenvalue of $\Tt^{\sca,E}$ touches the unit circle (in which case the Fredholm condition is violated). Hence  $\Ker(\HRs-E\,\one)$ is also independent of $\sca$ as long as $\sca^{-1}|z_L(E)|<1<\sca^{-1}|z_{L+1}(E)|$, {\it i.e.} for $s\in (|z_{L}(E)|,|z_{L+1}(E)|)$.
\hfill $\Box$

\vspace{.2cm}

In connection with Lemma~\ref{lem-scaIndependence}, let us note that $E\in\CM \setminus \LimitSet$, so $E\not\in\LimitSet$, always implies that there exists $s$ with $\Ind(\HRs-E\,\one)=\Wind^E(H^s)=0$. Hence  $\Outliers^R$ is characterized by the singularity of $\HRs-E\,\one$ for such $s$. Furthermore, as $\Ee^{\sca,E}_1=\Ee^{E}_{\sca}$ for $s\in (|z_{L}(E)|,|z_{L+1}(E)|)$, the proof of Lemma~\ref{lem-scaIndependence} also allows to rewrite the definition of the set $\Outliers^R$ merely in terms of spectral properties of the transfer matrix $\Tt^E$:
$$
\Outliers^R 
\,=\,
\big\{E\in\CM  \,:\,|z_L(E)|<|z_{L+1}(E)|\; \text{\rm and } 
\Ee^E_s\cap\Dd_0\not=\emptyset\; \text{\rm for }\sca\in (|z_{L}(E)|,|z_{L+1}(E)|)
\big\}
\;.
$$
From this representation one immediately deduces the following:

\begin{proposition}
\label{prop-L1Lambda}
For $L=1$, one has $\Outliers^R=\emptyset$.
\end{proposition}

\noindent {\bf Proof.} For $E\in\Outliers^R$ the $2\times 2$ transfer matrix $\Tt^E$ given in \eqref{eq-TraDef}  has two distinct eigenvalues. For $s$ satisfying $|z_{1}(E)|<s<|z_{2}(E)|$, the subspace $\Ee^E_s\subset\CM^2$ is $1$-dimensional and spanned by an eigenvector of $\Tt^E$. For $E\in\Outliers^R$, this eigenvector has to be constant multiple of $\binom{1}{0}$. However,  $\Tt^E$ has no eigenvector of this form as can be readily seen by looking at the lower component of the eigenvalue equation for $\Tt^E$.
\hfill $\Box$

\vspace{.2cm}

Proposition~\ref{prop-L1Lambda} together with \eqref{eq-PreSS} directly implies the next result which goes back to the work of Schmidt and Spitzer \cite{SS} (more precisely, it is the result $A=C$ therein).

\begin{corollary}
\label{coro-SchmidtSpitzer}
For $L=1$, one has 
\begin{equation}
\label{eq-SS}
\LimitSet \;=\;\bigcap_{\sca>0}\spec(\HRs)
\;.
\end{equation}
\end{corollary}

Another well-known result in the case $L=1$ states that $\LimitSet$ is connected \cite{Uhl}.

\begin{proposition}
\label{prop-LambdaDiscrete}
The set $\Outliers^R$ is discrete in $\CM$.
\end{proposition}

\noindent {\bf Proof.} Let us rewrite the definition of $\Outliers^R$ as
$$
\Outliers^R 
\,=\,
\big\{E\in\CM \setminus \LimitSet  \,:\,\exists\; \sca>0\; \mbox{\rm with } \Wind^E(H^\sca)=0\; \text{\rm and }\dim(\Ker( \HRs-E\,\one))\geq 1\big\}
\;.
$$
The condition $\Wind^E(H^\sca)=0$ is open in $E$ (and actually also in $\sca$). On the other hand, $\dim(\Ker( \HRs-E\,\one))\geq 1$ implies that $\dim\big(\Ker({\QFct}^{(1)}_{\sca}(E))\big)\geq 1$ by Corollary~\ref{coro-JKernelCalc}, so that
$$
\Outliers^R 
\,\subset\,
\big\{E\in\CM \setminus \LimitSet  \,:\,\exists\; \sca>0\; \mbox{\rm with } \Wind^E(H^\sca)=0\; \text{\rm and }\det({\QFct}^{(1)}_{\sca}(E))=0  \big\}
\;.
$$
But $E\mapsto\det({\QFct}^{(1)}_{\sca}(E))$ is analytic (except at points where the Fredholm condition is not satisfied) and not identically $0$. Hence its zeros form a discrete set and this implies that also $\Outliers^R$ is discrete.
\hfill $\Box$

\vspace{.2cm}

Up to now, only the right-sided Toeplitz operators $\HRs$ were considered. As $\LimitSet$ can also be expressed in terms of the $\HLs$ by \eqref{eq-Gamma0Char2}, let us also introduce 
\begin{equation}
\label{eq-scaIndependenceL}
\Outliers^L
\;=\; 
\big\{
E\in\CM \setminus \LimitSet  \;:\;\forall\; \sca>0 \mbox{ with } \Ind(\HLs-E\,\one)=0 \text{ one has } \Ker(\HLs-E\,\one)\not=\{0\}
\big\}
\;.
\end{equation}
Then all of the above properties directly transpose. In particular,
$$
\bigcap_{\sca>0}\spec(\HLs) \;=\; \LimitSet  \cup \Outliers^L
\;,
$$
and the equivalents of Lemma~\ref{lem-scaIndependence} and Propositions~\ref{prop-L1Lambda} and \ref{prop-LambdaDiscrete} hold. Note that it is possible (and not in contradiction to Proposition~\ref{prop-KernelCalc}) that $\Outliers^L\cap\Outliers^R\not=\emptyset$, but this is non-generic (it is generic within the set of chiral Hamiltonians though, see Section~\ref{sec-ZeroOutlier}). Finally, let us set
$$
\Outliers
\;=\;
\Outliers^L\cup\Outliers^R
\;.
$$
Then the proof of Proposition~\ref{prop-LambdaDiscrete} combined with Corollary~\ref{coro-JKernelCalc} shows that 
\begin{align}
\Outliers
&
\;=\;
\nonumber
\big\{E\in\CM \setminus \LimitSet  \,:\,\exists\; \sca>0\; \mbox{\rm with } \Wind^E(H^\sca)=0\; \text{\rm and }\det({\QFct}^{(1)}_{\sca}(E))=0  \big\}
\\
&
\;=\;
\big\{E\in\CM \setminus \LimitSet  \,:\,\exists\; \sca>0\; \mbox{\rm with } |z_L(E)|<\sca<|z_{L+1}(E)|\; \text{\rm and }\det({\QFct}^{(1)}_{\sca}(E))=0  \big\}
\;.
\label{eq-LambdaJ}
\end{align}
It will be shown in \eqref{eq-LambdaJ2} below that this coincides with the definition \eqref{eq-OutliersDef} given in the introduction. Let us point out that this set $\Outliers$ also appears in eq. (1.14) in \cite{Del} where it is denoted $G_0$.  In Widom's work \cite{Wid1} there is no notation for it, but it appears in Theorem~6.1 as $\det({\QFct}_{\sca}^{(1)})$ is the function denoted by $E$ in \cite{Wid1}. Widom also shows that $\det({\QFct}_{\sca}^{(1)})$ is equal to a certain Fredholm determinant, but this will not be of relevance in this work.

\subsection{Zero as spectral outlier of chiral Hamiltonians}
\label{sec-ZeroOutlier}

This short section is about the sets $\LimitSet$ and $\Outliers$ for block Hamiltonians \eqref{eq-HamGen} satisfying  the chiral symmetry \eqref{eq-ChiralSym}. In particular, $L$ is even and all the structural properties of the Hamiltonian and the transfer matrix discussed in Section~\ref{sec-ChiralOutliersHalf} hold. Hence, $\spec( \Tt^E ) = \spec(\Tt^{-E})$ and this implies that 
\begin{equation}
\label{eq-GammaLambdaSym}
\LimitSet 
\;=\; 
-\,\LimitSet
\;,
\qquad
\Outliers \;=\; -\,\Outliers
\;,
\end{equation}
namely both $\LimitSet$ and $\Outliers$ are reflection symmetric subsets of $\CM$. Hence the discrete set $\Outliers$ comes in pairs of points $(E,-E)$, except for the origin $0$ which is the only point that is invariant under reflection. It is hence of crucial interest whether $0\in\Outliers$ or $0\not\in\Outliers$. The following is Theorem~\ref{thm-ChiralIntro}.

\begin{proposition}
\label{prop-ZeroCriterion}
Let $H$ have the chiral symmetry \eqref{eq-ChiralSym} with off-diagonal entries $H_\pm$ as in \eqref{eq-HChrial} and suppose that $0\not\in\LimitSet$. If $\Wind^0(H^s_+)=-\Wind^0(H^s_-)\not=0$ for some  $s>0$, then $0\in\Outliers$.   
\end{proposition}

\noindent {\bf Proof.} By Proposition~\ref{prop-ChiralKernel}, the hypothesis implies that there exists an $s$ such that both $\Ker((H^s)^R)\not=\{0\}$ and $\Ker((H^s)^L)\not=\{0\}$ and, moreover, $\Ind(H^s )=\Wind^0(H^s)=0$. These facts together imply that $0\in\Outliers=\Outliers^L\cup\Outliers^R$ by Lemma~\ref{lem-scaIndependence} and \eqref{eq-scaIndependenceL}.
\hfill $\Box$

\subsection{Widom's determinant formula}
\label{sec-DetFormula}

\begin{theorem}
\label{theo-WidomFormula}
Let $E\not\in \Ff$. Then the characteristic polynomial is given by Widom's formula
\begin{equation}
\label{eq-WidomFormula}
{\det}_{NL}(H_N-E\,\one)
\;=\;
\sum_\IndSet G_I (E)^{N+1}\,q_I(E) 
\end{equation}
where the sum runs over all index sets $I\subset \{1,\ldots,2L\}$ of cardinality $L$ and the two functions are defined by
\begin{equation}
\label{eq-GDef}
G_\IndSet (E) \;=\; (-1)^L \det(R) \prod_{i\in\IndSet} \frac{1}{z_{i}(E)} 
\end{equation}
and
\begin{equation}
\label{eq-qIDef}
q_I(E)
\;=\;
{\det}_L\Big(
\oint_{\gamma_{\IndSet}^0}\frac{dz}{2\pi\imath\,z}\;(H(z)-E\,\one)^{-1}
\Big)
\;=\;
{\det}_L\Big(
\binom{0}{\one}^*
{\Rr}_\IndSet^E
\binom{\one}{0}\Big)
\;,
\end{equation}
where $\gamma_{\IndSet}^0$ is a path in $\CM$ which has a winding number $1$  around $0$ as well as  each  $z_{i}(E)$ for $i\in\IndSet$ and a winding number $0$ around all other eigenvalues of $\Tt^E$.
\end{theorem}

In contradistinction to Widom's work, in the present context it is possible to replace $\gamma_{\IndSet}^0$ in the definition of $q_I(E)$ by the path $\gamma_{\IndSet}$ which has a winding number $1$  around each  $z_{i}(E)$ for $i\in\IndSet$ and a winding number $0$ around all other eigenvalues of $\Tt^E$ as well as $0$. This holds because $z (H(z)-E\,\one)$ is a matrix-valued polynomial in $z$ and its eigenvalues are the same as the eigenvalues of $\Tt^E$ by \eqref{eq-TransferDet}, so that $0$ is not an eigenvalue due to the invertibility of $\Tt^E$. Let us also stress that, if $E$ is such that there are eigenvalues of same modulus, the ordering of the eigenvalues is ambiguous and hence the definition of $G_\IndSet (E)$ and $q_I(E)$ as well. However, Widom's formula does not inherit this ambiguity because it contains a sum over all index sets $\IndSet$ of cardinality $L$. Another way to circumvent the ambiguity is to choose one labelling at a point $E$ and then to analytically extend it to a neighborhood of $E$ (which is possible because $E\not\in\Ff$). 

\vspace{.2cm}

Let us stress that Theorem~\ref{theo-WidomFormula} is just a special case of Widom's Theorem~6.2 of \cite{Wid1} which also covers the case of non-invertible off-diagonal coefficients. The proof of Theorem~\ref{theo-WidomFormula} given in \cite{Wid1} is based on another formula from \cite{BSch} and is rather involved. Based on transfer matrix methods, it is here possible to provide a relatively short proof of Widom's formula. It will be shown below that this strategy also leads to a few new identities within the present restricted context. 

\vspace{.2cm}

Let us start out  by noting that
\begin{equation}
\label{eq-DetTransferRel}
{\det}_{NL}(H_N-E\,\one)
\;=\;
(-1)^{NL}\,{\det}_L(T)^{N}\,
{\det}_L
\Big(\binom{\one}{0}^*(\Tt^E)^N\binom{\one}{0}\Big)
\;.
\end{equation}
Indeed, one can start constructing all solutions at $E$ satisfying the Dirichlet boundary conditions at $1$ from \eqref{eq-SchrTra2} and then check whether any of these solution satisfying the Dirichlet boundary condition at $N$. Hence the multiplicity of the zero of the determinant on the r.h.s. is equal to the multiplicity of $E$ as an eigenvalue of $H_N$. Thus both sides of \eqref{eq-DetTransferRel} vanish exactly at the eigenvalues of $H_N$. Moreover, as both sides are polynomials of degree $NL$ in $E$, comparing the leading coefficient $E^{NL}$ immediately implies the identity. For the computation the last determinant, one can use the following lemma with $M=2L$. It is Lemma~6.3 from \cite{Wid1}, but again an alternative proof is provided.

\begin{lemma}
\label{lem-Rank1Det}
Let $A\colon \CM^M\to \CM^L$, $B\colon \CM^M \to \CM^M$ and $C\colon \CM^L\to \CM^M$ be linear maps with $B$ diagonalizable. Then there is a spectral decomposition $B = \sum_{i=1}^M b_i P_i$ where the $P_i$ are rank one idempotents onto the eigenvector associated to eigenvalue $b_i$ and
$$
{\det}_L(A B C) \;=\; \sum_{|I|=L} \Big(\prod_{i\in I} b_i\Big)\, {\det}_L \Big(\sum_{i\in I} A P_i C\Big)
$$
where the sum runs over subsets $I$ of $\{1,2,\dots,M\}$ of cardinality $|I|=L$.
\end{lemma}

\noindent {\bf Proof.} 
The existence of the spectral decomposition follows immediately from the diagonalizability of $B$.
Moreover, the exterior power $\bigwedge^L B \colon \bigwedge^L \CM^M \to \bigwedge^L \CM^M$ is also diagonalizable, since $B$ is diagonalizable, so for the same reason we have a decomposition $\bigwedge^L B = \sum_I b_I P_I $, where the $P_I$ are rank one projections onto the eigenvector associated to eigenvalue $b_I$ of $\bigwedge^L B$. Note that given a basis of eigenvectors $(v_i)_{i=1}^M$ of $B$ for $\CM^M$, the eigenvalues $b_I$ can be indexed by choices $I$ of subsets of $\{1,2,\dots,M\}$ of cardinality $L$ and can then exactly be given by $b_I = \prod_{i\in I}b_i$. Moreover, an eigenvector $v_I$ associated to $b_I$ can be given by $v_{i_1}\wedge v_{i_2}\wedge\dots\wedge v_{i_L}$ with $i_j\in I$ and $i_j<i_k$ if $j<k$. Hence 
$$
P_I 
\;=\; 
\bigwedge^L \sum_{i\in I} P_i
$$
which can be checked explicitly by evaluation on a basis of eigenvectors of $\bigwedge^L B$. Let $(e_i)_{i=1}^L$ be a basis of $\CM^L$. One finds
\begin{align*}
{\det}_L (A B C) &\,e_1 \wedge e_2 \wedge \dots \wedge e_L 
\;=\; 
(\bigwedge^L A \bigwedge^L B \bigwedge^L C)\, e_1 \wedge e_2 \wedge \dots \wedge e_L \\
 & \;=\; \Big(\bigwedge^L A \Big(\sum_I b_I P_I \Big) \bigwedge^L C\Big) \,e_1 \wedge e_2 \wedge \dots \wedge e_L \\
 & \;=\; \sum_I b_I \big(\bigwedge^L A P_I \bigwedge^L C\big)\, e_1 \wedge e_2 \wedge \dots \wedge e_L \\
 & \;= \;\sum_I \Big(\prod_{i\in I}b_i\Big) \big(\bigwedge^L A \bigwedge^L \sum_{i\in I} P_i \bigwedge^L C\big)\, e_1 \wedge e_2 \wedge \dots \wedge e_L \\
 & \;=\; \sum_{|I|=L} \Big(\prod_{i\in I}b_i\Big)\, {\det}_L\Big(\sum_{i\in I} A P_i C\Big) \,e_1 \wedge e_2 \wedge \dots \wedge e_L \;,
\end{align*}
which implies the claim.
\hfill $\Box$

\vspace{.2cm}

\noindent {\bf Proof} of Theorem~\ref{theo-WidomFormula}. First of all, the particular form \eqref{eq-TraDef} of the transfer matrix  allows to rewrite \eqref{eq-DetTransferRel} as
$$
{\det}_{NL}(H_N-E\,\one)
\;=\;
(-1)^{NL}\,{\det}_L(T)^{N+1}\,
{\det}_L
\Big(\binom{0}{\one}^*(\Tt^E)^{N+1}\binom{\one}{0}\Big)
\;.
$$
Next recall that the eigenvalues of the transfer matrix $\Tt^E$ are ${z}_i (E)$ which by the assumption $E\not\in\Ff$ are pairwise distinct. Then $\Tt^E$ is diagonalizable and an application of Lemma~\ref{lem-Rank1Det} gives
\begin{equation}
\label{eq-InterMed}
{\det}_{NL}(H_N-E\,\one)
\;=\;
(-1)^{NL}\,{\det}_L(T)^{N+1}\,
\sum_I \Big(\prod_{i\in I} {z}_i (E)^{N+1}\Big)
{\det}_L
\Big(\binom{0}{\one}^* {\Rr}_\IndSet^E \binom{\one}{0}\Big)
\;,
\end{equation}
where the sum is as in the statement of the theorem and ${\Rr}_\IndSet^E$ is the Riesz projection associated to the eigenvalues ${z}_i (E)$ with $i\in I$, namely
\begin{equation}
\label{eq-RI}
{\Rr}_\IndSet^E
\;=\;
\oint_{{\gamma}^0_\IndSet}\frac{dz}{2\pi\imath}\;(z\,\one-\Tt^E)^{-1}
\;,
\end{equation}
with  ${\gamma}^0_\IndSet$ also as above (which may here be replaced by $\gamma_\IndSet$ because $\Tt^E$ is invertible). As to the product of eigenvalues, let us use
$$
\Big(\prod_{i\in\IndSet}{z}_{i}(E)\Big)\Big(\prod_{i\in\IndSet^c}{z}_{i}(E)\Big)
\;=\;
{\det}_{2L}(\Tt^E)\;=\;\frac{{\det}_L(R)}{{\det}_L(T)}
\;,
$$
where $\IndSet^c=\{1,\ldots,2L\}\setminus \IndSet$ is the complementary set of $\IndSet$. Hence replacing the definition of  $G_\IndSet(E)$, one finds 
\begin{equation}
\label{eq-GG}
G_I(E)G_{I^c}(E)\;=\;{\det}_L(R)\,{\det}_L(T)
\;,
\end{equation}
so that
$$
{\det}_{NL}(H_N-E\,\one)
\;=\;
(-1)^{L}\,
\sum_I G_{\IndSet^c}(E)^{N+1}\,
{\det}_L
\Big(\binom{0}{\one}^* {\Rr}_\IndSet^E \binom{\one}{0}\Big)
\;.
$$
Due to
$$
\binom{0}{\one}^* \Rr_\IndSet^E \binom{\one}{0} \,+\, \binom{0}{\one}^* \Rr_{\IndSet^c}^E \binom{\one}{0}
\;=\;  \binom{0}{\one}^* \one_{2L} \binom{\one}{0} 
\;=\;  0
\;,
$$
this implies
$$
{\det}_{NL}(H_N-E\,\one)
\;=\;
\sum_\IndSet G_{\IndSet}(E)^{N+1}\,
{\det}_L
\Big(\binom{0}{\one}^* {\Rr}_{\IndSet}^E \binom{\one}{0}\Big)
\;,
$$
because the indices in $\sum_\IndSet$ and $\sum_{\IndSet^c}$ run through the same set. Finally replacing \eqref{eq-TransferResol} gives
$$
\binom{0}{\one}^*
{\Rr}_\IndSet^E
\binom{\one}{0}
\;=\;
\oint_{{\gamma}^0_\IndSet}\frac{dz}{2\pi\imath\,z}\;({H}(z)-E\,\one)^{-1}
\;.
$$
Taking the determinant shows the second equality in \eqref{eq-qIDef}, completing the proof.
\hfill $\Box$

\vspace{.2cm} 

The above strategy of proof can be slightly modified to obtain several similar identities in the present context. Let us spell out two such modifications. Starting directly from \eqref{eq-DetTransferRel}, one finds using first \eqref{eq-GG} and then \eqref{eq-TransferResol} that 
\begin{align}
{\det}_{NL}(H_N-E\,\one)
&
\;=\;
(-1)^{NL}\,{\det}_L(T)^{N}\,
\sum_I \Big(\prod_{i\in I} {z}_i (E)^{N}\Big)\,
{\det}_L
\Big(\binom{\one}{0}^* {\Rr}_\IndSet^E \binom{\one}{0}\Big)
\nonumber
\\
&
\;=\;
\sum_I G_{\IndSet^c}(E)^N\,
{\det}_L
\Big(\binom{\one}{0}^* {\Rr}_\IndSet^E \binom{\one}{0}\Big)
\nonumber
\\
&
\;=\;
\sum_I G_{\IndSet}(E)^N\,
{\det}_L
\Big(\one\,-\,\binom{\one}{0}^* {\Rr}_{\IndSet}^E \binom{\one}{0}\Big)
\nonumber
\\
&
\;=\;
\sum_I G_{\IndSet}(E)^N\,
{\det}_L\Big(
\one\,-\,T\,\QFct^{(0)}_{\IndSet}(E)
\Big)
\;,
\label{eq-WidomFormula2}
\end{align}
where 
$$
\QFct^{(j)}_{\IndSet}(E)
\;=\;
\oint_{{\gamma}^0_{\IndSet}}\frac{dz}{2\pi\imath\,z^j}\;({H}(z)-E\,\one)^{-1}
\;.
$$
By a similar calculation, one also finds
\begin{align}
{\det}_{NL}(H_N-E\,\one)
&
\;=\;
(-1)^{NL}\,\frac{{\det}_L(T)^{N+1}}{{\det}_L(-R)}\,
{\det}_L
\Big(\binom{0}{\one}^*(\Tt^E)^{N+2}\binom{0}{\one}\Big)
\nonumber
\\
&
\;=\;
\frac{1}{{\det}_L(-RT)}\,
\sum_I G_{\IndSet}(E)^{N+2}\,
{\det}_L\Big(
\one\,+\,\QFct^{(2)}_{\IndSet}(E)\,R
\Big)
\;.
\label{eq-WidomFormula3}
\end{align}

\vspace{.2cm}

One of the crucial facts about Widom's formula \eqref{eq-WidomFormula} explored in the next section is that for $E\not\in\LimitSet$ and in the limit $N\to\infty$ one of the summands is much larger than the others. In fact, it is clear from the definition \eqref{eq-GDef} that for $\IndSet_0=\{1,\ldots,L\}$ and any $\IndSet\not=\IndSet_0$ one has
$$
\Big|\frac{G_{\IndSet}(E)}{G_{\IndSet_0}(E)}\Big| 
\;=\; 
\Big|\frac{\prod_{i\in \IndSet_0} z_{i}(E)}{\prod_{j\in \IndSet} z_{j}(E)}\Big| 
\;<\; 1 \;,
$$
because the middle two eigenvalues have distinct moduli for $E\not\in\LimitSet$ and one always has $G_\IndSet(E)\not=0$. Moreover, the corresponding factors $\QFct^{(j)}_{\IndSet_0}(E)$ are connected to $\QFct_s^{(j)}(E)$ as defined in \eqref{eq-QDef}, namely for $E\not\in\LimitSet$  and $|z_L(E)|<s<|z_{L+1}(E)|$
\begin{equation}
\label{eq-QLink}
\QFct^{(0)}_{\IndSet_0}(E)
\;=\;
\QFct^{(0)}_{s}(E)
\;,
\qquad
\QFct^{(1)}_{\IndSet_0}(E)
\;=\;
\QFct^{(1)}_{s}(E)
\;,
\qquad
\QFct^{(2)}_{\IndSet_0}(E)
\;=\;
\QFct^{(2)}_{s}(E)
\;.
\end{equation}
Now equating \eqref{eq-WidomFormula} with \eqref{eq-WidomFormula2} and \eqref{eq-WidomFormula3}, then dividing by $G_{\IndSet_0}(E)^{N+1}$ and taking the limit $N\to\infty$, one finds for $s$ still satisfying $|z_L(E)|<s<|z_{L+1}(E)|$ that
$$
{\det}_L(\QFct^{(1)}_{s}(E))
\;=\;
\frac{1}{G_{\IndSet_0}(E)}\,
{\det}_L\Big(
\one\,-\,T\,\QFct^{(0)}_{s}(E)
\Big)
\;=\;
\frac{G_{\IndSet_0}(E)}{{\det}_L(-RT)}\,
{\det}_L\Big(\one-T\,\QFct^{(2)}_{s}(E)
\Big)
\;.
$$
These identities do not seem to be contained in Widom's works, but are likely restricted to the present set-up in which the Hypothesis A holds. This  also allows to rewrite \eqref{eq-LambdaJ} as 
\begin{align}
\Outliers
&
\;=\;
\big\{E\in\CM \setminus \LimitSet  \,:\,\exists\; \sca>0\; \mbox{\rm with } \Wind^E(H^\sca)=0\; \text{\rm and }{\det}_L(\one-T{\QFct}^{(0)}_{\sca}(E))=0  \big\}
\nonumber
\\
&
\;=\;
\big\{E\in\CM \setminus \LimitSet  \,:\,\exists\; \sca>0\; \mbox{\rm with } \Wind^E(H^\sca)=0\; \text{\rm and }{\det}_L(\one-T\,\QFct^{(2)}_{s}(E))=0 
\big\}
\nonumber
\\
&
\;=\;
\big\{E\in\CM \setminus \LimitSet  \,:\,q_{\IndSet_0}(E)=0  \big\}
\label{eq-LambdaJ2}
\;,
\end{align}
where the third equality follows because $E\in\CM\setminus \LimitSet$ implies that $|z_L(E)|<|z_{L+1}(E)|$ so that one can choose a scaling parameter $\sca>0$ such that $\Wind^E(H^\sca)=0$ and for such $\sca$ we have $q_{\IndSet_0}(E)=\det({\QFct}^{(1)}_{\sca}(E))$.

\subsection{Limit of the spectra of the finite-volume operators}
\label{sec-SpectraLimitFinite}

This section provides a more detailed description and discussion of Theorem~\ref{thm-LimitSpecIntro}. The following definition will be used to describe the asymptotic spectra.

\begin{definition}[{\it e.g.} \cite{BS}]
\label{def-LimitSets}
Let $(\Sigma_N)_{N\geq 1}$ be a sequence of subset $\Sigma_N\subset\CM$. Then the uniform and partial limit sets are respectively defined by
\begin{align*}
&
{\underset{N\to\infty}{\mbox{\rm u-}\!\lim}\;\Sigma_N}
\;=\;
\big\{ E\in\CM \,:\,\exists\; E_N \in\Sigma_N \;\mbox{\rm such that } E_N \to E  \big\} 
\;,
\\
&
{\underset{N\to\infty}{\mbox{\rm p-}\!\lim}\;\Sigma_N}
\;=\;
\big\{ E\in\CM \,:\,\exists\,(N_n)_{n\geq 1} \;\mbox{\rm and } E_{N_n} \in\Sigma_{N_n} \;\mbox{\rm such that } E_{N_n} \to E \;\mbox{\rm and }N_n\to\infty\big\} 
\;.
\end{align*}
\end{definition}

\begin{theorem}
\label{thm-LimitSpec}
Suppose that {{\rm Hypotheses A, B, C} and {\rm D}} hold. Then 
\begin{equation}
\label{eq-LimitSpec}
\partial\LimitSet \cup \Outliers
\;=\;
{\mbox{\rm closure}\Big(\underset{N\to\infty}{\mbox{\rm u-}\!\lim}}
\;\spec(H_N)\Big)
\;=\;
{\underset{N\to\infty}{\mbox{\rm p-}\!\lim}\;}
\spec(H_N)
\;.
\end{equation}
\end{theorem}

Clearly Theorem~\ref{thm-LimitSpec} implies Theorem~\ref{thm-LimitSpecIntro}.  The proof of Theorem~\ref{thm-LimitSpec} will be given at the end of Section~\ref{sec-EVConstruct}. In fact, in Sections~\ref{sec-EVConstructLambda} and \ref{sec-EVConstruct} will provide criteria for the construction of eigenstates near $E\in\Outliers$ and $E\in\LimitSet$ respectively. The criterion on $\LimitSet$ will invoke Hypothesis C in a crucial manner (more precisely, to show that the quotient in \eqref{eq-GQuotient} is not constant).   Let us here further elucidate Hypothesis C and D. It was already stated in Section~\ref{sec-Intro} that Hypothesis C holds generically. Example~\ref{ex-3Diag} further down provides a model that violates Hypothesis C. It is a nilpotent perturbation of a model with symmetries which enforce more than two of the eigenvalues $z_l(E)$ to be of equal modulus on an open set in $\CM$. First of all, let us prove that Hypothesis C implies that $\mathring{\LimitSet}=\emptyset$ which is actually Condition B in \cite{Wid1}:

\begin{lemma}
\label{lem-HypD}
{\rm Hypothesis  C} implies that $\LimitSet$ has empty interior, namely $\LimitSet=\partial\LimitSet$.
\end{lemma}

\noindent {\bf Proof.} Suppose that $\mathring{\LimitSet}\not=\emptyset$. Then for all $E\in\partial\LimitSet$ with vanishing distance $d(E,\mathring{\LimitSet})=0$ there would be an open neighborhood $U$ such that on the open set $U\cap\mathring{\LimitSet}\not=\emptyset$ one has $|z_L(E')|=|z_{L+1}(E')|$;  Lemma~\ref{lem-ConstantModulus} then implies that $z_L(E'')=c\,z_{L+1}(E'')$ for all $E''\in U$ and some constant $c\in\TM $ because all other eigenvalues are separated from the two middle ones by Hypothesis C and the continuity of the $z_l$; but then also $U\setminus \mathring{\LimitSet}$ lies in $\LimitSet$, in contradiction to $E\in\partial\LimitSet$.  
\hfill $\Box$

\vspace{.2cm}

Another result assuring empty interior of $\LimitSet$ is Lemma~\ref{lem-EmpInt}. Let us also recall from Lemma~\ref{lem-Boundary} that $\partial\LimitSet$ locally consists of a finite union of analytic arcs. The finite number of points that are excluded in Hypothesis C are precisely the intersection points of these analytic arcs. As to Hypothesis D for $q_{\IndSet_0}$, it actually only addresses the bounded components of $\CM\setminus\LimitSet$:

\begin{lemma}
$q_{\IndSet_0}$ {is not constantly zero on} the unbounded component of $\CM\setminus\LimitSet$.
\end{lemma}

\noindent {\bf Proof.} Due to the particular form of $\Tt^E$ given in \eqref{eq-TraDef}, one concludes that there is some constant $C>0$ such that for all $E$ large enough $|z_l(E)|\geq C\,|E|$ for all $l=L+1,\ldots 2L$ and, using the same bound for $\widetilde{\Tt}^E$ and invoking \eqref{eq-zSym}, $|z_l(E)|\leq \, \frac{1}{C |E|}$ for all $l=1,\ldots L$. Therefore $\gamma_{\IndSet_0}^0$ in the first formula in \eqref{eq-qIDef} can be chosen to be the unit circle $\TM $. 
Therefore,
$$
q_{\IndSet_0}(E)
\;=\;
\frac{(-1)^L}{E^L}\;
{\det}_L\Big(
\oint_{\TM }\frac{dz}{2\pi\imath\,z}\;\big(\one-\tfrac{H(z)}{E}\big)^{-1}
\Big)
\;=\;
\frac{(-1)^L}{E^L}\,+\,\Oo(E^{-(L+1)})
\;,
$$
because $\one-\frac{H(z)}{E}$ is invertible for $|E|$ large enough so that the integral is merely equal to $1$, up to corrections of higher order in $1/E$.
\hfill $\Box$

\vspace{.2cm}

Next let us provide the promised  example of a Hamiltonian for which Hypothesis C and D do not hold.

\begin{example}
\label{ex-3Diag}
{\rm
For $L=3$ choose
$$
R\,=\,
\begin{pmatrix}
r & 0 & 0 \\ 0 & -r & 0 \\ 0 & 0 & 2\,r
\end{pmatrix}
\;,
\qquad
V\,=\,
\begin{pmatrix}
0 & 0 & 0 \\ 0 & 0 & 0 \\ 0 & 0 & 0
\end{pmatrix}
\;,
\qquad
T\,=\,
\begin{pmatrix}
t & 0 & 0 \\ 0 & -t & 0 \\ 0 & 0 & \tfrac{t}{2}
\end{pmatrix}
\;,
$$
where $r,t$ are non-vanishing complex numbers. Then $H$ is a direct sum of three Hatano-Nelson models (see Section~\ref{sec-HNmodel}), and two of them merely differ by a sign. By Theorem~\ref{thm-LimitSpec}, the finite volume spectra converge to the union of the $\LimitSet$-sets of the three models with $L=1$ which are given by line segments (again Section~\ref{sec-HNmodel}). On the other hand, let us next show that $\LimitSet$ for $H$ has open interior.  The eigenvalues of the transfer matrix $\Tt^E$ can be computed as in Section~\ref{sec-HNmodel}, but here the only important point is that two merely differ by a sign, namely the analytic branches of these eigenvalues are of the form $z'_1(E)$, $z'_2(E)$, $-z'_1(E)$, $-z'_2(E)$, $2\, z'_1(E)$, $2\,z'_2(E)$. Hence one readily checks that $\LimitSet$ has open interior. This is not a contradiction to Theorem~\ref{thm-LimitSpec} though, because by Lemma~\ref{lem-HypD} Hypothesis C does not hold. Furthermore, also {Hypothesis D} is violated because for any $E$ in the interior of $\LimitSet$ the lowest three eigenvalues necessarily include all eigenvalues of one of the three direct summands, so that corresponding to this summand, the block-diagonal matrix ${\Rr}_\IndSet^E$ contains a $2\times 2$ identity matrix which by \eqref{eq-qIDef} implies that $q_{\IndSet_0}(E)=0$. All the above facts readily transpose to a wider class of models. For example, one can fill the (strict) upper triangles of $R$, $V$ and $T$ with arbitrary entries without modifying the sets $\LimitSet$ as well as the spectra.
}
\hfill $\diamond$
\end{example}

As already stressed above, the proofs in Sections~\ref{sec-EVConstructLambda} and \ref{sec-EVConstruct} explicitly construct eigenfunctions using Widom's formula. The aim of the following is to compare this to the standard procedure of constructing the solutions using the transfer matrices (which determines the fundamental solutions). The following result spells out what can be achieved near a point $E\in\LimitSet$. An analogous statement can be obtained for $E\in\Outliers$. 

\begin{proposition}
\label{prop-TransferEigenfct}
Let $E\in\LimitSet $. Then there is a normalized quasimode $\phi^E\in\CM^{NL}$ of $H_N$ such that
$$
\|(H_N-E\,\one)\phi^E\|
\;\leq\;
\frac{C}{\sqrt{N}}
\;,
$$
for some {constant $C$} independent of $N$. In particular,
$$
\|(H_N-E\,\one)^{-1}\|\;\geq\;\frac{\sqrt{N}}{C}
\;.
$$
\end{proposition}

\noindent {\bf Proof.} Let us choose the scaling $s$ in \eqref{eq-HamGenScale} such that the two middle eigenvalues of $\Tt^E$ both have modulus $1$. The bound will be shown for $S_NH_NS_N^{-1}$ which will simply be denoted by $H_N$ again. Then $s$ enters into the constant $C$. Let us start constructing the state $\phi^E$ from its values $\phi^E_{m},\phi^E_{m+1}\in\CM^L$ at sites $m$ and $m+1$ (say for $m=\frac{N}{2}$ if $N$ even) by choosing the vector $\binom{T\phi^E_{m+1}}{\phi^E_{m}}$ that lies both in the subspace of non-expanding directions of $\Tt^E$ and in the subspace of non-expanding directions of $(\Tt^E)^{-1}$. By hypothesis, these two subspaces are both of dimension $L+1$, so that the intersection is at least of dimension $2$. Then construct $\phi^E$ by application of the transfer matrices, both to the right and to the left of $m$. One has
$$
\sum_{n=1,\ldots,N}\|\phi^E_n\|^2\;\;\leq\;c\,N
$$
After normalization of $\phi^E_{m+1},\phi^E_{m}$ by $\sqrt{N}$, one obtains the desired state for which $\|\phi^E_N\|\leq cN^{-\frac{1}{2}}$ and $\|\phi^E_{1}\|\leq cN^{-\frac{1}{2}}$. This state satisfies the Schr\"odinger equation everywhere exactly, except at these two boundary sites $1$ and $N$. This leads to the desired bound. Then
$$
1\;=\;
\|\phi^E\|\;=\;
\|(H_N-E\,\one)^{-1}(H_N-E\,\one)\phi^E\|
\;\leq\;
\|(H_N-E\,\one)^{-1}\|\,\frac{C}{\sqrt{N}}
\;,
$$
implies the second bound.
\hfill $\Box$

\vspace{.2cm}

If now $H_N$ is normal, the spectral theorem implies $\|(H_N-E\,\one)^{-1}\|=\mbox{\rm dist}(E,\spec(H_N))^{-1}$ so that one can conclude from Proposition~\ref{prop-TransferEigenfct} that there is an eigenvalue in a neighborhood of $E$ of size of order $\Oo(N^{-\frac{1}{2}})$. However, for non-normal operators it is well-known \cite{TE} that $\|(H_N-E\,\one)^{-1}\|$ is much larger than $\mbox{\rm dist}(E,\spec(H_N))^{-1}$ and then it is not possible to use the approximate eigenfunctions constructed in Proposition~\ref{prop-TransferEigenfct} to conclude that $H_N$ has spectrum near $E$. This difficulty cannot be circumvented by general methods, and is addressed by Widom's formula in the next two sections.

\subsection{Construction of eigenfunctions close to $\Outliers$}
\label{sec-EVConstructLambda}

In this section it will be proved that the multiplicity of a  zero of $q_{\IndSet_0}$ determines the number of eigenvalues of $H_N$ that are attracted to it. Due to \eqref{eq-LambdaJ2} this proves the part of Theorem~\ref{thm-LimitSpec} involving the set $\Outliers$. As to notation, let us denote by $\text{\rm ord}_{f}(E)$ the order of the point $E\in\CM$ as zero of a holomorphic function $f$; if $E$ is not a zero of $f$,  then $\text{\rm ord}_f(E)=0$.

\begin{proposition}
\label{prop-LambdaLimit}
Assume that $q_{\IndSet_0}$ is not the zero function on $\mathbb{C}\setminus\LimitSet$. Then for all $E \in \mathbb{C} \setminus \LimitSet$ and all  $\rho >0$ small enough, there exists an $N_0 \in \mathbb{N}$ such that for all $N>N_0$
$$
\#\,\spec(H_N) \cap B_\rho (E)\; = \;\text{\rm ord}_{q_{\IndSet_0}}(E) \;.
$$
\end{proposition}

For the proof, let us start with a preparatory result that is also used in the next section. The eigenvalues $z_i(E)$ of the transfer matrix $\Tt^E$ are locally holomorphic on $\mathbb{C}\setminus\mathcal{F}$ up to relabeling as explained in the proof of Lemma \ref{lem-LimitSetNoIso}. Here locally holomorphic means holomorphic up to branch cuts. They are, however, not necessarily holomorphic at points in $\mathcal{F}$, but merely have a Puisseux expansion there. By a procedure as for the $z_i$ in the proof of Lemma \ref{lem-LimitSetNoIso}, we can assume that the $G_\IndSet$ are locally holomorphic on $\mathbb{C}\setminus\mathcal{F}$, but not necessarily at points in $\mathcal{F}$. The next lemma shows at which points in $\mathcal{F}$ the functions $G_\IndSet$ are nevertheless holomorphic. Note that this is in line with standard results on the analyticity of Riesz projections as derived in \cite{Kat}.

\begin{lemma}
\label{lem-AnalyticityGq}
Let $\IndSet\subset\{1,\ldots,2L\}$ consist of $L$ numbers and $E\in\mathbb{C}$ be such that for all $i\in I$ and $j\notin I$ one has $z_i(E) \neq z_j(E)$. Then $G_\IndSet$ and $q_\IndSet$  are holomorphic at $E$. In particular, $G_\IndSet$ and $q_\IndSet$  are locally holomorphic on $\CM\setminus\Ff$.
\end{lemma}

\noindent {\bf Proof.} 
Note that the  choice of $E$ assures that the path $\gamma_{\IndSet}^0$ as used in Theorem~\ref{theo-WidomFormula}  exists. Next
\begin{align*}
G_I (E) 
\;=\; & \exp\Log \Big( (-1)^L \,{{\det}_L (R)} \prod_{i\in\IndSet} \frac{1}{z_{i}(E)} \Big) \\
\;=\; & \exp \Big( \oint_{\gamma_{\IndSet}^0} \Log\Big( z^L \,{\det}_L (H(z)-E\,\one) \prod_{i\in\IndSet}(z-z_{i}(E))^{-1} \Big) \frac{dz}{2\pi \imath \,z}\Big) \\
\;=\; & \exp \Big( \oint_{\gamma_{\IndSet}^0} \Log \big({\det}_L (H(z)-E\,\one)\big) \frac{dz}{2\pi \imath\, z} + \oint_{\gamma_{\IndSet}^0} \Log\Big( z^L \prod_{i\in\IndSet}(z-z_{i}(E))^{-1} \Big) \frac{dz}{2\pi \imath \,z}\Big) \\
\;=\; & \exp \Big( \oint_{\gamma_{\IndSet}^0} \Log \big({\det}_L (H(z)-E\,\one)\big) \frac{dz}{2\pi \imath \,z} \Big) 
\;.
\end{align*}
From this expression it follows immediately that $G_\IndSet$ is holomorphic in $E$. The claim on $q_\IndSet$ follows similarly from \eqref{eq-qIDef}.
\hfill $\Box$

\begin{corollary}
\label{coro-Holomorph}
$G_{\IndSet_0}$ and $q_{\IndSet_0}$ are {locally} holomorphic on $\mathbb{C}\setminus\LimitSet$.
\end{corollary}

\noindent {\bf Proof.} For $E\in \mathbb{C}\setminus\LimitSet$ one has $|z_L(E)|<|z_{L+1}(E)|$ so for $E\in \mathcal{F}\setminus\LimitSet$ one has $z_i(E)=z_j(E)$ either for $i,j \geq L+1$ or $i,j \leq L$. The claim thus follows from Lemma~\ref{lem-AnalyticityGq}.
\hfill $\Box$

\vspace{.2cm}

\noindent {\bf Proof} of Proposition~\ref{prop-LambdaLimit}. 
For all $E \in \mathbb{C} \setminus \LimitSet$ and $\IndSet \neq \IndSet_0$, one has
$$
\Big|\frac{G_{\IndSet}(E)}{G_{\IndSet_0}(E)}\Big| 
\;=\; 
\Big|\frac{\prod_{i\in \IndSet_0} z_{i}(E)}{\prod_{i\in \IndSet} z_{i}(E)}\Big| 
\;<\; 1 \;.
$$
Using Widom's formula \eqref{eq-WidomFormula} and the fact that the functions $G_{\IndSet}$ are continuous, one deduces  that
\begin{equation}
\label{eq-fNdef}
\frac{\det(H_{N}-E\,\one)}{G_{\IndSet_0}(E)^{N+1}} 
\,-\,
q_{\IndSet_0}(E)
\;=\; 
\sum_{\IndSet\not=\IndSet_0}\Big(\frac{G_{\IndSet}(E)}{G_{\IndSet_0}(E)}\Big)^{N+1}q_\IndSet (E) 
\;\to\; 0
\qquad
\mbox{\rm as } N\to\infty
\;,
\end{equation}
uniformly on compact subset of $\mathbb{C}\setminus (\LimitSet \cup \Ff)$. Let us next analyze the convergence also on compact subsets of $\mathbb{C}\setminus \LimitSet$ which contain points in $\Ff$. For that purpose let $f_N$ denote the function on the l.h.s. of \eqref{eq-fNdef}. By  Corollary~\ref{coro-Holomorph} it is holomorphic in all $E\in\mathbb{C}\setminus\LimitSet$. Let next $E\in\mathcal{F}\setminus\LimitSet$ and choose $\rho>0$ such that the closed ball $B_{2\rho}(E)$ of size $2\rho$ around $E$ has trivial intersection with $\LimitSet$. Then for all $E'\in B_{\rho}(E)$, Cauchy's integral formula gives
$$
\big|f_N(E')\big|
\;=\;
\Big|\ointop_{\partial B_{2\rho}(E)}\frac{dz}{2\pi \imath}\,\frac{f_{N}(z)}{z-E'}\Big|
\;\leq\;
\frac{1}{2\pi\rho}\, \ointop_{\partial B_{2\rho}(E)}dz\,|f_{N}(z)|  
\;\leq\;
2\,\max_{z\in\partial B_{2\rho}(E)}|f_{N}(z)|
\;,
$$
which converges to $0$ by the above, uniformly in $E'$. In conclusion,
$$
\lim_{N\to\infty}\,
\frac{\det(H_{N}-E\,\one)}{G_{\IndSet_0}(E)^{N+1}} 
\;=\; 
q_{\IndSet_0}(E)
\;,
$$
uniformly on compact subsets of $\mathbb{C}\setminus \LimitSet$. Note that the functions $G_{\IndSet_0}$ and $q_{\IndSet_0}$ are locally holomorphic on $\mathbb{C}\setminus\LimitSet$ by Corollary~\ref{coro-Holomorph}. Moreover, ${\det}_{NL}(H_N-E\,\one)$ is polynomial in $E$ and thus an entire function. Since $G_{\IndSet_0}$ has no zeros because the transfer matrix is invertible for all $E$, it follows that also the expression ${\det}_{NL}(H_N-E\,\one) G_{\IndSet_0}(E)^{-N-1}$ is locally holomorphic in $E\in\mathbb{C}\setminus\LimitSet$. The statement of the proposition now follows from Hurwitz's theorem and the fact that the zeros of ${\det}_{NL}(H_N-E\,\one) G_{\IndSet_0}(E)^{-N-1}$ are exactly the eigenvalues of $H_N$.
\hfill $\Box$

\begin{remark}
\label{rem-EndJordan2}
{\rm
It is possible to generalize Proposition~\ref{prop-LambdaLimit} so that the {limit of the spectrum} of $H_N$ can be studied in cases where Hypotheses C and D do not hold. For that purpose, one can consider a generalized $\LimitSet$-set $\hat{\LimitSet} = \bigcup_{I,J}\partial \{E\in\CM:  |G_I(E)| = |G_J(E)|\}$. On the connected components of the complement of $\hat{\LimitSet}$ one has either $|G_I|<|G_J|$ or $|G_I|=|G_J|$. Now, there is an $I$ such that locally $q_I \neq 0$ and if one replaces $I_0$ with $I$ in \eqref{eq-fNdef}, then the limit again converges uniformly on compact subsets, but now possibly only for subsequences. Carrying out the argument as in the above proof, one obtains a similar result as in Proposition~\ref{prop-LambdaLimit}, but now on the set $\mathbb{C}\setminus\hat{\LimitSet}$ and possibly only for subsequences. Details will be provided elsewhere.
}
\hfill $\diamond$
\end{remark}

\subsection{Construction of eigenfunctions close to $\partial\LimitSet$}
\label{sec-EVConstruct}

The first lemma provides a criterion for an energy $E\in\partial\LimitSet$ to attract an eigenvalue of $H_N$. It is slightly more general than required for the proof of Theorem~\ref{thm-LimitSpec}.

\begin{lemma}
\label{lem-LocalCrit}
Let $E\in\CM\setminus\mathcal{F}$ be such that  $|z_{L-1}(E)|<|z_L(E)|=|z_{L+1}(E)|<|z_{L+2}(E)|$, and suppose that
$q_{I_0}(E)$ and $q_{I_1}(E)$ are non-zero and $G_{\IndSet_0}/G_{\IndSet_1}$ is locally non-constant around $E$, where $\IndSet_1=\{1,\ldots,L-1,L+1\}$ as in Section~\ref{sec-Intro}. Then there are constants $C>0$ and $n,N_0 \in\NM$ such that for all $N \geq N_0$ there is an $\epsilon\in\CM$ with $|\epsilon|\leq C \,N^{-\frac{1}{n}}$ and $E+\epsilon\in\spec(H_N)$. 
\end{lemma}

\noindent {\bf Proof.} Note that the supposed equality of the middle two eigenvalues implies that $E\in\LimitSet$. Let us start out by using Widom's formula \eqref{eq-WidomFormula} which implies that, if $E+\epsilon \notin \mathcal{F}$, then $E+\epsilon\in\spec(H_N)$ if and only if
\begin{equation}
\label{eq-HNEigenvalue}
0 
\;=\; 
\sum_\IndSet G_I (E+\epsilon)^{N+1} q_{\IndSet}(E+\epsilon)
\;.
\end{equation}
If $\epsilon = 0$ solves this equation, then it is the desired solution. In the following, it will hence be supposed that $\epsilon = 0$ does not solve \eqref{eq-HNEigenvalue}. Next using that $\epsilon\mapsto G_{\IndSet_0} (E+\epsilon)/G_{\IndSet_1} (E+\epsilon)$ is non-constant in a neighborhood of $\epsilon=0$, \eqref{eq-HNEigenvalue} can be rewritten to
$$
\Big( \frac{G_{\IndSet_1} (E)}{G_{\IndSet_0} (E)} \cdot \frac{G_{\IndSet_0} (E+\epsilon)}{G_{\IndSet_1} (E+\epsilon)} \Big)^{N+1}
\;=\;
-\,\frac{1}{q_{\IndSet_0}(E+\epsilon)}
\Big(\frac{G_{\IndSet_1} (E)}{G_{\IndSet_0} (E)} \Big)^{N+1}
\sum_{\IndSet\neq\IndSet_0} \Big(\frac{G_{\IndSet} (E+\epsilon)}{G_{\IndSet_1} (E+\epsilon)}\Big)^{N+1}\, q_{\IndSet}(E+\epsilon)
\;.
$$
Taking a (suitable branch of the) logarithm and dividing by $N+1$, one sees that \eqref{eq-HNEigenvalue} for $\epsilon\not=0$ 
(namely $E+\epsilon \in \spec(H_N)$) is equivalent to $g_N(\epsilon)=  \frac{1}{N+1}$ where
\begin{equation}
\label{eq-gNDef}
g_N (\epsilon) 
\;=\; 
\frac{\Log\Big( \frac{G_{\IndSet_1} (E)}{G_{\IndSet_0} (E)} \cdot \frac{G_{\IndSet_0} (E+\epsilon)}{G_{\IndSet_1} (E+\epsilon)} \Big)}
{\Log\Big( - q_{\IndSet_0}(E+\epsilon)^{-1}
\big(\frac{G_{\IndSet_1} (E)}{G_{\IndSet_0} (E)} \big)^{N+1}
\sum_{\IndSet\neq\IndSet_0} \big(\frac{G_{\IndSet} (E+\epsilon)}{G_{\IndSet_1} (E+\epsilon)}\big)^{N+1}\, q_{\IndSet}(E+\epsilon) \Big)}
\;.
\end{equation}
Let us stress that the function $g_N$ is well-defined and holomorphic on some connected, simply connected, open neighborhood of $\epsilon = 0$ because all $G_\IndSet$ and $q_{\IndSet}$ are, $q_{\IndSet_0}(E)$ is non-zero and the denominator is non-zero at $\epsilon = 0$ by the assumption that $\epsilon = 0$ does not solve \eqref{eq-HNEigenvalue}. Furthermore, let us note that $g_N(0) = 0$ and that $0$ is an isolated zero of $g_N$ since the denominator of $g_N$ around $\epsilon=0$ is bounded and $G_{\IndSet_0}/G_{\IndSet_1}$ is holomorphic and locally non-constant around $E$.

\vspace{.1cm}

Next, because $|z_{L-1}(E)|<|z_L(E)|=|z_{L+1}(E)|<|z_{L+2}(E)|$ and the $G_I$ are continuous, 
$$
\Big|
\frac{G_{\IndSet}(E+\epsilon)}{G_{\IndSet_1}(E+\epsilon)}
\Big| 
\;=\; 
\Big|
\frac{\prod_{i\in \IndSet_1} z_{i}(E+\epsilon)}{\prod_{i\in \IndSet} z_{i}(E+\epsilon)}
\Big| 
\;<\; 1 \;,
$$
for all $\IndSet \notin \{ \IndSet_0, \IndSet_1\}$ and all $\epsilon$ in a neighborhood of $0$. It follows that
$$
\lim_{N\to\infty}\,\sum_{\IndSet\neq\IndSet_0} \Big(\frac{G_{\IndSet} (E+\epsilon)}{G_{\IndSet_1} (E+\epsilon)}\Big)^{N+1} q_{\IndSet}(E+\epsilon)
\;= \;
q_{\IndSet_1}(E+\epsilon)
\;,
$$
uniformly on compact subsets of a neighborhood of $\epsilon = 0$. Since both $q_{\IndSet_0}$ and $q_{\IndSet_1}$ are continuous and non-zero at $E$ and since $G_{\IndSet_1}(E)/G_{\IndSet_0}(E) = z_L (E)/z_{L+1}(E) \in \mathbb{T}^1$, there exists an $M \in \mathbb{R}_{>0}$ and an $N_1 \in\mathbb{N}$ such that for all $N>N_1$ and all $\epsilon$ in an open neighborhood of $0$
$$
\Big| \Log\Big( - q_{\IndSet_0}(E+\epsilon)^{-1} 
\Big(\frac{G_{\IndSet_1} (E)}{G_{\IndSet_0} (E)} \Big)^{N+1}
\sum_{\IndSet\neq\IndSet_0} \Big(\frac{G_{\IndSet} (E+\epsilon)}{G_{\IndSet_1} (E+\epsilon)}\Big)^{N+1} q_{\IndSet}(E+\epsilon) \Big) \Big| 
\;<\; 
M \;.
$$
Moreover, the $G_I$ are holomorphic around $E$ since $E\notin\mathcal{F}$, and $G_{\IndSet_0}/G_{\IndSet_1}$ is not locally constant around $E$ by assumption. It follows that
\begin{equation}
\label{eq-GQuotient}
\frac{G_{\IndSet_1} (E)}{G_{\IndSet_0} (E)} \cdot \frac{G_{\IndSet_0} (E+\epsilon)}{G_{\IndSet_1} (E+\epsilon)}
\; = \; 
1 \,+\, c_n\, \epsilon^n \,+\, \mathcal{O}(\epsilon^{n+1})
\end{equation}
for some $n\in\mathbb{N}$ and non-zero $c_n\in\mathbb{C}$. Combining these results, one  finds that 
\begin{equation}
\label{eq-gNbound}
|g_N(\epsilon)|
\;>\;
\frac{1}{M} \Big| \Log
\Big( \frac{G_{\IndSet_1} (E)}{G_{\IndSet_0} (E)} \cdot \frac{G_{\IndSet_0} (E+\epsilon)}{G_{\IndSet_1} (E+\epsilon)} \Big) \Big|
\;=\;
\frac{1}{M} \,|c_n\,\epsilon^n (1+\mathcal{O}(\epsilon))|
\;>\;
\frac{c'}{M} \,|\epsilon|^n
\;,
\end{equation}
for some $c'>0$, all $N>N_1$ and all $\epsilon$ in a neighborhood of $0$.

\vspace{.1cm}

Next recall that $g_N$ is constructed in such a way that, whenever $g_N(\epsilon)$ exists and $E+\epsilon \notin \mathcal{F}$, then $g_N (\epsilon) = \frac{1}{N+1}$ if and only if $E+\epsilon \in \spec(H_N)$. Since $\mathcal{F}$ is finite and $E\notin\mathcal{F}$, one can now choose an $\epsilon$-neighborhood $V$ of $0$ on which the lower bound \eqref{eq-gNbound} holds and which, moreover, satisfies $(E+V)\cap\Ff=\emptyset$. As all $g_N$ are holomorphic around $\epsilon = 0$ and have $\epsilon = 0$ as an isolated zero, it follows from the open mapping theorem and the fact that $\frac{1}{N+1} \to 0$ as $N \to \infty$ that there exists an $N_2 \in \mathbb{N}$ such that for all $N>N_2$ the equations $g_N (\epsilon)=\frac{1}{N+1}$ have solutions $\epsilon_N$. Moreover, the lower bound \eqref{eq-gNbound} tells us that $\epsilon_N \to 0$ as $N\to\infty$. Because $V$ is a neighborhood of $0$, there thus is an $N_3\geq N_2$ such that for all $N>N_3$ one has $\epsilon_N \in V$. Using the lower bound \eqref{eq-gNbound} again one concludes that for all $N>\max\{ N_1,N_3\}$ 
$$
\frac{1}{N} 
\;>\; 
\frac{1}{N+1} 
\;=\; 
g_N(\epsilon_N) 
\;>\; 
\frac{c'}{M}\, |\epsilon_N|^n
\;,
$$
and $E+\epsilon_N \in \spec(H_N)$ since $\frac{1}{N+1} = g_N(\epsilon_N)$ and $\epsilon_N \in V$ such that $E+\epsilon_N\not\in\mathcal{F}$. 
\hfill $\Box$

\vspace{.2cm}

The previous lemma can under suitable conditions be used to prove that $\partial\LimitSet$ attracts spectrum of the Hamiltonians $H_N$.

\begin{proposition}
\label{prop-CriteriaGamma}
Suppose {\rm Hypotheses A, B, C} and {\rm D} hold. Then for all but finitely many $E \in \partial\LimitSet$ there are constants $C>0$ and $N_0 \in\NM$ such that for all $N \geq N_0$ there is an $\epsilon\in\CM$ with $|\epsilon|\leq \frac{C}{N}$ and $E+\epsilon\in\spec(H_N)$. 
\end{proposition}

\noindent {\bf Proof.} 
Note that the functions $q_{\IndSet}$ are all locally holomorphic on $\mathbb{C}\setminus\mathcal{F}$. Since $q_{\IndSet_0}$ and $q_{\IndSet_1}$ are not locally constantly zero on $\partial\LimitSet \setminus \mathcal{F}$, their zeros on this set are isolated. Because $\LimitSet$ is compact, it thus follows that $q_{\IndSet_0}$ and $q_{\IndSet_1}$ have only finitely many zeros on $\partial\LimitSet \setminus \mathcal{F}$. Moreover, $G_{\IndSet_0}(E)/G_{\IndSet_1}(E)= z_{L+1} (E)/z_{L}(E)$ is locally non-constant around all $E \in \partial\LimitSet \setminus \mathcal{F}$ because $E \in \partial\LimitSet$ and outside of $\LimitSet$ its modulus is less than $1$. Since $\mathcal{F}$ is finite, an application of the previous lemma now implies that for all but finitely many $E \in \partial\LimitSet$ there are constants $C>0$ and $n,N_0 \in\NM$ such that for all $N \geq N_0$ there is an $\epsilon\in\CM$ with $|\epsilon|\leq C\,N^{-\frac{1}{n}}$ and $E+\epsilon\in\spec(H_N)$. As the quotient on the l.h.s. of \eqref{eq-GQuotient} is equal to $ z_{L+1} (E)z_L(E+\epsilon)/(z_L(E)z_{L+1}(E+\epsilon)$, the proof of Lemma~\ref{lem-Boundary} shows that for all but finitely many $E$ one can take $n=1$ in the proof of Lemma~\ref{lem-LocalCrit}. 
\hfill $\Box$

\vspace{.2cm}

\noindent {\bf Proof} of Theorem~\ref{thm-LimitSpec}. First recall $\LimitSet=\partial\LimitSet$ by Lemma~\ref{lem-HypD}. Therefore Proposition~\ref{prop-LambdaLimit} implies that 
$$
\Big({\underset{N\to\infty}{\mbox{\rm u-}\!\lim}\;} \spec(H_N)\Big) \cap (\CM\setminus\partial\LimitSet) 
\;=\;
\Big({\underset{N\to\infty}{\mbox{\rm p-}\!\lim}\;} \spec(H_N)\Big) \cap (\CM\setminus\partial\LimitSet) 
\;=\;
\Outliers
\;.
$$
Furthermore, Proposition~\ref{prop-CriteriaGamma} shows that one has a sequence of eigenvalues converging to all but a finite number of points in $\partial\LimitSet$, hence $\partial\LimitSet \setminus \mbox{\rm u-}\!\lim_{N\to\infty} \spec(H_N)$ is finite. Since $\Outliers$ is discrete and $\LimitSet = \partial\LimitSet$ is closed and has no isolated points, it follows that
$$
{\mbox{\rm closure}\Big(\underset{N\to\infty}{\mbox{\rm u-}\!\lim}}
\;\spec(H_N)\Big)
\;=\;
\partial\LimitSet \cup \Outliers
\;.
$$
Note that for any sequence $(\Sigma_n)_{n\geq 1}$ of subsets its partial limit $\mbox{\rm p-}\!\lim_{n\to\infty} \Sigma_n$ is closed. Hence
$$
{\mbox{\rm closure}\Big(\underset{N\to\infty}{\mbox{\rm u-}\!\lim}}
\;\spec(H_N)\Big)
\;\subset\;
{\underset{N\to\infty}{\mbox{\rm p-}\!\lim}}
\;\spec(H_N)
\;\subset\;
\partial\LimitSet \cup \Outliers
\;,
$$
concluding the proof.
\hfill $\Box$

\section{Applications to non-hermitian quantum systems}
\label{sec-Applications}

This section applies the above results to several concrete models that have been studied in the physics literature. These models all describe quasi-one-dimensional non-hermitian quantum systems with periodic coefficients. In applications, the non-hermitian character typically results from either dissipative terms or external driving forces. The spectra of the infinite-volume operators and finite-volume approximations are computed numerically and are, as expected, often very different. This is illustrated in Section~\ref{sec-HNmodel} in the example of the Hatano-Nelson model \cite{HN} and one of its variants having next-nearest neighbor hopping. Furthermore the eigenfunctions of the finite-volume operators are typically localized at one or the other end of the system, an effect called the skin effect (see Section~\ref{sec-Skin}).  Next, Section~\ref{sec-Brillouin} briefly defines and discusses the generalized Brillouin zone. In the following Section~\ref{sec-TopoZero}, particular focus is on finite systems with topological invariants protected by a (chiral) symmetry. These invariants imply the existence of (approximate zero energy) bound states and this is again illustrated in a concrete model.

\subsection{Hatano-Nelson-like models}
\label{sec-HNmodel}

\begin{figure}
\centering
\includegraphics[width=5.3cm,height=5.7cm]{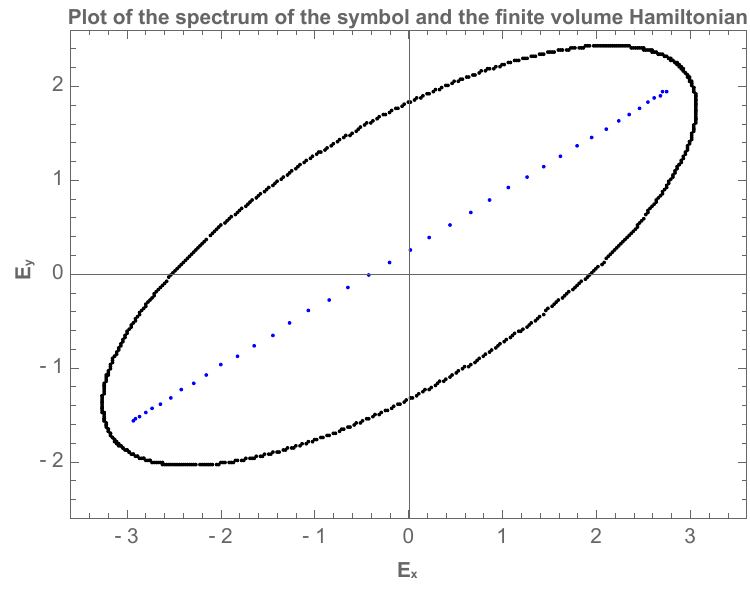}
\hspace{.1cm}
\includegraphics[width=5.3cm,height=5.7cm]{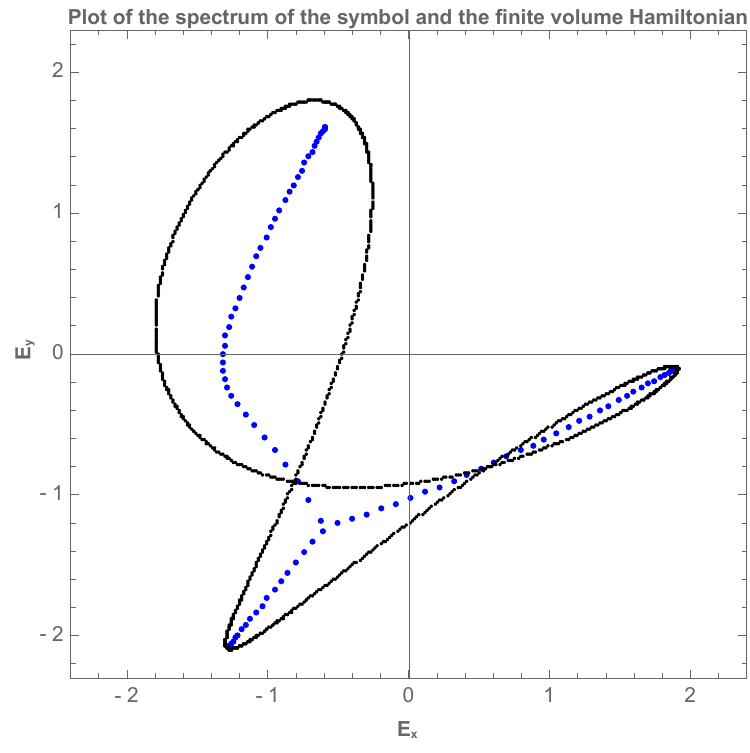}
\hspace{.1cm}
\includegraphics[width=5.3cm,height=5.7cm]{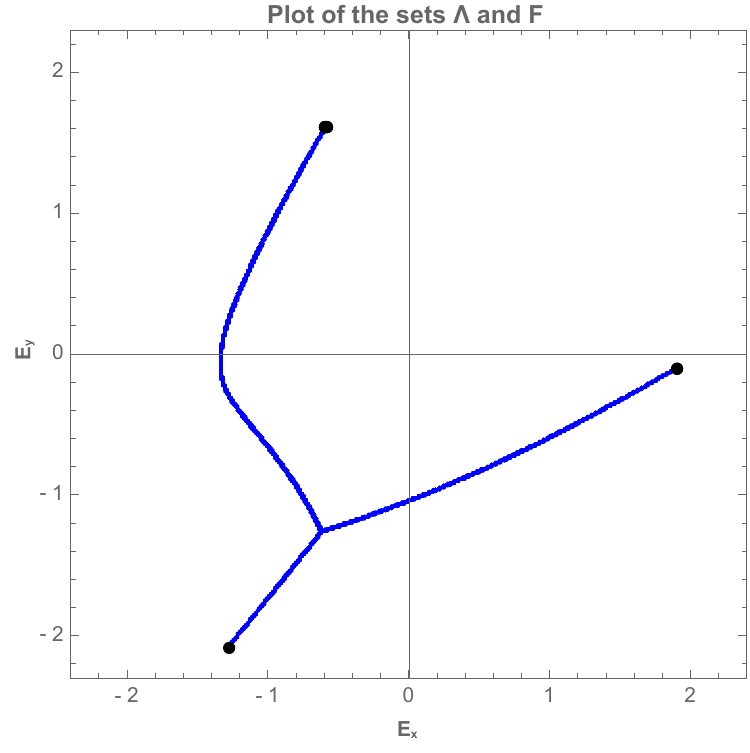}
\caption{\sl The first plot shows the numerical computation of the curve $\Sigma$ and the spectrum of the Hatano-Nelson model with $T=0.5+\imath$, $V=-0.1+0.2\,\imath$ and $R=2.5$, and $N=40$. The second plot provides the spectrum of the symbol and finite-volume Hamiltonian $H_N$ for the next-nearest neighbor version with $t_{-2} = 1.0$, $t_{-1} = -0.5\,\imath$, $t_0= -0.3 - 0.3\,\imath$, $t_1 = 1.0$ and $t_2=0.5\,\imath$ with $N=50$, and for the same parameters the last plot shows the set $\LimitSet$ as well as $\{E\in\LimitSet\,:\,z_L(E)=z_{L+1}(E)\} \subset \Ff\cap\LimitSet$ by larger dots (this latter set consists of merely the three end points of the segments).
}
\label{fig-HN}
\end{figure}

The Hatano-Nelson model \cite{HN} is the most simple strictly one-dimensional and translation invariant non-hermitian Hamiltonian, it is namely tridiagonal with constant scalar entries on the diagonal. This corresponds to the case $L=1$ with matrix entries $R,T\in\CM\setminus\{0\}$ and $V\in\CM$. As the symbol is scalar, one can readily write out the spectrum $\Sigma^\sca=\spec(H^\sca)$ of the periodic scaled Hamiltonian $H^\sca$:
\begin{equation}
\label{eq-SigmaSca}
\Sigma^\sca
\;=\;
\big\{\sca^{-1}R\,e^{-\imath k}+V+\sca\,T\,e^{\imath k}\;:\;k\in[-\pi,\pi)\big\}
\;.
\end{equation}
This set is an ellipse in the complex plane which, for the critical scaling parameter $\sca_c$ given by $\sca_c^2=|RT^{-1}|$, degenerates to a line segment. As the Schmidt-Spitzer result \eqref{eq-SS} applies in the scalar case considered here, one deduces
$$
\LimitSet
\;=\;
\Sigma^{\sca_c}
\;,
$$
namely $\LimitSet$ is a line segment. Of course, this can also be obtained directly from the definition \eqref{eq-Gamma0} of the set $\LimitSet$ as the set of energies where the middle two eigenvalues of the transfer matrix have the same modulus. Indeed, it follows from \eqref{eq-SigmaSca} for $\sca=\sca_c$ that each point $E\in\LimitSet$ is reached for two different values of $k$.  By \eqref{eq-TransferDet}, these two numbers on the unit circle are also the two eigenvalues $z_1(E),z_2(E)\in\TM $ of the transfer matrices $\Tt^{\sca_c,E}$, so that $E\in\LimitSet$. Yet another way to compute $\LimitSet$ starts from the formulas for the two eigenvalues of $\Tt^E$:
$$
z_{1/2}(E)
\;=\;
\frac{(E-V)T^{-1}}{2}\,\pm\,\sqrt{\frac{(E-V)^2T^{-2}}{4}\,-\,\frac{R}{T}}
\;.
$$
Then $\LimitSet=\{E\in\CM\,:\,|z_1(E)|=|z_2(E)|\}$ can be computed explicitly using the fact that for any complex number $\xi\in\CM$ one has the equivalence $\Re e(\xi)=0\,\Longleftrightarrow\,\xi^2\leq 0$ where the r.h.s. means that $\xi^2$ is real and non-positive. Elementary algebra then leads to
$$
\LimitSet
\;=\;
\left\{
E\in\CM\,:\,
\frac{(E-V)^2}{R\,T}\leq 4
\right\}
\;.
$$ 
This is yet another representation of $\LimitSet$ as a line segment. Note that this formula is intrinsically independent of scaling $(R,T)\mapsto (s^{-1}R,sT)$ and that for the well-known case of the discrete Laplacian where $R=T=1$ and $V=0$, one recovers $\LimitSet=\{E\in\RM\,:\,|E|\leq 2\}$.  Finally let us note that the set of outliers $\Outliers=\emptyset$ is empty by Proposition~\ref{prop-L1Lambda}. All of this is illustrated in the numerical example in Figure~\ref{fig-HN}.

\vspace{.2cm}

Next let us consider a non-hermitian scalar Hamiltonian with five diagonals, namely add next-nearest neighbor hopping terms. As described in Example~\ref{ex-ScalarBand} this model depends on five complex parameters $t_{-2},\ldots,t_2\in\CM$ with $t_{\pm 2}\not=0$ and can be cast in the form \eqref{eq-HamGen} with blocks of size $L=2$. The Schmidt-Spitzer result \eqref{eq-SS} applies still, but it does not allow to compute the set $\LimitSet$ as explicitly. One rather has to recourse to the definition \eqref{eq-Gamma0} for a numerical computation: for $E$ on a suitable grid of a region of the complex plane, one computes the eigenvalues of $\Tt^E$ and simply checks whether the middle two eigenvalues are of approximately the same modulus.  A first example is provided in Figure~\ref{fig-HN}, others in Figures~\ref{fig-Ser2} and \ref{fig-Ser2bis}. Let us mention that another numerical algorithm for the computation of $\LimitSet$ for scalar banded Toeplitz matrices (as described in Example~\ref{ex-ScalarBand}) is described and implemented in  the recent paper \cite{BGG}.

\subsection{The skin effect}
\label{sec-Skin}

For finite-volume non-hermitian Hamiltonians $H_N$, the eigenstates are generically localized either at the left or right boundary of the system, an effect that is called the {\it skin effect} in the physics literature (see \cite{ZZLC} for a physics review). The aim of this section is to provide a detailed understanding of the skin effect from a mathematical perspective. For the case of compressions of scalar banded Toeplitz matrices this was already analyzed in Chapter~12 of \cite{BG}, based on \cite{BGR}.

\vspace{.2cm}

Let us start by recalling that Theorem~\ref{thm-LimitSpec} shows that eigenvalues of $H_N$ accumulate on $\LimitSet$. Furthermore, by \eqref{eq-GammaWind}, one has that $\LimitSet$ is a subset of the union $\Sigma\cup \big\{ E\in\CM \,:\, \Wind^E(H)\neq 0\big\}$. Hence considering the decomposition of $\CM\setminus\Sigma$ into connected components, $\LimitSet$ will only lie in those components having non-vanishing winding number. This non-vanishing winding number in turn implies that generalized eigenspace $\Ee^E_1$ of all eigenvalues of the transfer matrix $\Tt^E$ with modulus less than $1$ has a dimension given by $\dim(\Ee^E_1)=L+\Wind^E(H)$, see \eqref{eq-E1Wind}. In particular, $\Wind^E(H)\neq 0$ implies that on $\LimitSet$ one either has at least $L+1$ contracting directions of $\Tt^E$ (for $\Wind^E(H)>0$) or $L+1$ expanding directions (for $\Wind^E(H)<0$), so that the middle two eigenvalues $z_L(E),z_{L+1}(E)$ of same modulus are hence of modulus smaller or larger than $1$ respectively. It is shown in Section~\ref{sec-EVConstruct} that the eigenfunctions of $H_N$ are constructed for energies close to $\LimitSet$ by using small perturbations of the two-dimensional subspace spanned by the eigenvectors of $z_L(E),z_{L+1}(E)$ (more precisely, the perturbations in energy and the subspace can indeed be done in such a way that the Dirichlet boundary conditions can be satisfied at both sides of the sample, see Section~\ref{sec-EVConstruct}). On the other hand, for eigenstates of $H_N$ at energies $E$ lying in a region of $\CM\setminus\Sigma$ with vanishing winding number (such energies can be either spectral outliers or points converging to $\LimitSet$ in the limit $N\to\infty$, but nevertheless still in a region with vanishing winding number for finite $N$), there are $L$ eigenvalues of $\Tt^E$ less than $1$ and $L$ others larger than $1$. Typically, a vector composed of two neighboring entries of an eigenstate is then neither in the increasing nor the decreasing subspace of $\Tt^E$, but rather a linear combination of two vectors from these spaces and hence a linear combination of a left skin and a right skin state. Two examples are given in the middle and right plots of Figure~\ref{fig-Ser2bisbis}. Finally for an eigenvalue $E$ of $H_N$ that accidentally falls on $\Sigma$, the corresponding eigenstate must have a contribution in the two middle eigenvalues of modulus $1$, namely essentially be of plane-wave nature. Let us summarize:

\begin{itemize}

\item {\it
For $E\in \spec(H_N)$ lying in a component of $\CM\setminus\Sigma$ with positive/negative winding number $\Wind^E(H)$, the eigenstate is localized at the left/right boundary (skin state). }

\item {\it For $E\in \spec(H_N)$ lying in a  region with vanishing winding number, the eigenstate need not be localized at one side of the sample (but it need not be a plane wave state either). }

\item {\it For $E\in \spec(H_N)\cap\Sigma$, the eigenstate is like a plane wave. }

\end{itemize}

\begin{figure}
\centering
\includegraphics[align=t,width=5.4cm,height=6.6cm]{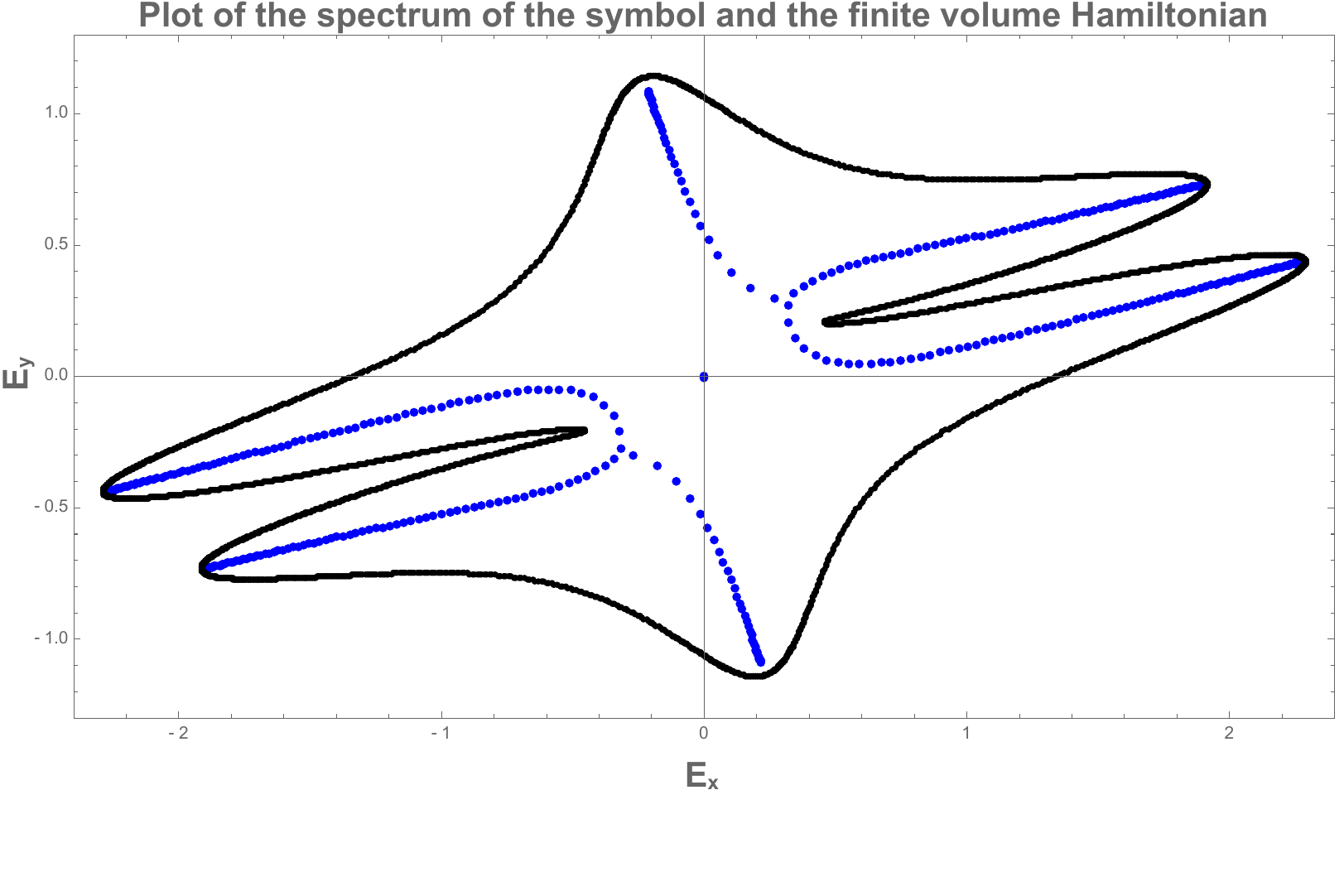}
\hspace{.1cm}
\includegraphics[align=t,width=5.4cm,height=6.6cm]{GammaFSer2updated.pdf}
\hspace{.1cm}
\includegraphics[align=t,width=5.4cm,height=6.0cm]{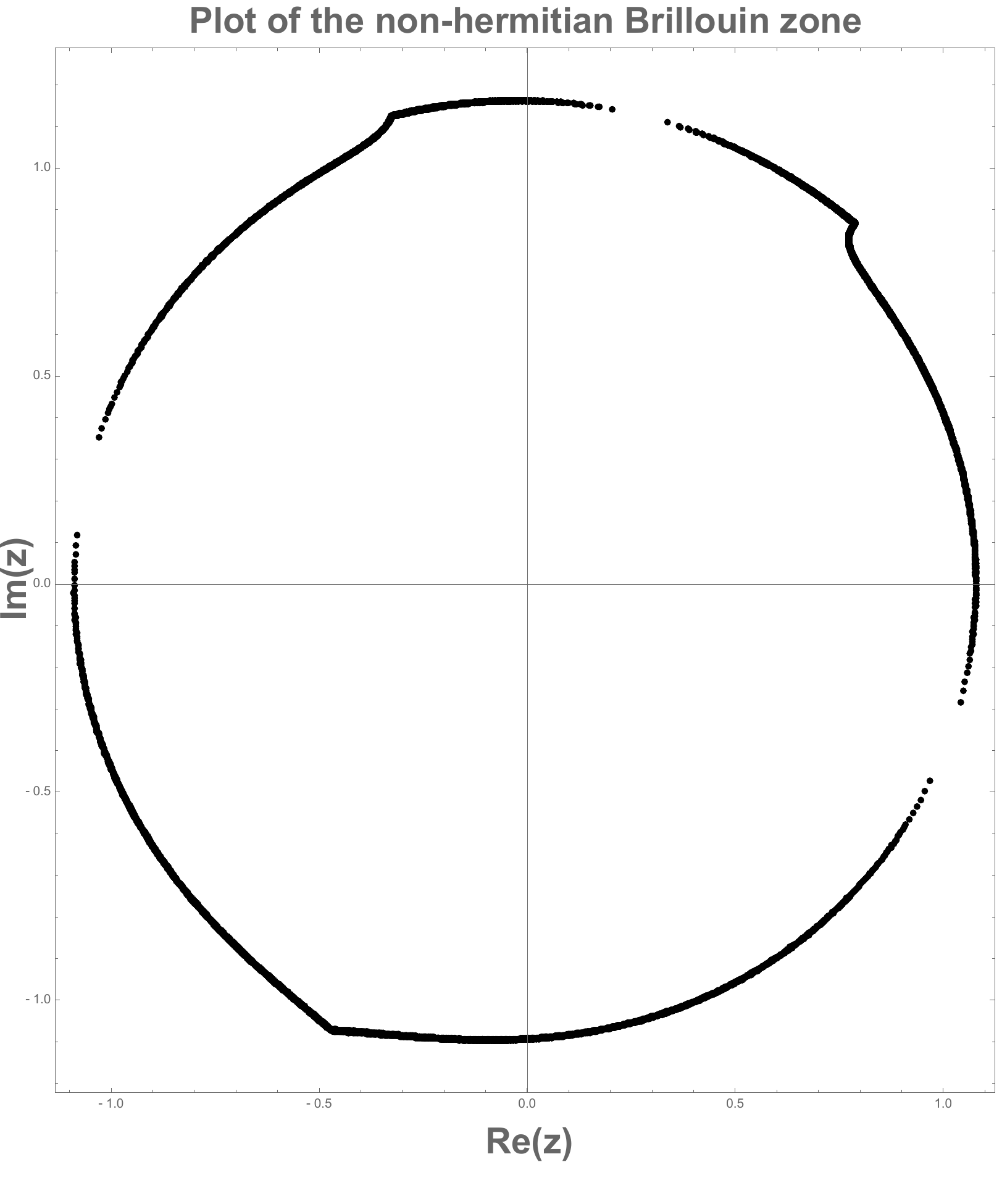}
\vspace{-1.0cm}
\caption{\sl 
Numerical analysis of the chiral Hamiltonian \eqref{eq-HamL2} with $v_+ = -0.1 - 0.5\,\imath$, $v_-= 0.7$, $t_+= 0.5 + \imath$, $t_-= 1=r_+$ and $r_-= 1.5 - 0.1\,\imath$. The left plot shows  the periodic spectrum (outer continuous curve) and the eigenvalues of the finite volume Hamiltonian of length $N=200$.  Note that there is a doubly degenerate (approximate) zero mode. The middle plot (same as in Figure~\ref{fig-Ser2Intro}) shows the  sets $\LimitSet$ and $\{E\in\LimitSet\,:\,z_L(E)=z_{L+1}(E)\} \subset \Ff\cap\LimitSet$  (method is described in the text). The right plot shows the non-hermitian Brillouin zone with holes stemming from numerical approximation and the abrupt changes of the eigenvalues of the transfer matrix along $\LimitSet$.
}
\label{fig-Ser2}
\end{figure}

Let us first illustrate these facts by looking at the middle plot in Figure~\ref{fig-HN}, hence corresponding to a model with five scalar diagonals. Clearly, there are three components $\CM\setminus\Sigma$ with non-vanishing winding number and whether it is positive or negative cannot be deduced from the plot which does not show the orientation of $k\mapsto\det(H(e^{\imath k})-E\,\one)$. The sign of the winding number in each component determines whether the skin states are left or right bound.  Moreover, there are two crosses of $\Sigma$ which also cross $\LimitSet$, and the corresponding eigenstates of $H_N$ near these crossings are plane wave like. As one changes the scaling, namely passes from $H$ to $H^s$, the curve $\LimitSet$ remains unchanged (and also the spectrum of $H^s_N$), but the curve $\Sigma^s=\spec(H^s)$ changes. In particular, the crosses of $\Sigma^s$ move along $\LimitSet$ as $s$ changes. Hence also  the left/right nature of the skin states changes as the cross moves along $\LimitSet$. Generically, eigenstates that are near $\Sigma$ have a less pronounced skin effect than those lying deep inside the components of $\CM\setminus\Sigma$ with non-vanishing winding number. This is not necessarily true though. For example, for the scalar Hatano-Nelson model (tridiagonal with scalar entries, see first plot in Figure~\ref{fig-HN}), the arguments in Section~\ref{sec-HNmodel} show that the middle two eigenvalues have a constant modulus on $\LimitSet$ so that the skin effect appears with a uniform rate throughout $\LimitSet$. Furthermore, for $H^{s_c}$ with critical scaling $s_c$, all states are plane-wave states of constant modulus. Further comments on the skin effect in Figures~\ref{fig-Ser2} to \ref{fig-Ser2bisbis} are given in Section~\ref{sec-TopoZero}.

\subsection{The generalized Brillouin zone}
\label{sec-Brillouin}

The generalized Brillouin zone (appearing in so-called non-Bloch band theory) is a popular object in the physics literature, {\it e.g.} \cite{YW,KD,YM,YZFH}. Even  though this work shows that it is not of relevance for the spectral analysis, let us define it as a mathematical object. Consider the map
\begin{equation}
\label{eq-BrillouinCurve}
\Upsilon=(\Upsilon_0,\Upsilon_1)\;:\;E\in\LimitSet\;\mapsto\;(z_L(E),z_{L+1}(E))\in\CM^2
\;,
\end{equation}
where the roots are chosen such that both $\Upsilon_0$ and $\Upsilon_1$ are continuous. Then the non-hermitian Brillouin zone is defined as
$$
\BM\;=\;\Upsilon_0(\LimitSet)\cup \Upsilon_1(\LimitSet)
\;.
$$
Numerically, this can readily be plotted from the data of $\LimitSet$. An example is given in Figure~\ref{fig-Ser2}.
In the hermitian case, $\LimitSet=\spec(H)$ and then $(z_L(E),z_{L+1}(E))\in\TM \times\TM $ for $E\in\LimitSet$. Hence the Brillouin zone is given by $\BM=\TM $, as usual. On the other hand, let us stress a difference with the hermitian case for which the transfer matrix is $\Ii$-unitary. This latter property implies that the eigenvalues of $\Tt^E$ lying on the unit circle have stability properties given by Krein theory (see \cite{SB} for a review in the context of transfer matrices). In particular, there can be open subsets of energies for which there are more than $2$ eigenvalues on $\TM $ (provided that $L>1$, of course). This is non-generic in the class of non-hermitian systems where there are typically only two eigenvalues of same modulus.

\subsection{Topological zero modes of finite-volume chiral Hamiltonians}
\label{sec-TopoZero}

This section considers a tridiagonal block Hamiltonian of the form \eqref{eq-HamGen} with $L=2$ and with symbol given by
\begin{equation}
\label{eq-HamL2}
H(z)
\;=\;
z^{-1}\,
\begin{pmatrix}
0 & r_+ \\ r_-& 0
\end{pmatrix}
\,+\,
\begin{pmatrix}
0 & v_+ \\ v_- & 0
\end{pmatrix}
\,+\,
z\,
\begin{pmatrix}
0 & t_+ \\ t_- & 0
\end{pmatrix}
\;,
\end{equation}
where $r_\pm,v_\pm,t_\pm\in\CM$ are parameters with $r_\pm \not= 0$ and $t_\pm\not=0$. The latter two conditions ensure that $R$ and $T$ are invertible so that Hypothesis A is satisfied. The model is called the non-hermitian SSH model and is widely studied in the physics literature, {\it e.g.} \cite{Lie,KD,BB}. Clearly $H(z)$ is off-diagonal and hence satisfies the chiral symmetry relation \eqref{eq-ChiralSym}. Therefore all the claims of Sections~\ref{sec-ChiralOutliersHalf} and \ref{sec-ZeroOutlier} hold for the Hamiltonian \eqref{eq-HamL2}. In particular, the off-diagonal entries define two scalar symbols $H_\pm(z)=r_\pm z^{-1}+ v_\pm+t_\pm z$ and one can, for any scaling parameter $s>0$, define two winding numbers $W_\pm(s)=\Wind^0(H^s_\pm)$ which appear in the criteria in Propositions~\ref{prop-ChiralKernel} and \ref{prop-ZeroCriterion} (as well as Theorem~\ref{thm-ChiralIntro}). Furthermore, the sets $\LimitSet$ and $\Outliers$ satisfy \eqref{eq-GammaLambdaSym}, namely they are reflection symmetric.

\begin{figure}
\centering
\includegraphics[width=5.3cm,height=6.4cm]{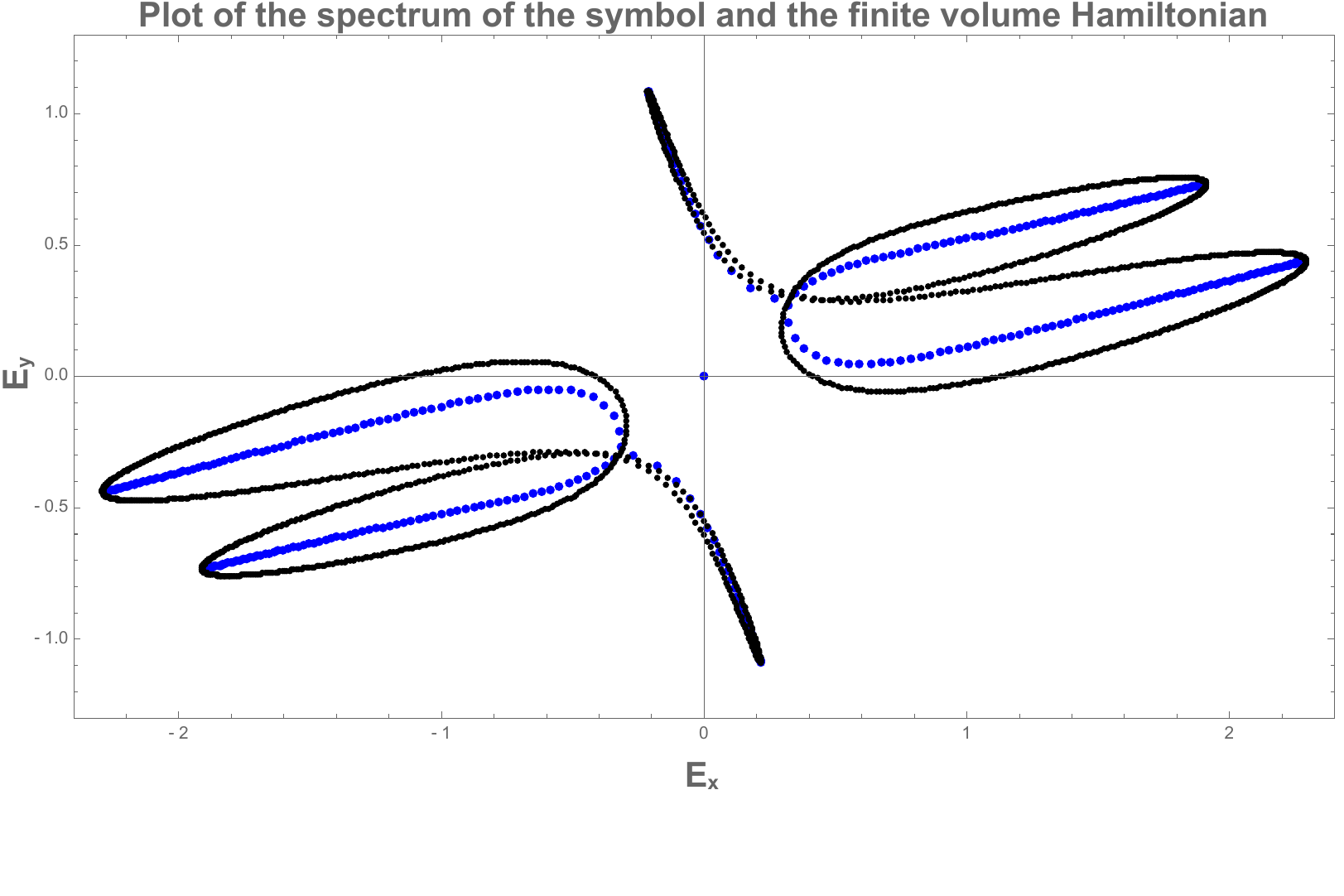}
\hspace{.1cm}
\includegraphics[width=5.3cm,height=6.4cm]{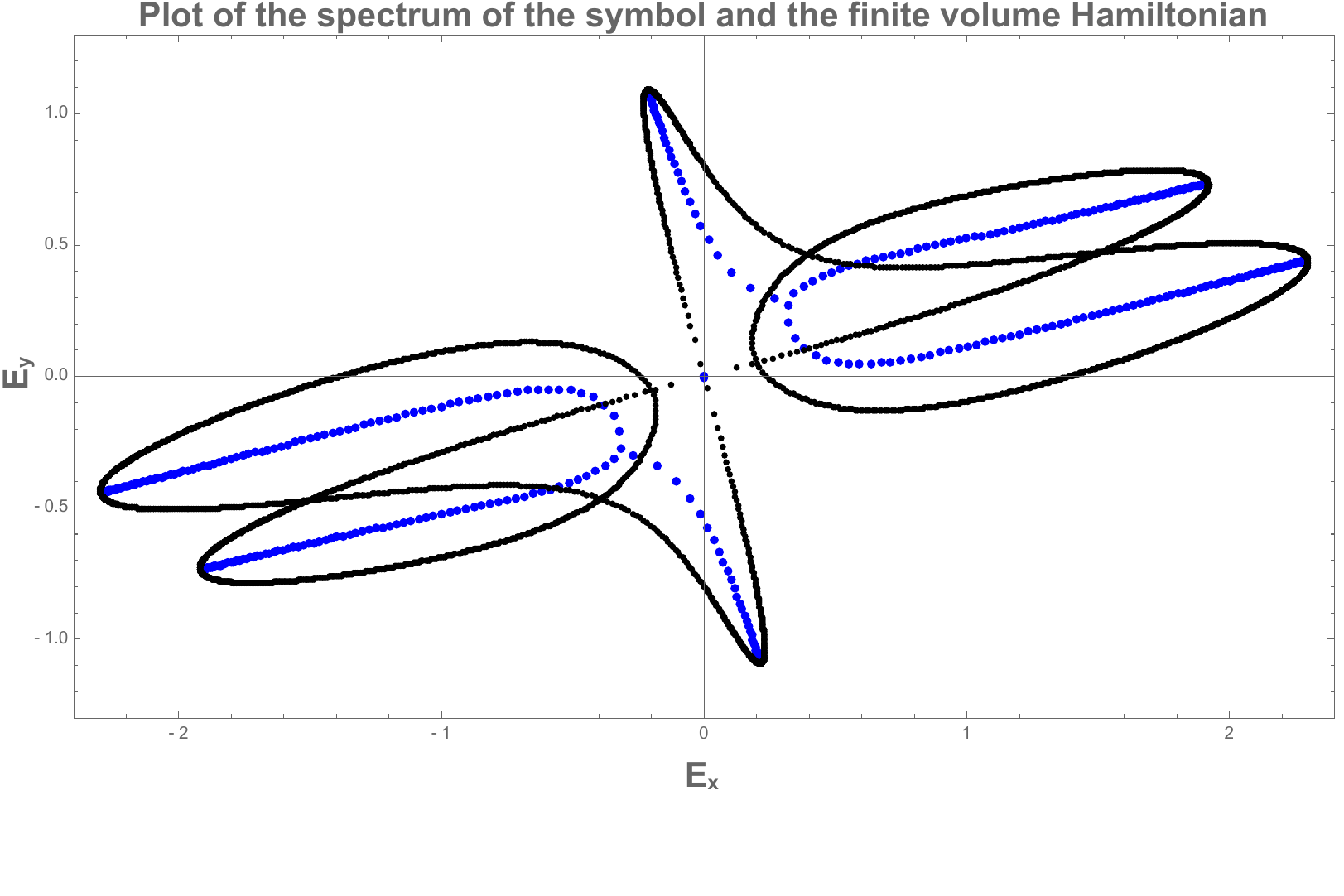}
\hspace{.1cm}
\includegraphics[width=5.3cm,height=6.4cm]{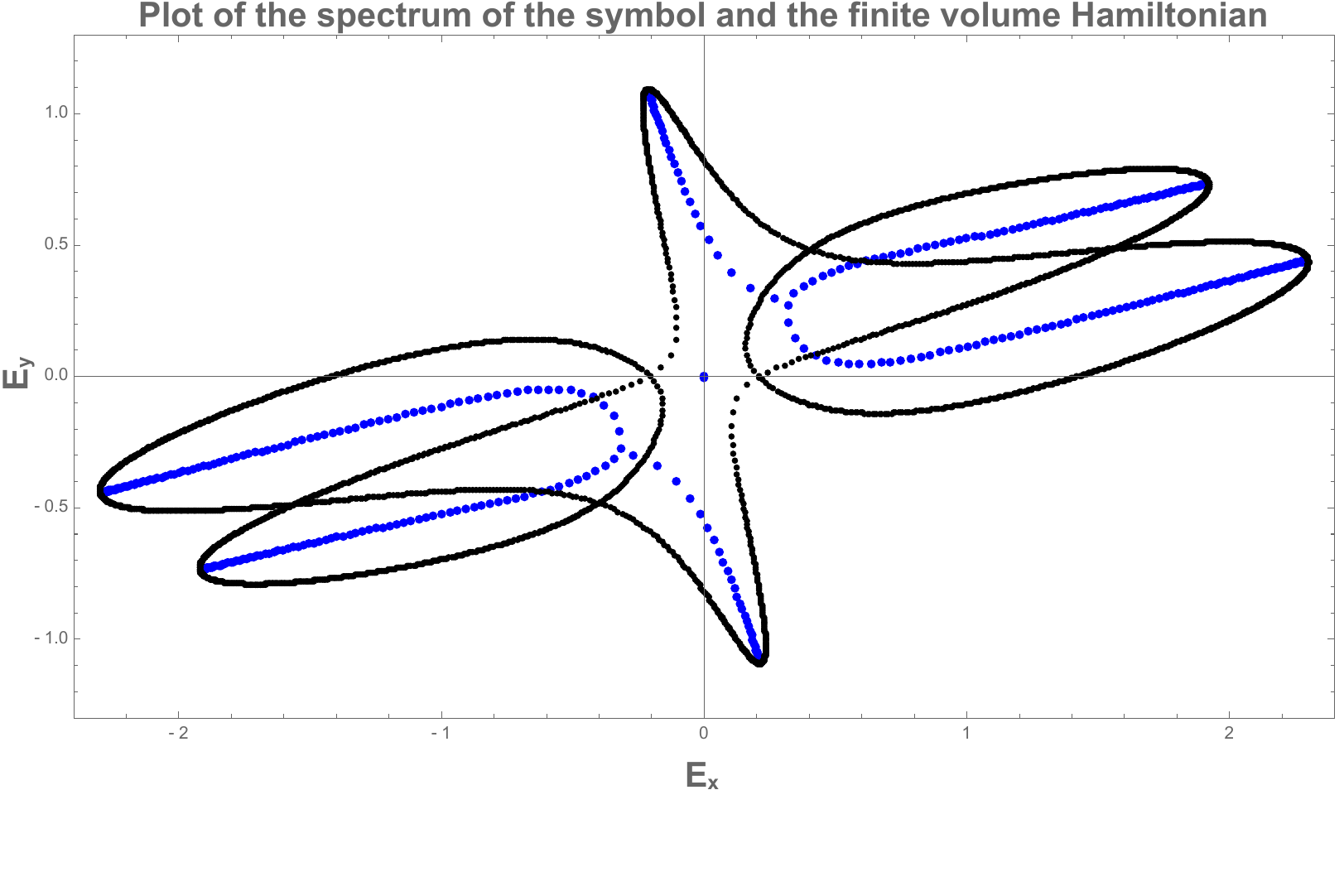}
\vspace{-1.0cm}
\caption{\sl Same as first plot in {\rm Figure~\ref{fig-Ser2}}, but with scaling $s=1.172$, $s=1.214$ and $s=1.22$. 
}
\label{fig-Ser2bis}
\end{figure}

\vspace{.2cm}

For a numerical illustration, particular values of the parameters are chosen as given in the capture of Figure~\ref{fig-Ser2}. The first plot shows the periodic spectrum $\Sigma=\spec(H)$ as well as the spectrum of a finite-volume approximation $H_N$ for $N=200$. Again one sees that these spectra differ considerably. However, the finite volume spectrum is well-approximated by the set $\LimitSet$ which can be computed numerically from merely the transfer matrix by the elementary method described at the end of Section~\ref{sec-HNmodel} (compute the spectrum of $\Tt^E$ for $E$ on some grid an check whether the middle two eigenvalues have approximately same modulus). The outcome is shown in the second plot of Figure~\ref{fig-Ser2}. These and all other plots in this paper were produced with Mathematica. The eigenvalue computations (as in the first plot of Figure~\ref{fig-Ser2}) seem to become numerically unstable for $N$ larger than about $300$ (the plot of $\LimitSet$, on the other hand, is very robust). From the numerical data of $\LimitSet$, one can also readily plot the generalized Brillouin zone simply by implementing \eqref{eq-BrillouinCurve}. This is shown in the third plot in Figure~\ref{fig-Ser2}. The second plot also shows that the set $\{E\in\LimitSet\,:\,z_L(E)=z_{L+1}(E)\} \subset \Ff\cap\LimitSet$ consists of merely six points given by the six ends of the graph. The local analysis of these endpoints is sketched in Remark~\ref{rem-EndJordan} where it is argued that they are linked to the appearance of Jordan blocks in the transfer matrix. In this context let us state that the branch points of the three legs are approximately at $E_\pm=\pm(0.33 + 0.30\,\imath)$ and are {\it not} in $\Ff$, but that numerics clearly show that there are rather three different eigenvalues of same modulus at these two complex energies (it can be argued that this is a generic scenario, but an analysis of such branch points involves the modulus of the eigenvalues and thus leaves the field of algebraic geometry, other than for the end points). 

\vspace{.2cm}

\begin{figure}
\centering
\includegraphics[width=5.3cm,height=6.4cm]{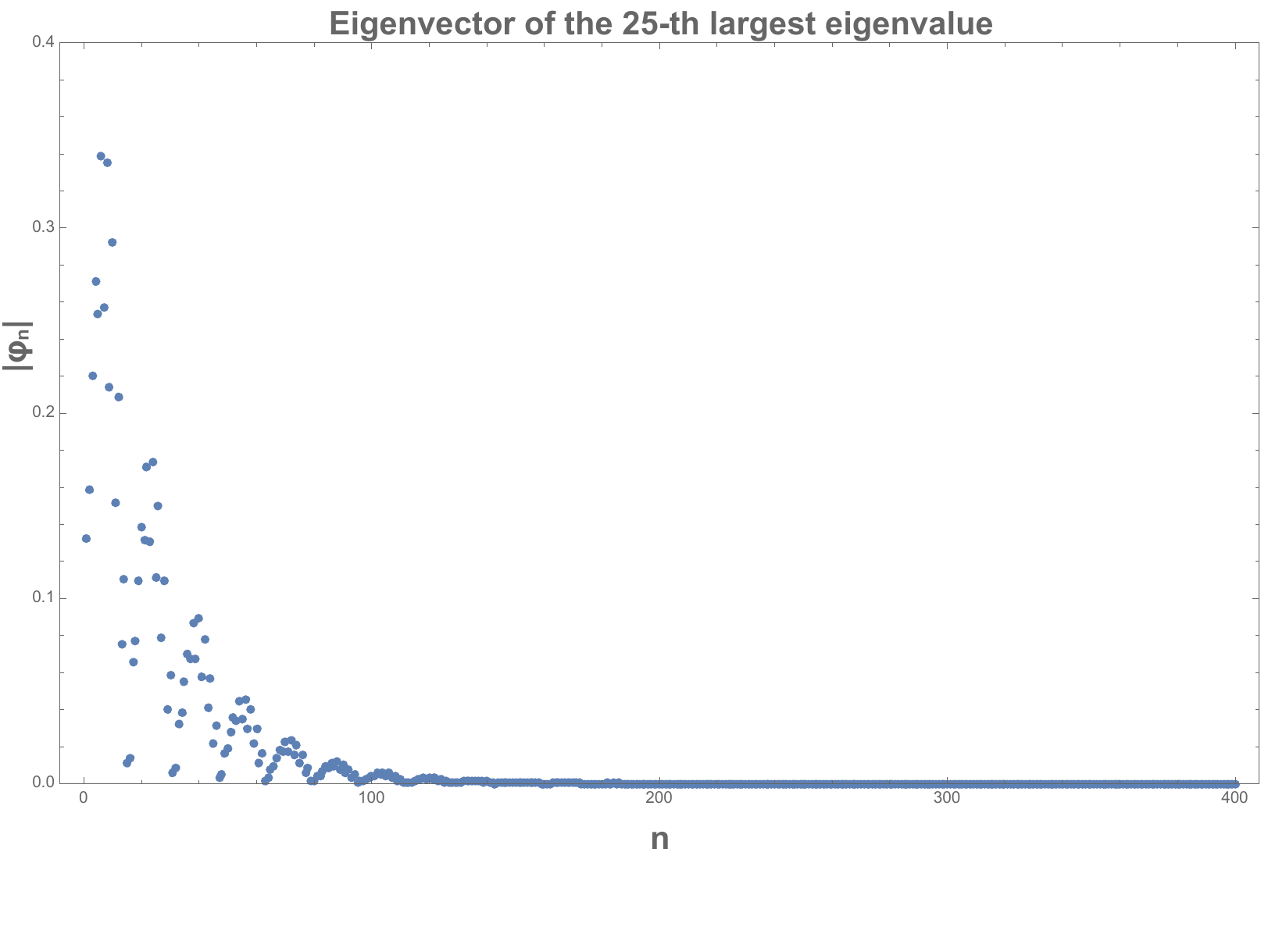}
\hspace{.1cm}
\includegraphics[width=5.3cm,height=6.4cm]{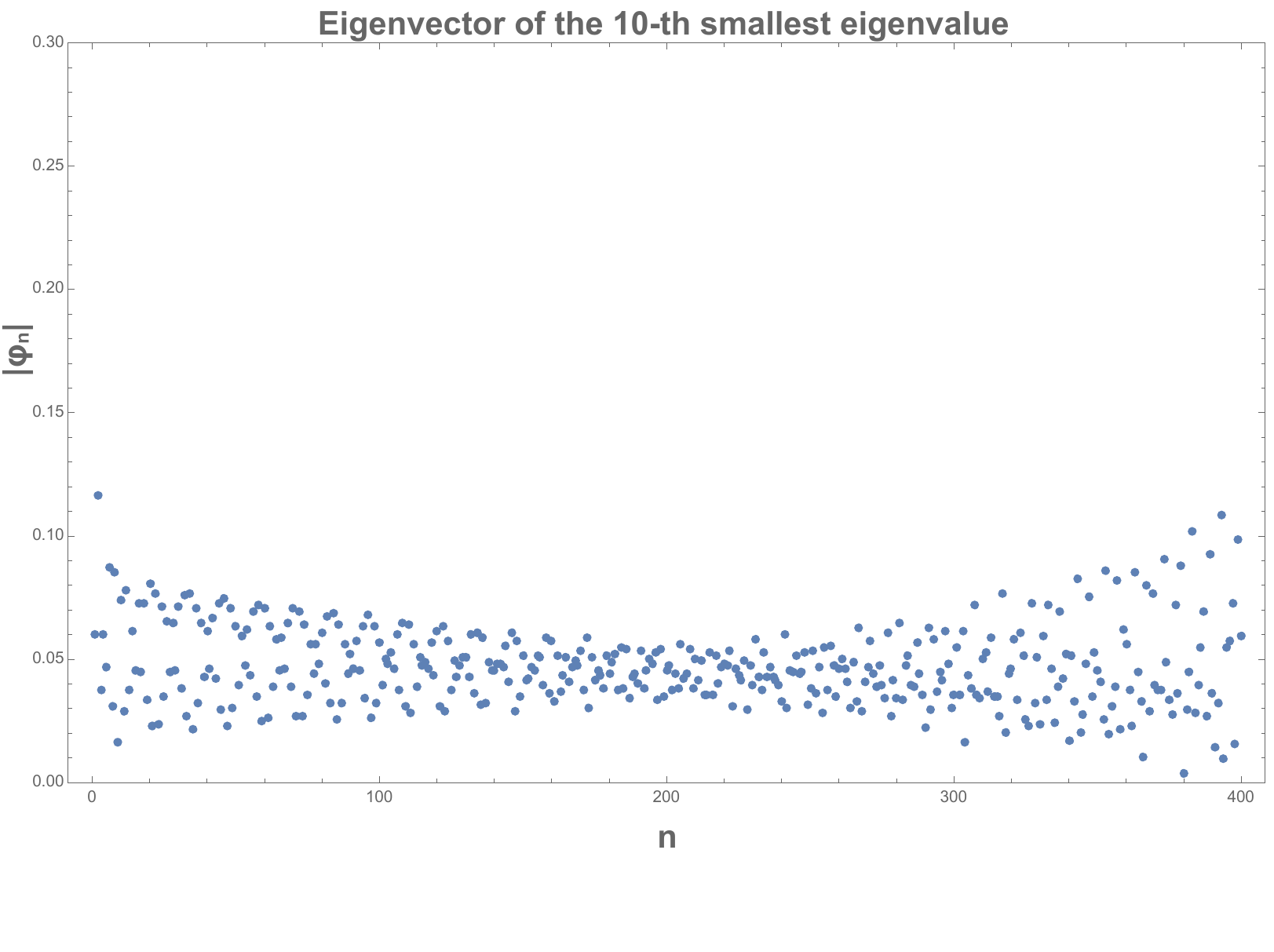}
\hspace{.1cm}
\includegraphics[width=5.3cm,height=6.4cm]{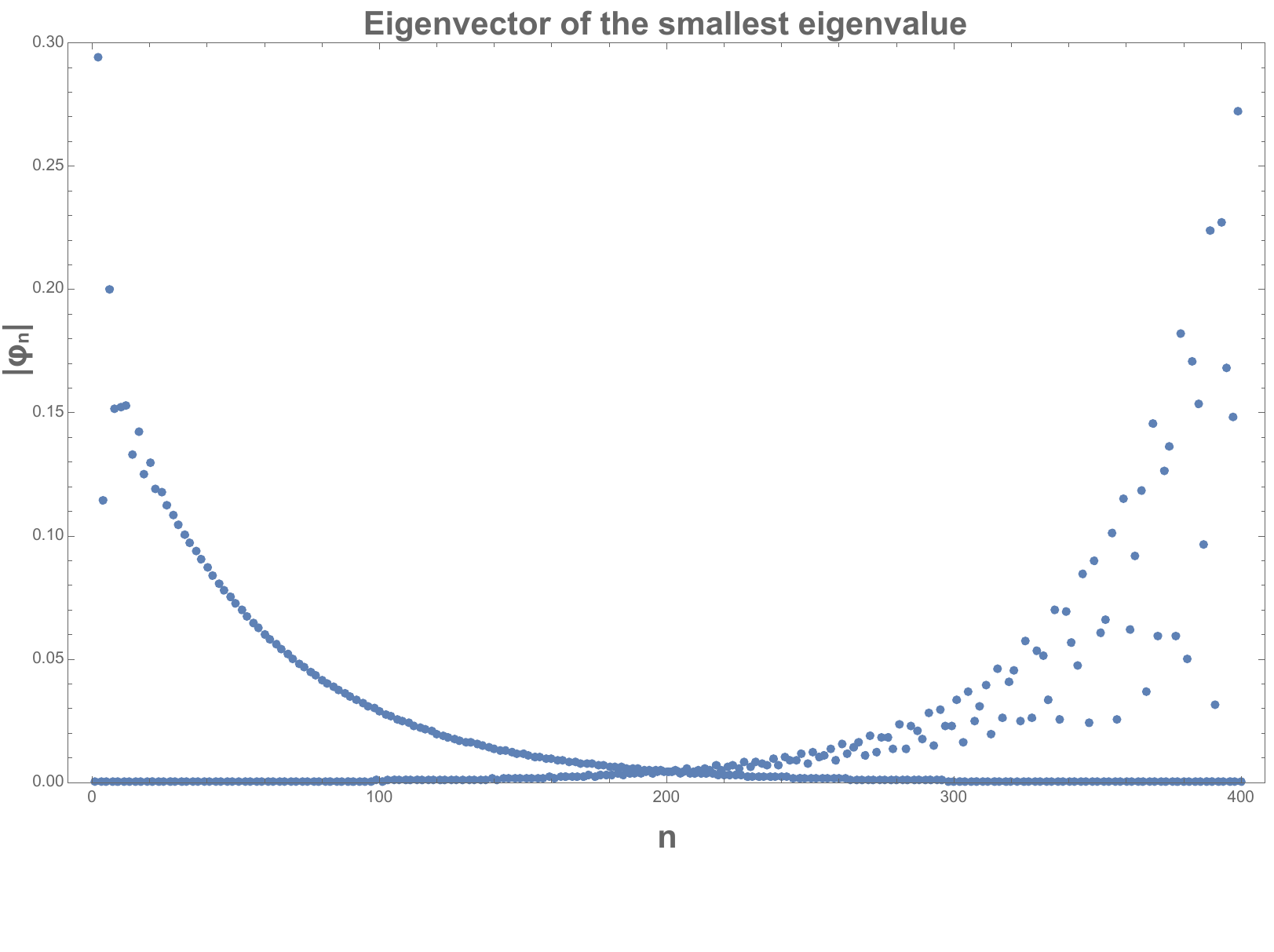}
\vspace{-1.0cm}
\caption{\sl Same model as {\rm Figure~\ref{fig-Ser2}}, but with scaling parameter $s=1.172$ so that $\LimitSet$ is given by first plot in {\rm Figure~\ref{fig-Ser2bis}}. The plots show the absolute value of the normalized eigenvector $\varphi$ for the 25th largest and the 10th smallest eigenvalues (at $-2.11-0.39\, \imath$ and $0.27 +0.30 \,\imath$ respectively) as well as the approximate zero modes. }
\label{fig-Ser2bisbis}
\end{figure}

Next let us consider varying the scaling parameter $s>0$. Figure~\ref{fig-Ser2bis} shows three plots of the scaled periodic spectrum $\Sigma^s=\spec(H^s)$, with a particular focus on a transition point around $s=1.214$. One clearly sees that the finite volume spectra and thus also the set $\LimitSet$ lie in regions with non-vanishing winding number for all $s$, except for the zero mode that will be discussed below. The eigenstates exhibit a skin effect, and as described in Section~\ref{sec-Skin} the sign of the winding number determines whether they are localized on the left or right side of the system. Plots of such states are unspectacular because they are are typically localized only about $50$ off the boundary. If $E$ is chosen somewhat close to $\Sigma^s$ though the states may leak over a $100$ sites into the sample (see first plot in Figure~\ref{fig-Ser2bisbis}). Furthermore, for states on $\Sigma^s$ or even energetic regions with vanishing winding number, the states may be extended throughout the system (as in the second plot in Figure~\ref{fig-Ser2bisbis}). 

\vspace{.2cm}

Finally let us address the zero modes. First of all, let us stress that in the first plot of Figure~\ref{fig-Ser2} as well as all plots in Figure~\ref{fig-Ser2bis} there are $2$ eigenvalues close to the origin. A simple state exactly at the origin is not compatible with the spectral symmetry $\spec(H_N)=-\spec(H_N)$ which follows directly from the chiral symmetry, because the Hilbert space is even dimensional and each eigenvalue $E$ with eigenstate $\phi$ has a partner $-E$ with eigenstate $K\phi$ (on the other hand, for the half-sided operators $H^R$ and $H^L$ such exact zero modes of multiplicity $1$ are allowed). Whether such a pair of approximate zero modes of $H_N$ exists can be predicted using Proposition~\ref{prop-ZeroCriterion}. One first determines a scaling parameter $s$ such that $\Wind^0(H^s)=0$ (one can always find such a scaling parameter uncovering the $0$). One choice is here given by {$\sca =1.172$}, see the first plot of Figure~\ref{fig-Ser2bis}. Then one computes numerically the two winding numbers and finds $W_\pm(s)=\Wind^0(H_\pm)=\pm 1$. In conclusion, $0\in\Outliers$. Consequently, Theorem~\ref{thm-LimitSpec} implies that $H^s_N$ has to have states close to $0$ for all $s>0$. The associated eigenstates $\phi$ and $K\phi$ only differ by signs, and hence their absolute value has the same profile. If $\Wind^0(H)\not=0$ as in the Figure~\ref{fig-Ser2} and the third plot in Figure~\ref{fig-Ser2bis}, the sign of the winding number $\Wind^0(H)$ determines again whether the zero mode is a left or right skin state. On the other hand, if $\Wind^0(H)=0$ as for $s =1.172$ in the first plot in Figure~\ref{fig-Ser2bis}, the transfer matrix $\Tt^E$ has $L$ eigenvalues of modulus smaller than $1$ and $L$ of modulus larger than $1$. Decomposing the eigenvector $\phi$ into one vector in the decaying and another in the expanding subspace, the relative proportion then determines skin nature of the approximate zero modes. Typically one part dominates and the zero mode is a skin state, but in the third plot in Figure~\ref{fig-Ser2bisbis} the scaling $s =1.172$ was chosen such that these zero modes are fairly balanced between left and right edge. 





\end{document}